\documentclass[a4paper,12pt]{article}
\usepackage[dvips]{graphics,graphicx}
\usepackage{calc,pstcol,multido,pst-node,amsgen,amsmath,amsfonts,amssymb,bm}
\usepackage{rotating}
\usepackage{tikz}
\usepackage{natbib}

\usetikzlibrary{positioning,shapes,arrows.meta}

\newcommand{\ci}{\perp\!\!\!\perp}
\newcommand{\Var}{ {\rm Var} }
\newcommand{\Cov}{ {\rm Cov} }

\newcommand{\thetahat}{ \hat{\theta}_1 }
\newcommand{\thetahattwo}{ \hat{\theta}_{{\rm CLW}2} }
\newcommand{\thetahatthr}{ \hat{\theta}_2 }
\newcommand{\tmleone}{ \hat{\theta}_{\rm TMLE1} }
\newcommand{\tmletwo}{ \hat{\theta}_{\rm TMLE2} }

\newcommand{\tauhat}{ \hat{\tau} }
\newcommand{\epsilonhat}{ \hat{\epsilon} }

\newcommand{\piBhat}{ \hat{\pi}^B }
\newcommand{\mhat}{ \hat{m} }

\newcommand{\mhatstar}{ \hat{m}^* }
\newcommand{\mhatextk}{ \hat{m}_k }
\newcommand{\underpiB}{ \underline{\pi}^B }
\newcommand{\underm}{ \underline{m} }
\newcommand{\underpiBhat}{ \underline{\hat{\pi}}^B }
\newcommand{\undermhat}{ \underline{\hat{m}} }

\newcommand{\piBtrue}{ \pi^B_0 }
\newcommand{\mtrue}{ m_0 }
\newcommand{\curlF}{ \mathcal{F}_J }
\newcommand{\curlG}{ \mathcal{G}_J }
\newcommand{\curlS}{ \mathcal{S} }
\newcommand{\convp}{ \xrightarrow{p} }
\newcommand{\Ybar}{ \bar{Y} }
\newcommand{\piBrate}{ c_{\pi} }
\newcommand{\mrate}{ c_m }
\newcommand{\chebysheveps}{ \Delta }
\newcommand{\truncM}{ \delta }
\newcommand{\Uthr}{ U^{\dagger} }

\begin{document}
\parindent=0pt
\parskip=5pt

\begin{center}
  {\Large \bf   Debiased machine learning for combining probability and nonprobability survey data}
  \vspace{.2cm}

  {\large \bf   Shaun R.\ Seaman$^1$, Tommy Nyberg$^{1,2}$, Andrew Copas$^3$ and Anne M.\ Presanis$^1$}

\end{center}

  $^1$ MRC Biostatistics Unit, University of Cambridge, East Forvie Building, University Forvie Site, Robinson Way, Cambridge, United Kingdom.

$^2$ Pathogen Dynamics Unit, Department of Genetics, University of Cambridge, Cambridge, UK.

$^3$ Institute for Global Health, University College London, London, UK.

email: shaun.seaman@mrc-bsu.cam.ac.uk

\section*{Abstract}

We consider estimation of the finite population mean $\Ybar$ of an outcome $Y$ using data from a nonprobability sample and auxiliary information from a probability sample.
Existing double robust (DR) estimators of this mean $\Ybar$ require estimation of two nuisance functions: the conditional probability of selection into the nonprobability sample given covariates $X$ that are observed in both samples, and the conditional expectation of $Y$ given $X$.
These nuisance functions can be estimated using parametric models, but the resulting estimator of $\Ybar$ will typically be biased if both models are misspecified.
It would therefore be advantageous to be able to use more flexible data-adaptive / machine-learning estimators of the nuisance functions.
We develop a general framework for the valid use of DR estimators of $\Ybar$ when the design of the probability sample uses sampling without replacement at the first stage and data-adaptive / machine-learning estimators are used for the nuisance functions.
We prove that several DR estimators of $\Ybar$, including targeted maximum likelihood estimators, are asymptotically normally distributed when the estimators of the nuisance functions converge faster than the $n^{1/4}$ rate and cross-fitting is used.
We present a simulation study that demonstrates good performance of these DR estimators compared to the corresponding DR estimators that rely on at least one correctly specified parametric model, and describe an application to sexual and reproductive health survey data.

\vspace{.1cm}

{\bf Keywords}: design-based inference; double robust; finite population; machine learning; sample surveys; targeted maximum likelihood.

\hspace{1cm}

\newpage
\setcounter{page}{1}

\section{Introduction}

Probability sampling methods are the gold standard for surveys.
They are designed to yield samples representative of the finite population from which they are drawn.
Nevertheless, there is considerable interest in using data from nonprobability samples, e.g.\ volunteer surveys, due to the low cost of collecting such data.
Such samples may, however, be unrepresentative, and so yield a biased estimate of the finite population mean $\Ybar$ of a variable $Y$ of interest.

Several methods have been proposed for using auxiliary covariate data from a probability sample (Sample A) to address nonrepresentation of a nonprobability sample (Sample B).
These methods fall into three classes, all of which use data on $Y$ observed in Sample B and covariates $X$ observed in both samples.
The first class, which includes inverse probability weighting (IPW) methods, involve estimating weights for the individuals in Sample B such that the weighted Sample B is representative of the population.
The second class, mass imputation methods, involve using the relation between $X$ and $Y$ in Sample B to impute the $Y$ values in Sample A.
The third class, double robust (DR) methods, combine IPW and mass imputation.
IPW methods require a consistent estimator of the conditional probability that an individual belongs to Sample B given $X$.
Mass imputation methods need a consistent estimator of the conditional expectation of $Y$ given $X$.
This conditional probability and expectation are called `nuisance' functions, because they are not of direct interest.
DR methods use estimators of both nuisance functions but only need one of these to be consistent.

A parametric model could be used for each of the nuisance functions, but the resulting estimator would typically be inconsistent if that model --- or, in the case of DR estimators, both models --- were misspecified.
There is therefore interest in using more flexible `data-adaptive' or `machine-learning' estimators of the nuisance functions.
Several researchers have done this.
\citet{FerriGarcia2020}, \citet{FerriGarcia2024} and \citet{Rueda2023,Rueda2024} use IPW and estimate the nuisance function using k-nearest neighbours, random forest or generalised boosting models (XGBoost).
\citet{CastroMartin2020} use IPW or mass imputation with random forest or XGBoost.
\citet{FerriGarcia2022} use IPW with classification and regression trees (CART).
\citet{CastroMartin2021} use IPW, mass imputation or a DR estimator with XGBoost.
\citet{Cobo2025} use a DR estimator with XGBoost, k-nearest neighbours or neural nets.
Some of these researchers apply kernel smoothing to the estimated inverse probability weights to reduce their variance.
\citet{WangGraubard2020,WangGraubard2022} developed a theoretical justification for this smoothing when logistic regression is used, but not when data-adaptive methods are used.
\citet{ChenYangKim2022} use mass imputation with kernel smoothing and show the resulting estimator of $\Ybar$ is asymptotically normal, but they note this method is subject to the curse of dimensionality.
\citet{Chlebicki2024} use mass imputation with predictive mean matching.
This provides some limited robustness to misspecification of a parametric model for the nuisance conditional expectation function, because it requires only that the true conditional expectation be a monotone function of the limiting value of the estimated conditional expectation.
Chlebicki et al.\ suggest that a data-adaptive method could be used to estimate this nuisance function but do not prove that the resulting estimator of $\Ybar$ would be $n^{1/2}$-consistent.

A closely related problem to estimation of $\Ybar$ for a finite population by combining $X$ data from a probability sample with $(X, Y)$ data from a nonprobability sample is estimation of $E(Y)$ for an infinite population when data $(X, Y)$ are independent and identically distributed (iid), $X$ is fully observed and $Y$ is missing at random given $X$.
IPW, regression imputation and DR methods exist for this problem.
The first two require consistent estimators of, respectively, the conditional probability that $Y$ is observed given $X$ and the conditional expectation of $Y$ given $X$, while DR methods need only one of these nuisance function estimators to be consistent.
For IPW and imputation, using data-adaptive methods to estimate the nuisance function can cause problems.
Data-adaptive methods typically yield nuisance function estimators that have slow convergence, which is inherited by the IPW or imputation estimator of $E(Y)$, causing it to have large finite-sample bias; confidence intervals can also have poor coverage \citep{Seaman2018}.
The DR approach (`debiased machine learning') can overcome this issue, because DR estimators do not inherit the slow convergence \citep{Ellul2025,Naimi2023,Hines2022,Chernozhukov2018}.
Given the similarity of the problem of estimating $E(Y)$ for an infinite population to that of estimating $\Ybar$ for a finite population, there is reason to be concerned that the same issue may well also apply to the latter problem.
In this article, we study the use of a DR estimator of $\Ybar$ when combining data from a probability sample and a nonprobability sample and using data-adaptive estimators of nuisance functions.
We aim to provide a theoretical justification for the validity of this approach.
An important difference from the problem of estimating $E(Y)$ with iid data is that when, as is common, the probability sample is obtained using sampling without replacement (SWOR), Sample A data are not independent.

We build on the work of \citet{ChenLiWu2020} (henceforth, CLW).
CLW developed a DR estimator for $\Ybar$ using parametric nuisance models.
\citet{YangKimSong2020} extended this to allow variable selection by smoothly clipped absolute deviation.
We generalise the work of CLW (and Yang et al.) to allow general data-adaptive estimation of the nuisance functions.
The article is structured thus.
Section~\ref{sect:dgm} describes our assumed data-generating mechanism and design-model-based inference framework.
Section~\ref{sect:vonMises} describes a DR estimator for $\Ybar$, its von Mises expansion, and cross-fitting.
In Section~\ref{sect:equal.probs} we consider the case where clusters (`primary sampling units') are sampled using simple random sampling without replacement (SRSWOR) and then individuals within clusters are sampled using an arbitrary sampling design that obeys the independence and invariance conditions \citep{Sarndal1992}.
Section~\ref{sect:unequal.probs} considers the more general case where cluster sampling probabilities vary.
Section~\ref{sect:DR2} describes a ratio-type DR estimator that does not need the population size to be known and which can be more efficient, and alternative targeted maximum likelihood estimators (TMLEs).
Possible data- adaptive estimators of nuisance functions are discussed in Section~\ref{sect:estimating.nuisance}.
Section~\ref{sect:simstudy} describes a simulation study that compares DR methods using data-adaptive estimators of nuisance functions with those using parametric models, and Section~\ref{sect:Natsal} describes an application to data from the third British National Survey of Sexual Attitudes and Lifestyles (Natsal-3) probability survey and four contemporaneous nonprobability surveys.
We end with a discussion.

\section{Assumed data-generating mechanism}

\label{sect:dgm}

Our assumed data-generating mechanism consists of a superpopulation model for the finite population and models for drawing Samples A and B from this population.
The superpopulation model generates a finite population of $J$ clusters thus.
Cluster sizes $N_1, \ldots, N_J$ are independently generated from distribution $f(n)$.
For each cluster $j$ ($j=1, \ldots, J$) independently, a $N_j \times p$ covariate matrix $(X_{j1}, \ldots, X_{j N_j})$ is generated conditionally on $N_j$ from distribution $f(x_1, \ldots, x_{N} \mid N=N_j)$.
For each individual $i=1, \ldots, N_j$ in each cluster $j=1, \ldots, J$, outcome $Y_{ji}$ is independently generated from distribution $f(y \mid X=X_{ji})$ with expectation $\mtrue (X_{ji}) = E(Y \mid X=X_{ji})$.
Let $\curlF = (N_1, \ldots, N_J, X_{11}, \ldots$, $X_{1N_1}, \ldots, X_{J1}, \ldots, X_{JN_J})$.
Our goal is to estimate $\Ybar = \sum_{j=1}^J \sum_{i=1}^{N_j} Y_{ji} \left/ \sum_{j=1}^J N_j \right.$, the population mean of $Y$.

As is usual in model-design-based inference, we shall consider repeated-sampling properties of an estimator of $\Ybar$ conditional on $\curlF$ \citep{ChenLiWu2020,Molina2001}.
Note that we have assumed the data-generating mechanism for $\curlF$ described above only for the asymptotics; this mechanism describes how the finite population grows as $J \rightarrow \infty$. 

The model for drawing Sample A is as follows.
Variable $R^A_{ji}$ will be a binary indicator that individual $i$ in cluster $j$ ($i=1, \ldots, N_j; \; j=1, \ldots, J$) is included in Sample A.
First, a sample of $M$ ($M<J$) clusters is drawn without replacement from the $J$ clusters, with the probability that cluster $j$ is included being proportional to $h(N_j, X_{j1}, \ldots, X_{jN_j})$ for some function $h$ of $N_j$ and $X_{j1}, \ldots, X_{jN_i}$.
An example would be $h(N_j, X_{j1}, \ldots, X_{jN_j}) = N_j$.
Let $R^C_j = 1$ if cluster $j$ is drawn; $R^C_j = 0$ otherwise.
If $R^C_j=0$, then $R^A_{ji}=0$ ($i=1, \ldots, N_j$).
If $R^C_j=1$, then $(R^A_{j1}, \ldots, R^A_{jN_j})$ is drawn according to some (possibly multistage) SWOR design that can depend on $N_j$ and $(X_{j1}, \ldots, X_{jN_j})$ but which, given these and $R^C_j = 1$, does not depend on $(Y_{j1}, \ldots, Y_{jN_j})$, $\{ (N_k, X_{k1}, \ldots, X_{kN_k}, Y_{k1}, \ldots, Y_{kN_k}): \; k=1, \ldots, j-1, j+1, \ldots J \}$ or $(R^C_1, \ldots, R^C_{j-1}, R^C_{j+1}, \ldots, R^C_J$).
This sampling of $(R^A_{j1}, \ldots, R^A_{jN_j})$ is done independently of sampling of $(R^A_{k1}, \ldots, R^A_{kN_k})$ within any other cluster $k$ with $R^C_k=1$.
Thus, we are assuming independence and invariance (see pp134--135 of~\citet{Sarndal1992}) for the sampling design.
Let $\pi^C_j = P(R^C_j = 1 \mid \curlF)$ and $\pi^{A \mid C}_{ji} = P(R^A_{ji} = 1 \mid R^C_j = 1, N_j, X_{j1}, \ldots, X_{jN_j})$.
Let $\pi^A_{ji} = \pi^C_j \times \pi^{A \mid C}_{ji}$ denote the first-order inclusion probability for individual $i$ in cluster $j$.

The model for drawing Sample B is as follows.
Let $R^B_{ji}$ be a binary indicator that individual $i$ in cluster $j$ is in Sample B.
Let $\piBtrue (X)$ be a function of $X$.
Given $\curlF$, each $R^B_{ji}$ is sampled from Bernoulli distribution with probability parameter $\piBtrue (X_{ji})$ independently of the other $R^B_{ji}$'s, $\{ (Y_{ji}, R^A_{ji}): j=1, \ldots, J; i=1, \ldots, N_j \}$ and $R^C_1, \ldots, R^C_J$.

Now relabel the $n = n (\curlF) = \sum_{j=1}^J N_j$ individuals in the population so that, for each $j=1, \ldots, J$, the $N_j$ individuals in cluster $j$ are labelled as individuals $\sum_{k=1}^{j-1} N_k + 1, \sum_{k=1}^{j-1} N_k + 2, \ldots, \sum_{k=1}^j N_k$.
Let $D_i$ ($1 \leq D_i \leq J$) denote the index of the cluster to which individual $i$ belongs.
We can now write the population mean as $\Ybar = n^{-1} \sum_{i=1}^n Y_i$.

The observed data are the $X$ and $\pi^A$ values of individuals in Sample A, the $X$ and $Y$ values of individuals in Sample B, and the index of the cluster to which each of these individuals belongs, i.e.\
$
\{ (R^A_i, R^A_i X_i, R^A_i \pi^A_i, R^A_i D_i, R^B_i, R^B_i X_i, R^B Y_i, R^B_i D_i): \; i=1, \ldots, n \}.
$
Note that our notation suggests we know whether an individual in Sample A also appears in Sample B, and vice versa.
In fact, this is unnecessary; this notation is used only for convenience.
We shall though need to know, for each individual in Sample A and each in Sample B, whether they belong to the same cluster; this is for the cross-fitting procedure (see Section~\ref{sect:vonMises}).

We shall consider asymptotic repeated-sampling distributions of several DR estimators of $\Ybar$ conditional on $\curlF$ in an asymptotic framework in which $J \rightarrow \infty$ and $M \rightarrow \infty$ with $M / J$ converging to a constant.
Thus, we do not need to distinguish between $M \rightarrow \infty$ and $J \rightarrow \infty$.
The models for drawing Samples A and B from the population do not change as $M \rightarrow \infty$.

For simplicity, we have assumed Sample A does not use stratification.
However, a simple extension allows for this.
The $J$ clusters are divided into strata, distributions $f(n)$ and $f(x_1, \ldots, x_N \mid N=N_j)$ may depend on the cluster's stratum, and $X$ includes the stratum index.
For Sample A, clusters are sampled without replacement in each stratum separately.

\section{DR estimator and its von Mises expansion}

\label{sect:vonMises}


It is instructive to consider the simpler context of estimating $E(Y)$ for an infinite population using a sample of $n$ iid individuals with $X$ observed for all $n$ individuals, $Y$ observed for only some of them, and these data being missing at random.
In this context, a DR estimator for $E(Y)$ is $\hat{\theta}_{\rm iid} (\hat{\pi}_{\rm iid}, \mhat_{\rm iid})$, where
$
\hat{\theta}_{\rm iid} (\pi_{\rm iid}, m_{\rm iid}) =
n^{-1} \sum_{i=1}^n m_{\rm iid} (X_i)
+
R_i \{ Y_i - m_{\rm iid} (X_i) \} / \pi_{\rm iid} ( X_i ),
$
$R$ is an indicator that $Y$ is observed, and $\hat{\pi}_{\rm iid} (X)$ and $\mhat_{\rm iid} (X)$ are estimators of $\pi_{\rm iid0} (X) = P(R=1 \mid X)$ and $m_{\rm iid0} (X) = E(Y \mid X)$.
\citet{Hines2022} describe how a von Mises expansion can be used to show that if $\hat{\pi}_{\rm iid}$ and $\mhat_{\rm iid}$ converge to $\pi_{\rm iid0}$ and $m_{\rm iid0}$ sufficiently fast as $n \rightarrow \infty$, then $\hat{\theta}_{\rm iid} (\hat{\pi}_{\rm iid}, \mhat_{\rm iid}) = \hat{\theta}_{\rm iid} (\pi_{\rm iid0}, m_{\rm iid0}) + o_p(n^{-1/2})$, and so $\hat{\theta}_{\rm iid} (\hat{\pi}_{\rm iid}, \mhat_{\rm iid})$ has the same asymptotic normal distribution as $\hat{\theta}_{\rm iid} (\pi_{\rm iid0}, m_{\rm iid0})$.
The required convergence rates of $\hat{\pi}_{\rm iid}$ and $\mhat_{\rm iid}$ are slower than the parametric $n^{1/2}$ rate, which allows use of data-adaptive estimators, provided that these satisfy the Donsker condition or cross-fitting is used.

  
We now obtain an analogous result for our sample survey setting, making use of cross-fitting.
Randomly partition the set of $n$ individuals into $K$ (e.g.\ $K=5$) subsets, called folds.
Sections~\ref{sect:equal.probs} and~\ref{sect:unequal.probs} describe how this is done.
Let $\curlS_k \subset \{1, \ldots, n\}$ denote the set of indices of individuals belonging to fold $k$ ($k=1, \ldots, K$), and let $n_k = | \curlS_k |$ denote the number of these individuals.
Let $\curlS_{-k}$ denote the indices of the individuals not in fold $k$ (so, $\curlS_k \cup \curlS_{-k} = \{ 1, \ldots, n \}$). 
Let $\mhat_k$ be an estimator of $\mtrue$ calculated using only the data $\{ (X_i, Y_i): R^B_i=1 \mbox{ and } i \in \curlS_{-k} \}$.
Let $\piBhat_k$ be an estimator of $\piBtrue$ obtained using only the data $\{ (R^A_i, R^B_i, R^A_i \pi^A_i, R^A_i X_i, R^B_i X_i): i \in \curlS_{-k} \}$.
We discuss possible estimators in Section~\ref{sect:estimating.nuisance}. 
Write $\underpiB = (\pi^B_1, \ldots, \pi^B_K)$, $\underm = (m_1, \ldots, m_K)$, $\underpiBhat = (\piBhat_1, \ldots, \piBhat_K)$ and $\undermhat = (\mhat_1, \ldots, \mhat_K)$.

We shall consider the DR estimator $\thetahat (\underpiBhat, \undermhat)$ of $\Ybar$, where
\[
\thetahat (\underpiB, \underm)
=
\frac{1}{n} \sum_{k=1}^K \sum_{i \in \curlS_k} U_i (\pi^B_k, m_k)
\]
\[
U_i (\pi^B, m)
=
\frac{ R^A_i }{ \pi^A_i } m(X_i)
+
\frac{ R^B_i }{ \pi^B(X_i) } \{ Y_i - m(X_i) \}.
\]
We shall use $\thetahat (\pi^B, m)$ as shorthand for $\thetahat \big( (\pi^B, \ldots, \pi^B), (m, \ldots, m) \big)$.
Note that $\thetahat$ is analogous to $\hat{\theta}_{\rm iid}$, but with the weighted mean of $m(X)$ in Sample A substituting for the unknown mean of $m(X)$.
We assume that $\piBhat_k$ and $\mhat_k$ satisfy the following condition.

{\it Condition C1: There exist $\piBrate > 0$ and $\mrate > 0$ such that $\piBrate + \mrate = 1$,
\begin{equation}
E_X \left[ \left\{ \frac{ \piBtrue (X) }{ \piBhat_k (X) } - 1 \right\}^2 \right]
= o_p ( M^{- \piBrate} ),
\label{eq:convrate.piBhat}
\end{equation}
\begin{equation}
  \mbox{and} \hspace{0.6cm}
  E_X \left[ \{ \mhatextk (X) - \mtrue (X) \}^2 \right]
= o_p ( M^{- \mrate} ).
\label{eq:convrate.mhat}
\end{equation}
}

Condition C1 allows $\piBhat_k$ and $\mhat_k$ to converge more slower than parametric estimators \citep{Hines2022}, e.g.\ it is satisfied when both $\piBhat_k$ and $\mhat_k$ converge faster than the $M^{1/4}$ rate.

We shall exploit the following von Mises expansion of $\thetahat (\underpiBhat, \undermhat) - \Ybar$:
\begin{eqnarray}
  \thetahat (\underpiBhat, \undermhat) - \Ybar
  & = &
  \frac{1}{n} \sum_{i=1}^n \{ U_i (\piBtrue, \mtrue) - \Ybar \}
  \nonumber \\
  && +
  \sum_{k=1}^K \frac{n_k}{n}
  \frac{1}{n_k} \sum_{i \in \curlS_k} \left[
    \{ U_i (\piBhat_k, \mhat_k) - U_i (\piBtrue, \mtrue) \}
    \right.
  \nonumber \\
  && \hspace{3cm} -
   \left. 
   E \left\{ U_i (\piBhat_k, \mhat_k) - U_i (\piBtrue, \mtrue) \mid \curlF, \curlS_k, \piBhat_k, \mhat_k \right\}
  \right]
  \nonumber \\
  && +
   \sum_{k=1}^K \frac{n_k}{n}
  \frac{1}{n_k} \sum_{i \in \curlS_k}
  E \left\{ U_i (\piBhat_k, \mhat_k) - U_i (\piBtrue, \mtrue) \mid \curlF, \curlS_k, \piBhat_k, \mhat_k \right\}.
  \label{eq:vonMises}
\end{eqnarray}

The first term on the right-hand side of equation~(\ref{eq:vonMises}) is just $\thetahat (\piBtrue, \mtrue) - \Ybar$.
The second and third terms will be referred to as, respectively, the `empirical process term' and `remainder term'.
We shall prove that these two terms are $o_p (M^{-1/2})$ if Condition C1 holds.
It then follows that $\thetahat (\underpiBhat, \undermhat) - \Ybar$ has the same asymptotic distribution as $\thetahat (\piBtrue, \mtrue) - \Ybar$, i.e.\
%
\begin{eqnarray}
  \sqrt{M} \{ \thetahat (\underpiBhat, \undermhat) - \Ybar \}
  & = &
  \sqrt{M} \{ \thetahat (\piBtrue, \mtrue) - \Ybar \} + o_p(1)
  \nonumber \\
  & = & 
  \sqrt{M} \frac{1}{n} \sum_{k=1}^K \sum_{i \in \curlS_k}
    \frac{ R^A_i }{ \pi^A_i } \mtrue (X_i)
    \nonumber \\
    && +
    \sqrt{M} \frac{1}{n} \sum_{k=1}^K \sum_{i \in \curlS_k}
    \frac{ R^B_i }{ \piBtrue (X_i) } \{ Y_i - \mtrue (X_i) \}
    - \sqrt{M} \Ybar + o_p(1).
  \nonumber \\
  && 
  \label{eq:DR1.expansion}
\end{eqnarray}

Formulae for the asymptotic variance of $\thetahat (\piBtrue, \mtrue) - \Ybar$ and an estimator of this variance are given by CLW and, in more detail, by \citet{Seaman2025} (specifically, Seaman et al.'s formulae for $\mu_{\rm KH1}$ when the model for $\piBtrue$ is correctly specified).
Subject to regularity conditions, $\thetahat (\underpiBhat, \undermhat) - \Ybar$ is asymptotically normally distributed \citep{ChauvetVallee2020}.

An important complication when proving the empirical process and remainder terms are $o_p (M^{-1/2})$ is that, in general, $\piBhat_k$ is not independent of the data in fold $k$.
This is because $\piBhat_k (X)$ is calculated using the $R^A_i$ values of individuals in $\curlS_{-k}$, and $\sum_{i \in \curlS_k} U_i (\piBhat_k, \mhat_k)$ is calculated using the $R^A_i$ values of individuals in $\curlS_k$, and $\{ R^A_i: \; i \in \curlS_{-k} \}$ and $\{ R^A_i: \; i \in \curlS_k \}$ are not independent when clusters are selected using SWOR.
For example, the number of clusters with $R^C_j=1$ in fold $k$ must equal $M$ minus the number of clusters with $R^C_j=1$ in the remaining folds.
We address this complication in Sections~\ref{sect:equal.probs} and~\ref{sect:unequal.probs}.

\section{Simple random sampling without replacement of clusters}

\label{sect:equal.probs}

First, consider the case of SRSWOR of clusters (so $\pi^C_1 = \ldots = \pi^C_J$).
As described in Section~\ref{sect:dgm}, individuals within sampled clusters are sampled using some (possibly multistage) sampling design.
Suppose $M$ and $J$ are both integer multiples of $K$; the case where $M$ and/or $J$ are not multiples of $K$ is considered as a special case in Section~\ref{sect:unequal.probs}.

Randomly partition the $M$ sampled clusters evenly into $K$ folds and, likewise, the $J-M$ unsampled clusters evenly into the $K$ folds, with all such partitions being equally probable.
Each fold contains $M/K$ sampled clusters and $(J-M)/K$ unsampled clusters.
%
This partitioning method ensures that the conditional distribution of $\{ R^A_i: \; i \in \curlS_k \}$ given $\curlF$, $\curlS_k$ and $\{ R^A_i: \; i \in \curlS_{-k} \}$ is the same as the distribution of $\{ R^A_i: \; i \in \curlS_k \}$ given $\curlF$ and $\curlS_k$, and corresponds to SRSWOR of $M/K$ clusters from the $J/K$ clusters in fold $k$ followed by the original second-stage sampling mechanism for individuals within clusters.

Recall that $Y_i$ is assumed to be conditionally independent of $R^A_{i'}$, $R^C_j$ and $Y_{i''}$ given $\curlF$ for all $i$, $i'$ and $j$ and all $i'' \neq i$, and that $R^B_i$ is assumed to be conditionally independent of $R^A_{i'}$, $R^C_j$, $Y_{i'}$ and $R^B_{i''}$ given $\curlF$ for all $i$, $i'$ and $j$ and all $i'' \neq i$.
Because the choice of folds does not depend on $Y$ or $R^B$ values, these independences also hold conditional on $\curlF$ and $\curlS_k$.
%

All this implies that the data $\{ (R^A_i, R^B_i, R^A_i \pi^A_i, R^A_i X_i, R^B_i X_i, R^B_i Y_i) : i \in \curlS_k \}$ on fold $k$ are conditionally independent of the data $\{ (R^A_i, R^B_i, R^A_i \pi^A_i, R^A_i X_i, R^B_i X_i, R^B_i Y_i) : i \in \curlS_{-k} \}$ on the remaining folds given $\curlF$ and $\curlS_k$.
Hence, since $\piBhat_k$ and $\mhat_k$ are calculated using only data $\{ (R^A_i, R^B_i, R^A_i \pi^A_i, R^A_i X_i, R^B_i X_i, R^B_i Y_i) : i \in \curlS_{-k} \}$, the data $\{ (R^A_i, R^B_i, R^A_i \pi^A_i, R^A_i X_i, R^B_i X_i, R^B_i Y_i) : i \in \curlS_k \}$ are independent of $\piBhat_k$ and $\mhat_k$ given $\curlF$ and $\curlS_k$.
In Appendix~\ref{sect:appendix.simple} we show that this ensures that the empirical process and remainder terms are $o_p (M^{-1/2})$ if Condition C1 holds.

\section{Varying cluster sampling probabilities}

\label{sect:unequal.probs}


Now consider the more general case where there are $L$ distinct values of the cluster sampling probability $\pi^C_j$, which we denote as $\pi^{C(1)}, \ldots, \pi^{C(L)}$.
SRSWOR of clusters is the special case with $L=1$.
Assume $L$ is small relative to $M$.
Let $J^{(l)} = \sum_{j=1}^J I(\pi^C_j = \pi^{C(l)})$ and $M^{(l)} = M^{(l)} (R^C_1, \ldots, R^C_J) = \sum_{j=1}^J R^C_j \; I(\pi^C_j = \pi^{C(l)})$ denote the numbers of clusters and sampled clusters with sampling probability $\pi^{C(l)}$ ($l=1, \ldots, L$).
Clearly, $\sum_{l=1}^L J^{(l)} = J$ and $\sum_{l=1}^L M^{(l)} = M$.
Assume the cluster sampling mechanism satisfies the symmetry condition
$
P(R^C_1 = r_1, \ldots, R^C_1 = r_J \mid \curlF)
=
P(R^C_1 = r_1', \ldots, R^C_1 = r_J' \mid \curlF)
$
for all values $(r_1, \ldots, r_J)$ and $(r_1', \ldots, r_J')$ of $(R^C_1, \ldots, R^C_J)$ such that
$
M^{(l)} (r_1, \ldots, r_J) = M^{(l)} (r_1', \ldots, r_J')
\hspace{0.2cm}
\forall \; l=1, \ldots, L.
$
This is satisfied by, for example, conditional Poisson sampling, Sampford sampling, Pareto sampling and randomised systematic sampling \citep{Bondesson2006,Grafstrom2009}.

For each $l=1, \ldots, L$, randomly partition the $M^{(l)}$ sampled clusters with $\pi^C_j = \pi^{C(l)}$ evenly into $K$ folds.
For each $l$, this partition begins by randomly choosing a set of $K \times \lfloor M^{(l)} / K \rfloor$ sampled clusters ($\lfloor x \rfloor$ denotes the integer part of $x$) and evenly partitioning these between the folds.
This leaves fewer than $K$ sampled clusters so far unassigned to folds.
Each of these is randomly assigned to the $K$ folds in such a way that no fold receives more than one of them.
Then the unsampled clusters are assigned to these same $K$ folds in the same way.

Let $J^{(l)}_k$ and $M^{(l)}_k$ denote, the numbers of, respectively, clusters and sampled clusters with $\pi^C_j = \pi^{C(l)}$ in fold $k$.
Let $M^{(.)}_k = (M^{(1)}_k, \ldots, M^{(L)}_k)$, $J^{(.)}_k = (J^{(1)}_k, \ldots, J^{(L)}_k)$ and $M^{(.)} = (M^{(.)}_1, \ldots, M^{(.)}_K)$.
In an abuse of notation, let $\pi^C_i$ denote the sampling probability of the cluster to which individual $i$ belongs ($\pi^C_j$ still denotes the sampling probability of cluster $j$).

As in Section~\ref{sect:equal.probs}, $\mhat_k$ is calculated using the $Y$ and $X$ values of all individuals with $R^B=1$ not in fold $k$, and $\piBhat_k$ uses data $\{ (R^B_i, R^B_i X_i): i \notin \curlS_k \}$ on all individuals not in fold $k$.
However, unlike in Section~\ref{sect:equal.probs}, where $\piBhat_k$ was calculated using data $\{ (R^A_i, R^A_i \pi^A_i, R^A_i X_i): i \notin \curlS_k \}$ on all individuals not in fold $k$, here we only use data $(R^A, R^A \pi^A, R^A X)$ on individuals in Sample A who belong to a {\it subset} of the clusters not in fold $k$.
We refer to this subset of clusters as the `active' subset.
As explained below, the reason for only using this subset is to ensure that information about the $R^A$ values of individuals in fold $k$ provided by the value of $\piBhat_k$ vanishes asymptotically.

To choose the active subset, first choose a number $0 < \truncM < 1$ and define
\begin{equation}
  C^{(l)} = \lfloor \pi^{C(l)} (1 - \truncM) (J^{(l)} - J^{(l)}_k) \rfloor
  \wedge (M^{(l)} - M^{(l)}_k)
  \hspace{1cm} (l=1, \ldots, L)
  \label{eq:active}
\end{equation}
(here, $a \wedge b$ denotes the minimum of $a$ and $b$).
For each $l=1, \ldots, L$, randomly subsample $C^{(l)}$ of the $M^{(l)} - M^{(l)}_k$ sampled clusters with $\pi^C_j = \pi^{C(l)}$ not in fold $k$.
The set of $\sum_{l=1}^L M^{(l)} - M^{(l)}_k$ subsampled clusters is the active subset.
To compensate for this subsampling, we temporarily multiply the $\pi^A$ values of individuals with $\pi^C = \pi^{C(l)}$ not in fold $k$ by
\begin{equation}
\frac{ \lfloor \pi^{C(l)} (1 - \truncM) (J^{(l)} - J^{(l)}_k) \rfloor }
     { \pi^{C(l)} (J^{(l)} - J^{(l)}_k) }
     \label{eq:compensate.active}
\end{equation}
when calculating $\piBhat_k$.

The motivation for calculating $\piBhat_k$ using $(R^A, R^A X)$ values only of individuals in the active clusters is that when $M$ is large, $C^{(l)}$ is very likely to equal $\lfloor \pi^{C(l)} (1 - \truncM) (J^{(l)} - J^{(l)}_k) \rfloor$.
If $C^{(l)}$ were sure to equal $\lfloor \pi^{C(l)} (1 - \truncM) (J^{(l)} - J^{(l)}_k) \rfloor$ for all $l$, then knowing $\piBhat_k$ would tell us nothing about the $(R^A, R^B, Y)$ values of individuals in fold $k$ given $\curlF$ and $\curlS_k$.
Hence, information about $R^A$ values of individuals in fold $k$ provided by the value of $\piBhat_k$ vanishes asymptotically.

When $M$ is large, $\lfloor \pi^{C(l)} (1 - \truncM) (J^{(l)} - J^{(l)}_k) \rfloor \approx (M^{(l)} - M^{(l)}_k) (1 - \truncM)$.
So, the price paid for this noninformativeness of $\piBhat_k$ when $M$ is large is the discarding of about $100 \truncM$\% of the individuals with $R^A_i=1$ when calculating $\piBhat_k$.
This motivates us to choose $\truncM$ close to zero, e.g.\ $\truncM = 0.01$.

In Appendix~\ref{sect:appendix.unequal} we prove that when the folds and active subset are chosen in the way described above, the empirical process and remainder terms in the von Mises expansion (equation~(\ref{eq:vonMises})) are $o_p (M^{-1/2})$ if Condition C1 is satisfied.

In the special case of $L=1$ (SRSWOR), we can instead choose the active subset to be a random subset of $M - \lceil M / K \rceil$ of the $M - M_k$ sampled clusters not in fold $k$ ($\lceil x \rceil$ denotes the smallest integer such that $\lceil x \rceil \geq x$).
This ensures noninformativeness of $\piBhat_k$, because there must be at least $M - \lceil M / K \rceil$ sampled clusters not in fold $k$, and ensures that at most $K-1$ sampled clusters are excluded from the active subset.
To compensate for the subsampling, we temporarily multiply the $\pi^A$ values of individuals not in fold $k$ by $(M - \lceil M / K \rceil) / (M - M/K)$ when calculating $\piBhat_k$.
Clearly, if $M$ is an integer multiple of $K$, then all sampled clusters not in fold $k$ are in the active subset, i.e.\ there is no subsampling.

In practice, the number of distinct cluster sampling probabilities will often not be small.
In that case, we propose dividing the $J$ clusters into a small number $L$ of sets according to the ranks of their sampling probabilities $\pi^C_j$.
For example, one might divide them by quartile of $\pi^C_j$ into $L=4$ sets.
When subsampling the active subset from the $l$th set of clusters and compensating for this subsampling, replace $\pi^{C(l)}$ in expressions~(\ref{eq:active}) and~(\ref{eq:compensate.active}) by the mean $\pi^C_j$ value of clusters in set $l$.
This is what we do in the simulation study of Section~\ref{sect:simstudy} and the Natsal-3 application of Section~\ref{sect:Natsal}.
Provided that the values of $\pi^C_j$ do not vary greatly within each of the $L$ sets, the results in Appendix~\ref{sect:appendix.unequal} should be approximately valid.

If the population is divided into strata, assignment of clusters to the $K$ folds and selection of active subsets are done separately in each stratum.

\section{Ratio estimator and TMLE}

\label{sect:DR2}

In the context where parametric nuisance models are used, CLW also proposed the estimator
\begin{equation*}
\thetahattwo = \thetahattwo (\piBhat, \mhat)
= \frac{ 1 }{ \hat{n}^A } \sum_{i=1}^n \frac{ R^A_i }{ \pi^A_i } \mhat(X_i)
+
\frac{ 1 }{ \hat{n}^B } \sum_{i=1}^n \frac{ R^B_i }{ \piBhat (X_i) } \{ Y_i - \mhat(X_i) \}
\end{equation*}
where $\hat{n}^A = n^{-1} \sum_{i=1}^n R^A_i / \pi^A_i$ and $\hat{n}^B = n^{-1} \sum_{i=1}^n R^B_i / \piBhat (X_i)$.
This estimator does not require the population size $n$ to be known.
It may also be more efficient than $\thetahat$, for the same reason that the Hajek estimator may be more efficient than the Horwitz-Thompson estimator.
In CLW’s simulation study with parametric nuisance models, $\thetahattwo$ did indeed have smaller variance than $\thetahat$.

Here we propose instead the ratio estimator
\begin{equation}
  \thetahatthr
  = \thetahatthr (\underpiBhat, \undermhat)
  = \frac{n}{ \hat{n}^A } \thetahat (\underpiBhat, \undermhat)
  = \frac{ 1 }{ \hat{n}^A } \sum_{k=1}^K \sum_{i \in \curlS_k} \left[
  \frac{ R^A_i }{ \pi^A_i } \mhat_k (X_i)
  +
  \frac{ R^B_i }{ \piBhat_k (X_i) } \{ Y_i - \mhat_k (X_i) \}
  \right],
  \label{eq:ratio.estimator}
\end{equation}
which also does not require $n$ to be known.
In Appendix~\ref{sect:appendix.DR2}, we show that if Condition C1 is satisfied,
\begin{eqnarray}
\sqrt{M} \{ \thetahatthr (\underpiBhat, \undermhat) - \Ybar \}
& = &
\sqrt{M} \frac{1}{n} \sum_{k=1}^K \sum_{i \in \curlS_k}
\frac{R^A_i}{\pi^A_i} \left\{ \mtrue (X_i) - \bar{m}_0 \right\}
\nonumber \\
&& +
\sqrt{M} \frac{1}{n} \sum_{k=1}^K \sum_{i \in \curlS_k}
\frac{R^B_i}{\piBtrue (X_i)} \{ Y_i - \mtrue(X_i) \}
\nonumber \\
&& +
\sqrt{M} (\bar{m}_0 - \Ybar) + o_p(1),
\label{eq:DR3.expansion}
\end{eqnarray}
where $\bar{m}_0 = n^{-1} \sum_{i=1}^n \mtrue (X_i)$, and that
\[
\sqrt{M} \thetahatthr = \sqrt{M} \thetahattwo
- \sqrt{M} \frac{ 1 }{ n } \sum_{i=1}^n \left\{ \frac{ R^B_i }{ \piBtrue (X_i) } - 1 \right\} (\bar{m}_0 - \Ybar)
+ o_p (1).
\]
The mean-zero term $\sqrt{M} n^{-1} \sum_{i=1}^n \left\{ R^B_i / \piBtrue (X_i) - 1 \right\} ( \bar{m}_0 - \Ybar )$ is $O_p (1)$ and does not vanish asymptotically.
However, when the population is large compared to Sample B, its variance is small compared to the second term on the right-hand side of equation~(\ref{eq:DR3.expansion}).
Hence, in this case, $\thetahatthr \approx \thetahattwo + o_p (M^{-1/2})$, i.e.\ $\thetahatthr$ is approximately asymptotically equivalent to $\thetahattwo$.

The asymptotic variance of $\thetahatthr - \Ybar$ and an estimator of this variance is given by \citet{Seaman2025} (specifically, their formulae for $\mu_{\rm KH2}$ when the $\piBtrue$ model is correctly specified).

A popular alternative to estimator $\hat{\theta}_{\rm iid}$ for iid data is the TMLE.
This involves modifying an initial estimate $\mhat_{\rm iid}$ of $E(Y \mid X)$ to $\mhat^*_{\rm iid}$ in such a way that the regression imputation estimator $n^{-1} \sum_{i=1}^n \mhat_{\rm iid}^* (X_i)$ of $E(Y)$ can be used even when $\mhat_{\rm iid}$ is obtained using a data-adaptive method.
We now describe an analogous TMLE method for our situation.

Define $\mhat_k (x; \epsilon_k)$ as
\begin{equation}
\mhat_k (x; \epsilon_k) =
\mhat_k (x) + \epsilon_k \; \frac{ 1 }{ \piBhat_k (X_i) }
\label{eq:TMLE.linear}
\end{equation}
for some $\epsilon_k$.
Alternatively, if $Y$ is bounded by zero and one, i.e.\ $P(0 < Y < 1) = 1$, then $\mhat_k (x; \epsilon_k)$ can be defined by either equation~(\ref{eq:TMLE.linear}) or
\begin{equation}
\mhat_k (x; \epsilon_k) =
\frac{ \exp \{ \mbox{logit } \mhat_k (x) + \epsilon_k / \piBhat_k (X_i) \} }
     { 1 + \exp \{ \mbox{logit } \mhat_k (x) + \epsilon_k / \piBhat_k (X_i) \} }.
\label{eq:TMLE.logit}
\end{equation}
Define $\mhatstar_k (x) = \mhat_k (x; \epsilonhat_k)$, where $\epsilonhat_k$ is the solution to the estimating equation
\begin{equation}
\sum_{i \in \curlS_k} \frac{ R^B_i }{ \piBhat_k (X_i) }
\{ Y_i - \mhat_k (X_i; \epsilonhat_k) \}
= 0.
\label{eq:epsilon.esteq}
\end{equation}
%

If $\mhat_k (x; \epsilon_k)$ is defined by equation~(\ref{eq:TMLE.linear}), $\epsilonhat_k$ can be calculated by fitting a linear regression model with no intercept and only the covariate $1 / \piBhat$ to the outcome $Y - \mhat_k (X)$ in individuals with $R^B=1$ in fold $k$.
If $\mhat_k (x; \epsilon_k)$ is defined by equation~(\ref{eq:TMLE.logit}), $\epsilonhat_k$ can be calculated by fitting a logistic regression model with offset $\mbox{logit } \mhat_k (X)$, no intercept and only the covariate $1 / \piBhat$ to the outcome $Y$ in those same individuals.

It follows from equation~(\ref{eq:epsilon.esteq}) that
\[
\tmleone \equiv
\frac{1}{n} \sum_{k=1}^K \sum_{i \in \curlS_k} U_i (\piBhat_k, \mhatstar_k) =
\frac{1}{n} \sum_{k=1}^K \sum_{i \in \curlS_k} \frac{ R^A_i }{ \pi^A_i } \; \mhatstar_k (X_i).
\]
This mass imputation estimator $\tmleone$ is our TMLE estimator.
In Appendix~\ref{sect:appendix.tmle} we show it is asymptotically equivalent to $\thetahat$, i.e.\ $\tmleone = \thetahat + o_p (M^{-1/2})$, if Condition C1 holds.

We also define the corresponding TMLE estimator $\tmletwo = \tmleone \times n / \hat{n}^A$, which has the same asymptotic distribution as $\thetahatthr$, i.e.\ $\tmletwo = \thetahatthr + o_p (M^{-1/2})$.

\section{Calculating $\piBhat_k$ and $\mhat_k$}

\label{sect:estimating.nuisance}

So far, we have been agnostic about how to calculate $\piBhat_k$ and $\mhat_k$, except to specify that they only use, respectively, data $\{ (R^A_i, R^B_i, R^A_i \pi^A_i, R^A_i X_i, R^B_i X_i): i \notin \curlS_k \}$ and data $\{ (R^B_i, R^B_i X_i, R^B_i Y_i): i \notin \curlS_k \}$ not in fold $k$ (and when calculating $\piBhat$, using only Sample A individuals in active clusters and compensating for this by scaling the $\pi^A_i$'s).
Estimating $\mtrue (X) = E(Y \mid X)$ is straightforward, because $Y_1, \ldots, Y_n$ are assumed to be independent given $X_1, \ldots, X_n$.
This is a standard problem for which many data-adaptive methods could be used.
In Section~\ref{sect:simstudy} we use XGBoost and Highly Adaptive LASSO (HAL); in Section~\ref{sect:Natsal}, a Superlearner.

Estimating $\piBtrue (X) = P(R^B=1 \mid X)$ is more complicated, because, although $R^B_1, \ldots, R^B_n$ are independent given $X_1, \ldots, X_n$, we only observe $X$ for individuals with $R^A=1$ or $R^B = 1$.
CLW, working with parametric nuisance models, discuss pseudo-maximum likelihood, approximate pseudo-maximum likelihood, calibration and the method of \citet{KimHaziza2014}.
The log pseudo-likelihood is defined as
\begin{equation}
  \sum_{i=1}^n R^B_i \; \log \pi^B (X_i) - \left( \frac{ R^A_i }{ \pi^A } - R^B_i \right) \log \{ 1 - \pi^B (X_i) \}.
  \label{eq:pseudo.loglike}
\end{equation}
This is the log likelihood for a binomial regression of $R^B$ on $X$ in the population of $n$ individuals, using the individuals in Sample A weighted by their inverse sampling probabilities to represent the population.
\citet{FerriGarcia2022} (see also~\citet{ChuBeaumont2019}) use the pseudo-likelihood together with CART to estimate $\piBtrue (X)$.
In Section~\ref{sect:simstudy}, we use expression~(\ref{eq:pseudo.loglike}) as an objective function (to be maximised) for XGBoost.

The approximate log pseudo-likelihood is expression~(\ref{eq:pseudo.loglike}) omitting the $R^B_i \log \{ 1 - \pi^B (X_i) \}$ term.
When Sample B is small relative to the population, this term will be negligible compared to the other terms in~(\ref{eq:pseudo.loglike}).
The approximate pseudo-likelihood has the advantage that it can be used quite generally with data-adaptive methods that predict a binary outcome and allow for weights.
In Section~\ref{sect:simstudy}, we use it with HAL.
We accept CLW's criticism of the approximate pseudo-likelihood when Sample B is not a small fraction of the population, but we focus in Sections~\ref{sect:simstudy} and~\ref{sect:Natsal} on scenarios where it is.
Such scenarios are common in practice.

\section{Simulation study}

\label{sect:simstudy}

We simulated one finite population with $J=1000$ clusters.
Cluster $j$ contains $H_{jq}$ households of $q$ individuals ($q=1, 2, 3$), where $H_{jq}$ is negatively binomially distributed with mean 100 and variance 400.
Hence, the expected cluster size is 600 individuals and expected population size $n$ is 600,000.
For each individual $i$ in cluster $j$, continuous variables $X_{1i}$ and $X_{2i}$ and binary variables $X_{3i}$ and $X_{4i}$ were generated independently, with expectations that depend on the number of households in the cluster, the size of individual's household, and a cluster-level random effect.
See Appendix~\ref{sect:appendix.simstudy.dgm} for details.
This population was used for all simulations.

For each simulated dataset, $Y_1, \ldots, Y_n$ were generated independently from a normal distribution with mean depending on $X = (X_1, X_2, X_3, X_4)$ and variance 1.
Sample A was drawn using a three-stage sampling design.
$M$ clusters were selected using Sampford or random systematic sampling, with probability of selecting cluster $j$ being proportional to the number of households in cluster $j$.
From each selected cluster, $n_{\rm house}$ households were selected using SRSWOR.
Then one individual was sampled from each selected household.
Thus, Sample A contained $M \times n_{\rm house}$ individuals.
Sample B was drawn as described in Section~\ref{sect:dgm}.

We considered 48 scenarios.
In Scenarios 1--12, $\mtrue (X)$ is linear in $X$ and $\piBtrue (X)$ follows a logistic regression model with only main effects.
Specifically, $\mtrue (X) = X_1 + X_2 + 2 X_3 + X_4$ and $\mbox{logit } \piBtrue (X) = \alpha_{\rm int} + 0.5 X_1 + X_2 + 0.5 X_3 + X_4$.
In Scenarios 13--24, $\mtrue (X)$ and $\piBtrue (X)$ each include an interaction and a quadratic term.
Specifically, $\mtrue (X) = 0.5 X_1 + 0.5 X_2 + 2 X_3 + X_4 + 2 X_1 X_3 + X_2^2$ and $\mbox{logit } \piBtrue (X) = \alpha_{\rm int} + 0.25 X_1 + 0.5 X_2 + 0.5 X_3 + X_4 + X_1 X_3 + 0.5 X_2^2$.
The value of $\alpha_{\rm int}$ was chosen to achieve a desired expected size $E(n^B)$ of Sample B.
Each of these two sets of 12 scenarios corresponded to two choices of $M$ (150 and 50), two choices of $n_{\rm house}$ (20 and 5), and three choices of $E(n^B)$ (5000, 1000 and 500).
Scenarios 1b--24b are the same as 1--24 except that $Y$ is binary.
In Scenarios 1b--12b, $\mbox{logit } \mtrue (X) = \beta_{\rm int} + X_1 + X_2 + 2 X_3 + X_4$, and in Scenarios 13b--24b, $\mbox{logit } \mtrue (X) = \beta_{\rm int} + 0.5 X_1 + 2 X_3 + X_4 + 2 X_1 X_3 + X_2^2$, with $\beta_{\rm int}$ chosen to make $P(Y=1)=0.1$.

Twenty-five estimators were applied to each of 400 simulated datasets for each scenario.
The first nine were: simple unweighted mean of $Y$ in Sample A [Naive]; Horwitz-Thompson and Hajek estimators that only use Sample A [HT, Hajek]; $\thetahat$, $\thetahattwo$, $\thetahatthr$, $\tmleone$ and $\tmletwo$ using parametric nuisance models [DR1, DR2clw, DR2, TMLE1, TMLE2]; and $\thetahat$ with parametric nuisance models whose parameters are estimated using the method of Kim and Haziza [KH].
The next four were $\thetahat$, $\thetahatthr$, $\tmleone$ and $\tmletwo$ using HAL to estimate $\piBtrue$ and $\mtrue$ and using cross-fitting [DR1.hal5, DR2.hal5, TMLE1.hal5, TMLE.hal5].
Another four used XGBoost instead of HAL [DR1.xgb5, DR2.xgb5, TMLE1.xgb5, TMLE.xgb5].
The final eight used HAL or XGBoost but without cross-fitting [DR1.hal1, DR2.hal1, TMLE1.hal1, TMLE.hal1, DR1.xgb1, DR2.xgb1, TMLE1.xgb1, TMLE.xgb1].
Of these 25 estimators, DR1, DR2clw and KH were described by CLW.

The parametric model used for $\mtrue$ was a linear regression (if $Y$ was continuous) or logistic regression (if binary) model with main effects for $X_1$, $X_2$, $X_3$ and $X_4$.
The parametric model for $\piBtrue$ was a logistic regression, also with main effects.
Hence, these models are correctly specified in Scenarios 1--12 and 1b--12b, and misspecified in Scenarios 13--24 and 13b--24b.
For DR1, DR2clw, TMLE1 and TMLE2, we estimated the standard error (SE) and calculated 95\% confidence intervals (CIs) using variance estimators that are valid when the model for $\piBtrue$ (but not necessarily $\mtrue$) is correctly specified \citep{Seaman2025}.
For methods using cross-fitting, we used $K=5$ folds, set $\truncM=0.01$ and used $L=4$ sets of clusters based on quartiles of the distribution of $\pi^C_j$ (as described at the end of Section~\ref{sect:unequal.probs}).

\begin{figure}[p]
\begin{center}
\includegraphics[width=1\textwidth,page=1]{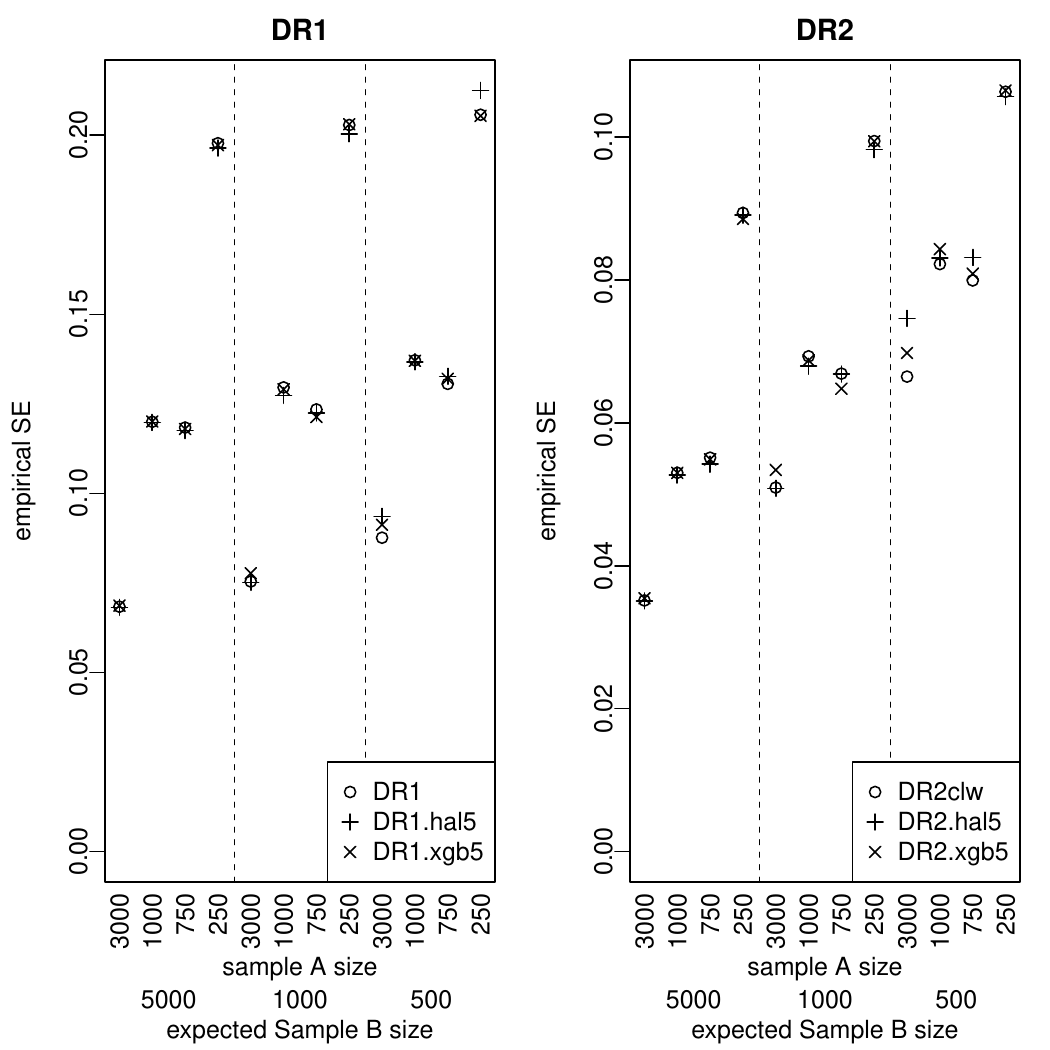}
\hspace*{-5mm}
\end{center}
\caption{Left: Empirical standard error of $\thetahat$ when used with parametric models (DR1), HAL (DR1.hal5) or XGBoost (DR1.xgb5) in Scenarios 1--12.  
Right: Empirical standard error of $\thetahattwo$ when used with parametric models (DR2clw) or $\thetahatthr$ when used with HAL (DR2.hal5) or XGBoost (DR2.xgb5).}
\label{fig:compareSEs}
\end{figure}

Full results are given in Appendix~\ref{sect:appendix.simstudy.results}.
Here, we highlight the main findings, beginning with the continuous $Y$ scenarios.
In Scenarios 1--4, where $E(n^B) = 5000$, all estimators except Naive were approximately unbiased with CIs that had approximately correct coverage.
Estimators that used HAL or XGBoost with cross-fitting had very similar SE to the corresponding estimators that used parametric nuisance models (Figure~\ref{fig:compareSEs}).
The Hajek, DR2 and TMLE2 estimators were more efficient than the corresponding HT, DR1 and TMLE1 estimators; the SEs of the former were approximately half that of the former.
Findings for Scenarios 5--12, where $E(n^B) = 1000$ or 500, were similar, although CI coverage of DR2.xgb5 and/or TMLE.xgb5 fell to 88\% in Scenarios 6 and 10, where $M=150$ and $n_{\rm house}=5$.

\begin{figure}[p]
\begin{center}
\includegraphics[width=1\textwidth,page=2]{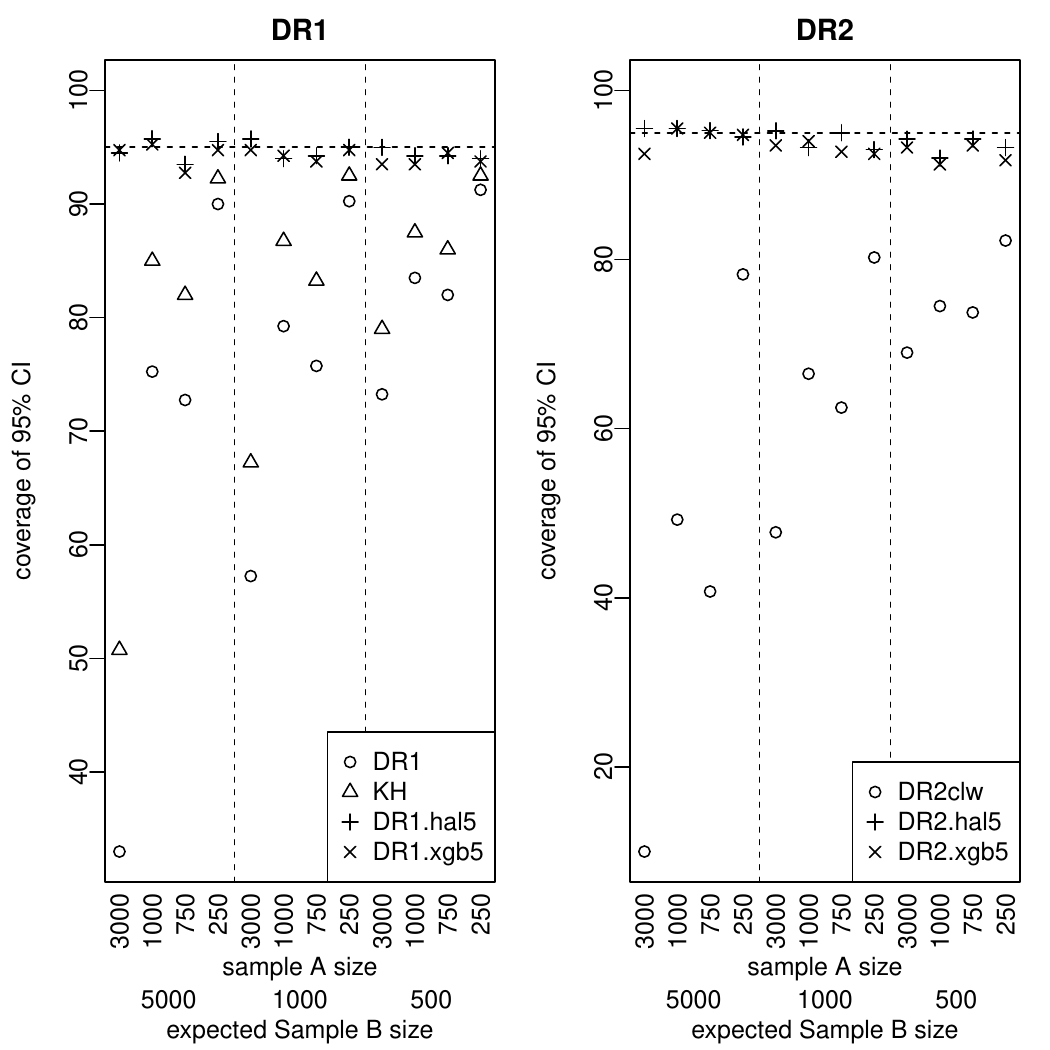}
\hspace*{-5mm}
\end{center}
\caption{Left: Coverage of 95\% confidence interval for $\Ybar$ in Scenarios 13--24 when $\thetahat$ is used with parametric nuisance models (DR1), Kim and Haziza's method (KH), HAL (DR1.hal5) or XGBoost (DR1.xgb5).  Right: Coverage when $\thetahattwo$ or $\thetahatthr$ is used with parametric nuisance models (DR2clw) or $\thetahatthr$ used with HAL (DR2.hal5) or XGBoost (DR2.xgb5).}
\label{fig:compare.cover}
\end{figure}

In Scenarios 13--24, the estimators using parametric nuisance models (DR1, DR2clw, TMLE1, TMLE2 and KH) were biased and had CIs with undercoverage (Figure~\ref{fig:compare.cover}).
Coverage was particularly bad (in one case, as low as 10\%) when Samples A and B were both large.
Estimators using HAL or XGBoost had much smaller bias, especially in Scenarios 13--15, where $E(n^B)=5000$ and the size of Sample A was at least 750; here bias was almost zero.
CI coverage for these estimators was approximately 95\% and never dropped below 90\%.

Using HAL or XGBoost without cross-fitting had virtually no effect on the empirical SE (Figure~\ref{fig:empSE}).
Omitting the cross-fitting step tended to decrease bias slightly for HAL, but to increase bias slightly for XGboost (Figure~\ref{fig:bias}).
It also tended to reduce the mean estimated standard error for DR2 with XGBoost (Figure~\ref{fig:estimatedSE}).
The net effect was that omitting cross-fitting tended to cause CI coverage to deteriorate for XGBoost in many of the scenarios, especially for DR2 and TMLE2 (see Figure~\ref{fig:coverage}).
This deterioration of coverage is in line with the findings of \citet{Ellul2025}, although they found no obvious effect on bias.

We also calculated $\thetahatthr$ using parametric nuisance models.
As expected, biases and SEs were very similar to those of DR2clw.
Over the 24 scenarios, the mean (maximum) absolute difference in their biases was 0.0011 (0.0036); the ratio of their SEs varied from 0.99 to 1.01.

Findings for Scenarios 1b--24b were similar, except that omitting the cross-fitting step tended slightly to increase bias not only when using XGboost, but also when using HAL (Figure~\ref{fig:bias.b}).
Nevertheless, using cross-fitting with XGboost (and to a less degree, HAL) still improved coverage of confidence intervals in most scenarios (Figure~\ref{fig:coverage.b}), due to its causing a slight increase in mean estimated standard error (Figure~\ref{fig:estimatedSE.b}).

\section{Application to Natsal-3 data}

\label{sect:Natsal}

We demonstrate the proposed methods using the Natsal-3 and four nonprobability surveys that used a selection of questions from the Natsal-3 questionnaire.
Natsal-3 was an address-based probability survey of 16--74 year-olds in Britain, carried out in 2010--2012  \citep{Erens2014natsal3paf}.
It was collected stratified on region, and involved sampling postcode areas within regions, households within sampled postcode areas, and then one individual from each sampled household.
The nonprobability surveys, carried out in 2012, targeted 18--44 year-olds \citep{Copas2020natsal3web}.
We estimated the prevalence of the outcome ``having had sex with a same-gender partner at least once in the past five years'' in the subpopulation of men aged 18--44.
There were 3628 and 4074 men aged 18--44 in the probability and nonprobability samples, respectively.
The (unweighted) outcome prevalence in the nonprobability sample was 10.9\%.

We applied estimators $\thetahatthr$ and $\tmletwo$ with 5-fold cross-fitting.
For the covariates $X$, we used 14 socio-demographic, sexual identity, health, attitudinal and behavioural variables, as well as region.
To estimate $\pi^B$ and $m$, we used a Superlearner with library that included logistic regression, LASSO, generalised additive model, HAL and XGBoost.
Also we calculated $\thetahattwo$ and $\tmletwo$ using parametric logistic regression models for $\pi^B$ and $m$.
These models included main effects for the covariates, and were fitted using either maximum pseudo likelihood for $\pi^B$ and maximum likelihood for $m$, or the method of Kim and Haziza.
Item nonresponse was addressed using the R package {\tt mice} for multiple imputation.

As the probability survey covered the population of men and women aged 16--74 and we are estimating prevalence only in men aged 18--44, a minor extension of our methods was needed.
Essentially, this involves replacing $Y_i$ by $S_i Y_i$, where $S_i = 1$ if individual $i$ is a man aged 18--44 and $S_i=0$ otherwise, and $\hat{n}_A$  by $\sum_{i=1}^n R^A_i S_i / \pi^A_i$ (see Appendix~\ref{sect:domain.estimation} for details).

Table \ref{tab:NS3results} shows the prevalence estimates.
All were similar (around 3.6\%), suggesting that at least one of the parametric nuisance models was fairly well specified.
Estimated SEs were also similar.
In this illustrative example, the outcome was also measured in the probability survey, which allows the DR estimates to be compared with the Hajek estimate that uses only the outcome data from the probability sample.
This Hajek estimate was slightly smaller (3.2\%) than the DR estimates, but the size of the difference was only approximately one SE.

\section{Discussion}

\label{sect:discussion}

We have described a debiased machine learning approach for estimating the mean of $Y$ from a nonprobability sample when auxiliary information on $X$ is available from a probability sample.
This method yields an asymptotically normally distributed estimator with standard error that can estimated as easily as the standard error of a Horwitz-Thompson estimator.

We have used cross-fitting.
For the related problem of estimating $E(Y)$ for an infinite population using iid data, an alternative to cross-fitting is to rely on the Donsker condition \citep{Hines2022}.
This limits the choice of data-adaptive estimators of $P(R=1 \mid X)$ and $E(Y \mid X)$ to those that satisfy this condition.
Nevertheless, it would be interesting to establish whether the Donsker condition would suffice in our setting.

Some researchers have previously used DR estimators with data-adaptive estimation of nuisance functions for combining probability and nonprobability survey data, e.g.~\citet{CastroMartin2021} and~\citet{Cobo2025}.
However, they did not establish validity of this approach.
Indeed~\citet{Cobo2025} comment that ``there are hardly any results on the theoretical properties of variance estimators when more complex machine learning models and techniques are used.''
Moreover, these researchers used nuisance estimators that may not satisfy the Donsker condition and without using cross-fitting.

We have used subsampling of an `active subset' of clusters when calculating $\piBhat$.
This should have no effect on asymptotic efficiency, and in our simulation study we observed no loss of efficiency.
However, if efficiency loss is a concern, it could be mitigated by using repeated cross-fitting \citep{Chernozhukov2018}.
As with other debiased machine learning estimators using cross-fitting, repeated cross-fitting also has the advantage of reducing the Monte Carlo variation induced by the randomness of the process of partitioning units (in our case, clusters) among folds, although at the cost of increasing computation time.

As discussed in Section~\ref{sect:equal.probs}, if SRSWOR is used to select the $M$ clusters for Sample A and $M$ is an integer multiple of $K$, there is no need to subsample an active subset.
Appendix~\ref{sect:appendix.conditional.poisson} shows this is true also when the clusters are sampled using conditional Poisson sampling.

When the population of clusters is stratified, assignment of clusters to the $K$ folds is done within strata.
This is unproblematic when the probability sample uses SRSWOR of clusters within strata or when many clusters are sampled from each stratum.
When cluster sampling probabilities within strata vary and few clusters are sampled from each stratum, there may be two problems: the subsampling of active subsets of clusters could lead to substantial loss of information for estimating $\piBhat$; and the asymptotic property that $\piBhat_k$ is uninformative about $R^A$ values in fold $k$ may be a poor approximation.
The scenario of sampling few clusters per stratum is worthy of further research.

Several researchers have considered the problem of efficiently combining an IPW or DR estimator of the mean of $Y$ with a Horwitz-Thompson or Hajek estimator based on Sample A alone when $Y$ is observed in both samples.
\citet{GaoYang2023} consider the DR estimator $\thetahat$ and use parametric nuisance models.
\citet{Seaman2025} discuss this approach and extend it to $\thetahattwo$ and IPW estimators and domain estimation, again using parametric nuisance models.
\citet{Rueda2023} used an IPW estimator with random forest or XGBoost to estimate the nuisance function.
Since, when Condition C1 is satisfied, the debiased machine learning estimators of $\Ybar$ we have presented here have the same asymptotic distribution as the corresponding DR estimators of $\Ybar$ that use correctly specified parametric nuisance models, the combining methods of Seaman et al.\ (2025) should apply.

Finally, \citet{Dagdoug2026} consider debiased machine learning using (only) a probability sample in which $Y$ has been measured.
They study model-assisted estimation of $\Ybar$ using data-adaptive estimation of $m(X)$, and imputation of missing at random $Y$.

\subsection*{Acknowledgements}

This work is supported by the National Survey of Sexual Attitudes and Lifestyles (Natsal) grant from the Wellcome Trust (212931/Z/18/Z), with contributions from the Economic and Social Research Council (ESRC) and the National Institute for Health Research (NIHR).
We thank the Natsal team: Pam Sonnenberg and Cath Mercer, Soazig Clifton who leads the Natsal methodology work package, and Elise Paul, Anika Knuppel, Luke Muschialli and Claudia Gascoyne for data management and cleaning of the Natsal-3 and non-probability datasets used here.
For the purpose of open access, the authors have applied a Creative Commons Attribution (CC BY) licence to any Author Accepted Manuscript version arising.

\subsection*{Data Availability}

De-identified individual-level participant data from Natsal-3 are available via the UK Data Archive (serial number SN 7799). Datasets can be accessed through registration with the UK Data Service. Data from the online survey study are available on reasonable request.

\subsection*{ORCID}

Shaun R. Seaman https://orcid.org/0000-0003-3726-5937

\begin{table}[p]
  \centering
  \begin{tabular}{lccc}
    Estimator & Estimate & SE & 95\% CI \\
    \hline
    $\thetahatthr$ with SL      & 0.0353 & 0.0032 & 0.0291 -- 0.0416 \\
    $\tmletwo$ with SL          & 0.0354 & 0.0032 & 0.0292 -- 0.0416 \\
    $\thetahattwo$ with LR      & 0.0364 & 0.0033 & 0.0299 -- 0.0429 \\
    $\tmletwo$ with LR          & 0.0364 & 0.0033 & 0.0299 -- 0.0429 \\
    $\thetahattwo$ with KH      & 0.0356 & 0.0033 & 0.0291 -- 0.0421 \\
    Hajek                       & 0.0322 & 0.0034 & 0.0256 -- 0.0389 \\
  \end{tabular}
  \vspace{.8cm}
  \caption{Prevalence estimates of the self-reported outcome ``having had sex with a same-gender partner at least once in the past five years'' in men aged 18--44 years in 2010--2012, obtained using Natsal-3 data.
    `SL' means nuisance functions estimated by Superlearner; `LR' means logistic regression nuisance models fitted by maximum pseudo likelihood and maximum likelihood; `KH' means logistic regression nuisance models fitted by method of Kim and Haziza (2014); `Hajek' means Hajek estimator using data only from the probability survey.}
  \label{tab:NS3results}
\end{table}

\begin{figure}[p]
\begin{center}
\includegraphics[width=1\textwidth]{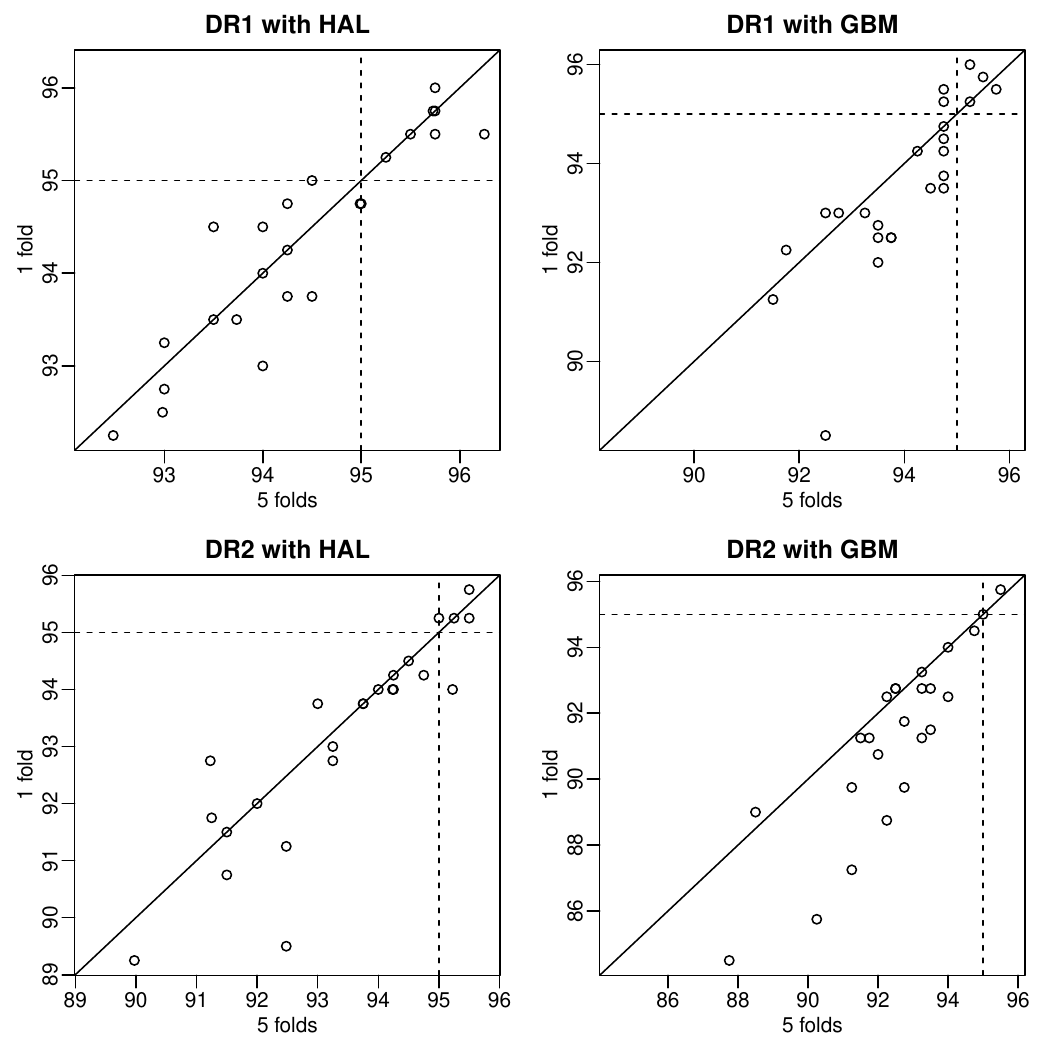}
\hspace*{-5mm}
\end{center}
\caption{Coverage of 95\% confidence interval for $\Ybar$ when $\thetahat$ (top) or $\thetahatthr$ (bottom) is used with HAL (left) or XGBoost (right).  Each plot compares coverage when 5-fold cross-fitting is used versus when no cross-fitting is used (1 fold).  Each dot corresponds to one of Scenarios 1--24.  Broken lines indicate perfect 95\% coverage.}
\label{fig:coverage}
\end{figure}

\clearpage

\newpage

\begin{center}
{\LARGE \bf   Supplementary Materials for `Debiased machine learning for combining probability and non-probability samples'}
\vspace{.2cm}

{\large \bf Shaun R.\ Seaman, Tommy Nyberg, Andrew Copas and Anne M.\ Presanis}
\end{center}

\renewcommand\thesection{A\arabic{section}}
\renewcommand\thesubsection{A\arabic{section}.\arabic{subsection}}
\renewcommand\thetable{A\arabic{table}}
\renewcommand\thefigure{A\arabic{figure}}    
\renewcommand\theequation{A\arabic{equation}}    
\setcounter{section}{0}
\setcounter{table}{0}    
\setcounter{figure}{0}    
\setcounter{equation}{0}

\section{Proof that empirical process and remainder terms are $o_p(M^{-1/2})$ for SRSWOR of clusters}

\label{sect:appendix.simple}

\subsection{Empirical process term}

We now introduce the notation
\begin{eqnarray*}
  \curlG
  & = &
  (R^A_{11}, \ldots, R^A_{1N_1}, \ldots, R^A_{J1}, \ldots, R^A_{JN_J}, R^B_{11}, \ldots, R^B_{1N_1}, \ldots, R^B_{J1}, \ldots, R^B_{JN_J},
  \\
  && \hspace{.2cm}
  Y_{11}, \ldots, Y_{1N_1}, \ldots, Y_{J1}, \ldots, Y_{JN_J}).
\end{eqnarray*}
We shall sometimes use this to help the reader to know what with respect to which random variables an expectation or variance is being taken.

Consider the empirical process term for the $k$th fold ($k=1, \ldots, K$), i.e.\
\begin{eqnarray}
  &&
\frac{1}{n_k} \sum_{i \in \curlS_k} \left[
    \{ U_i (\piBhat_k, \mhat_k) - U_i (\piBtrue, \mtrue) \}
    \right.
  \nonumber \\
  && \hspace{1.8cm} -
   \left. 
   E_{\curlG} \left\{ U_i (\piBhat_k, \mhat_k) - U_i (\piBtrue, \mtrue) \mid \curlF, \curlS_k, \piBhat_k, \mhat_k \right\}
   \right].
\label{eq:empirical.process}
\end{eqnarray}

By the Law of Total Probability, we have
\begin{eqnarray}
  &&
  P \left( \; \left| \frac{1}{n_k} \sum_{i \in \curlS_k} \{ U_i (\piBhat_k, \mhat_k) - U_i (\piBtrue, \mtrue) \} \right. \right.
  \nonumber \\
  && \hspace{2.2cm} - \left. \left.
  E_{\curlG} \left[ \frac{1}{n_k} \sum_{i \in \curlS_k} \{ U_i (\piBhat_k, \mhat_k) - U_i (\piBtrue, \mtrue) \} \mid \curlF, \curlS_k, \piBhat_k, \mhat_k \right] \right| > \chebysheveps \mid \curlF \right)
  \nonumber \\
  && =
  E_{\curlS_k, \piBhat_k, \mhat_k} \left\{
  P \left( \; \left| \frac{1}{n_k} \sum_{i \in \curlS_k} \{ U_i (\piBhat_k, \mhat_k) - U_i (\piBtrue, \mtrue) \} \right. \right. \right.
  \nonumber \\
  && \hspace{0.4cm} \left. \left. -
  E_{\curlG} \left[ \frac{1}{n_k} \sum_{i \in \curlS_k} \{ U_i (\piBhat_k, \mhat_k) - U_i (\piBtrue, \mtrue) \} \mid \curlF, \curlS_k, \piBhat_k, \mhat_k \right] \bigg| > \chebysheveps \mid \curlF, \curlS_k, \piBhat_k, \mhat_k \right)
  \mid \curlF
  \right\}.
  \nonumber \\
  &&
  \label{eq:empirical.law.total.prob}
\end{eqnarray}

By Chebeshev's Inequality, we have, for any $\chebysheveps>0$ and $\curlF$, $\curlS_k$, $\piBhat_k$ and $\mhat_k$,
\begin{eqnarray}
  &&
     P \left( \; \left| \frac{1}{n_k} \sum_{i \in \curlS_k} \{ U_i (\piBhat_k, \mhat_k) - U_i (\piBtrue, \mtrue) \} \right. \right.
     \nonumber \\
     && \hspace{1cm} - \left.
  E_{\curlG} \left[ \frac{1}{n_k} \sum_{i \in \curlS_k} \{ U_i (\piBhat_k, \mhat_k) - U_i (\piBtrue, \mtrue) \mid \curlF, \curlS_k, \piBhat_k, \mhat_k \} \right] \bigg| > \chebysheveps \mid \curlF, \curlS_k, \piBhat_k, \mhat_k \right)
  \nonumber \\
&& \hspace{1cm} \leq
  \frac{ \Var_{\curlG} \left[ \frac{1}{n_k} \sum_{i \in \curlS_k} \{ U_i (\piBhat_k, \mhat_k) - U_i (\piBtrue, \mtrue) \} \mid \curlF, \curlS_k, \piBhat_k, \mhat_k \right] }{ \chebysheveps^2 }
  \label{eq:Chebeshev}
\end{eqnarray}

Now,
\begin{eqnarray*}
  U (\piBhat_k, \mhat_k) - U (\piBtrue, \mtrue)
  & = &
     R^B Y \left\{ \frac{1}{ \piBhat_k (X) } - \frac{1}{ \piBtrue (X) } \right\}
       + \frac{ R^A }{ \pi^A } \{ \mhat_k (X) - \mtrue (X) \}
     \nonumber \\     
  &&
     - R^B \left\{ \frac{ \mhat_k (X) }{ \piBhat_k (X) } - \frac{ \mtrue (X) }{ \piBtrue (X) } \right\}
\end{eqnarray*}

As we argued in Section~\ref{sect:equal.probs}, $(R^B_i, Y_i)$ is conditionally independent of $R^B_{i'}$, $Y_{i'}$, $R^A_{i''}$ and $R^C_j$ for all $i$, all $i' \neq i$, and all $i''$ and $j$ given $\curlF$, $\curlS_k$, $\piBhat_k$ and $\mhat_k$.
Hence,
\begin{eqnarray}
  &&
  \Var_{\curlG} \left[ \frac{1}{n_k} \sum_{i \in \curlS_k} \{ U_i (\piBhat_k, \mhat_k) - U_i (\piBtrue, \mtrue) \} \mid \curlF, \curlS_k, \piBhat_k, \mhat_k \right]
  \nonumber \\
  && =
  \Var_{\curlG} \left( \frac{1}{n_k} \sum_{i \in \curlS_k} \left[ R^B_i Y_i \left\{ \frac{1}{ \piBhat_k (X_i) } - \frac{1}{ \piBtrue (X_i) } \right\}
      - R^B_i \left\{ \frac{ \mhat_k (X_i) }{ \piBhat_k (X_i) } - \frac{ \mtrue (X_i) }{ \piBtrue (X_i) } \right\} \right] \mid \curlF, \curlS_k, \piBhat_k, \mhat_k \right)
  \nonumber \\
  && \hspace{0.4cm} +
  \Var_{\curlG} \left[ \frac{1}{n_k} \sum_{i \in \curlS_k} \frac{ R^A_i }{ \pi^A_i } \{ \mhat_k (X_i) - \mtrue (X_i) \}  \mid \curlF, \curlS_k, \piBhat_k, \mhat_k \right].
  \label{eq:variance.decomposition}
\end{eqnarray}

If we can show that the conditional expectation given $\curlF$ of each the two variances in expression~(\ref{eq:variance.decomposition}) is $o_p(M^{-1})$, then it will follow from equation~(\ref{eq:empirical.law.total.prob}) that the empirical process term for fold $k$ (i.e.\ expression~(\ref{eq:empirical.process})) is $o_p (M^{-1/2})$.
We shall now look at these two variances in turn.

Consider the first variance in expression~(\ref{eq:variance.decomposition}).
As we argued in Section~\ref{sect:equal.probs}, $R^B_1, Y_i, \ldots, R^B_n, Y_n$ are conditionally independent of $\curlS_k$, $\piBhat_k$ and $\mhat_k$, and of each other, given $\curlF$.
Hence,
\begin{eqnarray}
  &&
  \Var_{\curlG} \left( \frac{1}{n_k} \sum_{i \in \curlS_k} \left[ R^B_i Y_i \left\{ \frac{1}{ \piBhat_k (X_i) } - \frac{1}{ \piBtrue (X_i) } \right\}
      - R^B_i \left\{ \frac{ \mhat_k (X_i) }{ \piBhat_k (X_i) } - \frac{ \mtrue (X_i) }{ \piBtrue (X_i) } \right\} \right] \mid \curlF, \curlS_k, \piBhat_k, \mhat_k \right)
  \nonumber \\
  && \hspace{0.3cm} =
  \frac{1}{n_k^2} \sum_{i \in \curlS_k} \left\{ \frac{1}{ \piBhat_k (X_i) } - \frac{1}{ \piBtrue (X_i) } \right\}^2
  \Var (R^B_i Y_i \mid \curlF, \curlS_k, \piBhat_k, \mhat_k)
  \nonumber \\
  && \hspace{.5cm} +
  \frac{1}{n_k^2} \sum_{i \in \curlS_k} \left\{ \frac{ \mhat_k (X_i) }{ \piBhat_k (X_i) } - \frac{ \mtrue (X_i) }{ \piBtrue (X_i) } \right\}^2 \Var ( R^B_i  \mid \curlF, \curlS_k, \piBhat_k, \mhat_k )
  \nonumber \\
  && \hspace{.5cm} -
  \frac{2}{n_k^2} \sum_{i \in \curlS_k} \left\{ \frac{1}{ \piBhat_k (X_i) } - \frac{1}{ \piBtrue (X_i) } \right\}
  \left\{ \frac{ \mhat_k (X_i) }{ \piBhat_k (X_i) } - \frac{ \mtrue (X_i) }{ \piBtrue (X_i) } \right\}
  \Cov (R^B_i Y_i, R^B_i \mid \curlF, \curlS_k, \piBhat_k, \mhat_k)
  \nonumber \\
  && \hspace{0.3cm} =
  \frac{1}{n_k^2} \sum_{i \in \curlS_k} \left\{ \frac{1}{ \piBhat_k (X_i) } - \frac{1}{ \piBtrue (X_i) } \right\}^2
  \Var (R^B_i Y_i \mid \curlF)
  \nonumber \\
  && \hspace{.5cm} +
  \frac{1}{n_k^2} \sum_{i \in \curlS_k} \left\{ \frac{ \mhat_k (X_i) }{ \piBhat_k (X_i) } - \frac{ \mtrue (X_i) }{ \piBtrue (X_i) } \right\}^2 \Var (R^B_i  \mid \curlF)
  \nonumber \\
  && \hspace{.5cm} -
  \frac{2}{n_k^2} \sum_{i \in \curlS_k} \left\{ \frac{1}{ \piBhat_k (X_i) } - \frac{1}{ \piBtrue (X_i) } \right\}
  \left\{ \frac{ \mhat_k (X_i) }{ \piBhat_k (X_i) } - \frac{ \mtrue (X_i) }{ \piBtrue (X_i) } \right\}
  \Cov (R^B_i Y_i, R^B_i \mid \curlF)
  \nonumber \\
  && =
  \frac{1}{n_k^2} \sum_{i \in \curlS_k} \left\{ \frac{1}{ \piBhat_k (X_i) } - \frac{1}{ \piBtrue (X_i) } \right\}^2
  [ \piBtrue (X_i) \; \{ 1 - \piBtrue (X_i) \} \; \{ \mtrue(X_i) \}^2 + \piBtrue (X_i) \; \Var (Y_i \mid X_i) ]
  \nonumber \\
  && \hspace{.5cm} +
  \frac{1}{n_k^2} \sum_{i \in \curlS_k} \left\{ \frac{ \mhat_k (X_i) }{ \piBhat_k (X_i) } - \frac{ \mtrue (X_i) }{ \piBtrue (X_i) } \right\}^2 \piBtrue (X_i) \; \{ 1 - \piBtrue (X_i) \}
  \nonumber \\
  && \hspace{.5cm} -
  \frac{2}{n_k^2} \sum_{i \in \curlS_k} \left\{ \frac{1}{ \piBhat_k (X_i) } - \frac{1}{ \piBtrue (X_i) } \right\}
  \left\{ \frac{ \mhat_k (X_i) }{ \piBhat_k (X_i) } - \frac{ \mtrue (X_i) }{ \piBtrue (X_i) } \right\}
  \piBtrue (X_i) \; \{ 1 - \piBtrue (X_i) \} \; \mtrue (X_i).
  \nonumber \\
   &&
  \label{eq:first.variance}
\end{eqnarray}
We see that for any fixed functions $\piBhat_k$ and $\mhat_k$, expression~(\ref{eq:first.variance}) is $O_p (M^{-1})$.

Now consider the second variance in expression~(\ref{eq:variance.decomposition}).
As we argued in Section~\ref{sect:equal.probs}, the distribution of $\{ R_i^A: i \in \curlS_k \}$ given $\curlF, \curlS_k, \piBhat_k, \mhat_k$ corresponds to SRSWOR of $M/K$ clusters from the $J/K$ clusters in $\curlS_k$ followed by the original second-stage sampling of individuals within clusters.
Hence, for any fixed function $\mhat_k$,
\begin{equation}
  \Var_{\curlG} \left[ \frac{1}{n_k} \sum_{i \in \curlS_k} \frac{ R^A_i }{ \pi^A_i } \{ \mhat_k (X_i) - \mtrue (X_i) \}  \mid \curlF, \curlS_k, \piBhat_k, \mhat_k \right]
  \label{eq:second.variance}
\end{equation}
is the variance of the Horwitz-Thompson estimator of the population mean of $\mhat_k (X) - \mtrue (X)$ in individuals in $\curlS_k$ given $\curlF$ when the clusters in $\curlS_k$ are selected by SRSWOR and individuals within clusters are sampled according to the original second-stage sampling procedure.
Subject to some regularity conditions, this variance is $O_p(M^{-1})$ (Proposition 2 of \citet{ChauvetVallee2020}).

It follows that, if $\piBhat_k \xrightarrow{p} \piBtrue$ and $\mhat_k \xrightarrow{p} \mtrue$ as $M \rightarrow \infty$, then expressions~(\ref{eq:first.variance}) and~(\ref{eq:second.variance}), and hence (\ref{eq:variance.decomposition}), are $o_p (M^{-1})$, as required.

\subsection{Remainder term}


We can write the remainder term for fold $k$ as
\begin{eqnarray}
  &&
  \frac{1}{n_k} \sum_{i \in \curlS_k}
  E_{\curlG} \left\{ U_i (\piBhat_k, \mhat_k) - U_i (\piBtrue, \mtrue) \mid \curlF, \curlS_k, \piBhat_k, \mhat_k \right\}
  \nonumber \\
  && \hspace{0.5cm} =
  \frac{1}{n_k} \sum_{i \in \curlS_k} E \left[ \frac{R^B_i}{\piBhat_k (X_i)} Y_i + \left\{ \frac{R^A_i}{\pi^A_i} - \frac{R^B_i}{\piBhat_k (X_i)} \right\} \mhat_k (X_i) \mid \curlF, \curlS_k, \piBhat_k, \mhat_k \right]
  \nonumber \\
  && \hspace{1cm} -
  \frac{1}{n_k} \sum_{i \in \curlS_k} E \left[ \frac{R^B_i}{\piBtrue (X_i)} Y_i + \left\{ \frac{R^A_i}{\pi^A_i} - \frac{R^B_i}{\piBtrue (X_i)} \right\} \mtrue (X_i) \mid \curlF, \curlS_k, \piBhat_k, \mhat_k \right]
  \nonumber \\
  && \hspace{0.5cm} =
  \frac{1}{n_k} \sum_{i \in \curlS_k} \frac{ \piBtrue(X_i) }{ \piBhat_k (X_i) } \mtrue(X_i) + \left\{ \frac{ E(R^A_i \mid \curlF, \curlS_k, \piBhat_k, \mhat_k) }{\pi^A_i} - \frac{ \piBtrue(X_i) }{ \piBhat_k (X_i) } \right\} \mhat_k (X_i)
  \nonumber \\
  && \hspace{1cm} -
  \frac{1}{n_k} \sum_{i \in \curlS_k} \frac{ E(R^A_i \mid \curlF, \curlS_k, \piBhat_k, \mhat_k) }{\pi^A_i} \; \mtrue (X_i).
  \label{eq:remainder}
\end{eqnarray}

As argued in Section~\ref{sect:equal.probs}, the distribution of $\{ R_i^A: i \in \curlS_k \}$ given $\curlF, \curlS_k, \piBhat_k, \mhat_k$ corresponds to SRSWOR of $M/K$ clusters from the $J/K$ clusters in $\curlS_k$ followed by the original second-stage sampling of individuals within clusters.
Hence, $E(R^A_i \mid \curlF, \curlS_k, \piBhat_k, \mhat_k) = \pi^A_i$.
Hence, expression~(\ref{eq:remainder}) reduces to
\begin{eqnarray}
  &&
  \frac{1}{n_k} \sum_{i \in \curlS_k} \frac{ \piBtrue(X_i) }{ \piBhat_k (X_i) } \mtrue(X_i) + \left\{ 1 - \frac{ \piBtrue(X_i) }{ \piBhat_k (X_i) } \right\} \mhat_k (X_i)
 - \frac{1}{n_k} \sum_{i \in \curlS_k} \mtrue (X_i)
 \nonumber \\
 && \hspace{0.5cm} =
- \frac{1}{n_k} \sum_{i \in \curlS_k} \left\{ \frac{ \piBtrue (X_i) }{ \piBhat_k (X_i) } - 1 \right\} \{ \mhat_k (X_i) - \mtrue (X_i) \}
\label{eq:remainder.reduces.to} \\
  && =
- E \left[ \left\{ \frac{ \piBtrue (X) }{ \piBhat_k (X) } - 1 \right\} \{ \mhat_k (X) - \mtrue (X) \} \right]
  \nonumber \\
&& \hspace{0.5cm}
-
\left[ \frac{1}{n_k} \sum_{i \in \curlS_k} \left\{ \frac{ \piBtrue (X_i) }{ \piBhat_k (X_i) } - 1 \right\} \{ \mhat_k (X_i) - \mtrue (X_i) \}
  \right.
  \nonumber \\
  && \hspace{1.2cm} - \left.
  E \left[ \left\{ \frac{ \piBtrue (X) }{ \piBhat_k (X) } - 1 \right\} \{ \mhat_k (X) - \mtrue (X) \}
  \right] \right].
\label{eq:square.bracket.bit}
\end{eqnarray}

For fixed functions $\piBhat_k$ and $\mhat_k$, the term
\begin{eqnarray*}
  &&
\left[ \frac{1}{n_k} \sum_{i \in \curlS_k} \left\{ \frac{ \piBtrue (X_i) }{ \piBhat_k (X_i) } - 1 \right\} \{ \mhat_k (X_i) - \mtrue (X_i) \}
  \right.
 \nonumber \\
 && \hspace{0.5cm}
  - \left.
  E \left[ \left\{ \frac{ \piBtrue (X) }{ \piBhat_k (X) } - 1 \right\} \{ \mhat_k (X) - \mtrue (X) \}
  \right] \right]
\nonumber
\end{eqnarray*}
in expression~(\ref{eq:square.bracket.bit}) is $O_p(M^{-1/2})$.
Hence, if $\piBhat_k \convp \piBtrue$ or $\mhat_k \convp \mtrue$, then it is $o_p(M^{1/2})$.
This leaves only the term
\[
- E \left[ \left\{ \frac{ \piBtrue (X) }{ \piBhat_k (X) } - 1 \right\} \{ \mhat_k (X) - \mtrue (X) \} \right]
\]
in expression~(\ref{eq:square.bracket.bit}).
By the Cauchy-Schwarz Inequality, the absolute value of this is less than or equal to
\[
  \sqrt{
    E \left\{ \frac{ \piBtrue (X) }{ \piBhat_k (X) } - 1 \right\}^2
    \times
    E \{ \mhat_k (X) - \mtrue (X) \}^2.
  }
\]

So, if Condition C1 holds, then expression~(\ref{eq:remainder.reduces.to}) is $o_p(M^{-1/2})$, as required.

\section{Proof that empirical process and remainder terms are $o_p(M^{-1/2})$ for SWOR of clusters with unequal probabilities}

\label{sect:appendix.unequal}

\subsection{Empirical process term}

Consider the empirical process term for fold $k$.
Equations~(\ref{eq:empirical.law.total.prob}) and~(\ref{eq:Chebeshev}) still apply when clusters are selected by SWOR with unequal probabilities.
Consider the variance in expression~(\ref{eq:Chebeshev}).
\begin{eqnarray}
  &&
\Var_{\curlG} \left[ \frac{1}{n_k} \sum_{i \in \curlS_k} \{ U_i (\piBhat_k, \mhat_k) - U_i (\piBtrue, \mtrue) \} \mid \curlF, \curlS_k, \piBhat_k, \mhat_k \right]
\nonumber \\
  && \hspace{.2cm} =
E \left( \Var_{\curlG} \left[ \frac{1}{n_k} \sum_{i \in \curlS_k} \{ U_i (\piBhat_k, \mhat_k) - U_i (\piBtrue, \mtrue) \} \mid \curlF, \curlS_k, \piBhat_k, \mhat_k, M^{(.)}_k \right] \mid \curlF, \curlS_k, \piBhat_k, \mhat_k \right)
\nonumber \\
  && \hspace{.3cm} +
\Var \left( E_{\curlG} \left[ \frac{1}{n_k} \sum_{i \in \curlS_k} \{ U_i (\piBhat_k, \mhat_k) - U_i (\piBtrue, \mtrue) \} \mid \curlF, \curlS_k, \piBhat_k, \mhat_k, M^{(.)}_k \right] \mid \curlF, \curlS_k, \piBhat_k, \mhat_k \right).
\nonumber \\
&&
\label{eq:towerformulaM}
\end{eqnarray}
If we can show that the conditional expectation given $\curlF$ of each of the two summands in expression~(\ref{eq:towerformulaM}) is $o_p(M^{-1})$, then we shall have shown that expression~(\ref{eq:empirical.process}), the empirical process term for fold $k$, is $o_p (M^{-1/2})$.
We shall look at the two terms in expression~(\ref{eq:towerformulaM}) in turn.

Consider the first of the two terms in expression~(\ref{eq:towerformulaM}) and look at the inner variance term, i.e.\
\[
\Var_{\curlG} \left[ \frac{1}{n_k} \sum_{i \in \curlS_k} \{ U_i (\piBhat_k, \mhat_k) - U_i (\piBtrue, \mtrue) \} \mid \curlF, \curlS_k, \piBhat_k, \mhat_k, M^{(.)}_k \right].
\]
Analogously to equation~(\ref{eq:variance.decomposition}), we have
\begin{eqnarray}
  &&
  \Var_{\curlG} \left[ \frac{1}{n_k} \sum_{i \in \curlS_k} \{ U_i (\piBhat_k, \mhat_k) - U_i (\piBtrue, \mtrue) \} \mid \curlF, \curlS_k, \piBhat_k, \mhat_k, M^{(.)}_k \right]
  \nonumber \\
  && =
  \Var_{\curlG} \left( \frac{1}{n_k} \sum_{i \in \curlS_k} \left[ R^B_i Y_i \left\{ \frac{1}{ \piBhat_k (X_i) } - \frac{1}{ \piBtrue (X_i) } \right\}
    - R^B_i \left\{ \frac{ \mhat_k (X_i) }{ \piBhat_k (X_i) } - \frac{ \mtrue (X_i) }{ \piBtrue (X_i) } \right\} \right]
  \right.
  \nonumber \\
  && \hspace{2.5cm}
  \left. \mid \curlF, \curlS_k, \piBhat_k, \mhat_k, M^{(.)}_k \right)
  \nonumber \\
  && \hspace{0.4cm} +
  \Var_{\curlG} \left[ \frac{1}{n_k} \sum_{i \in \curlS_k} \frac{ R^A_i }{ \pi^A_i } \{ \mhat_k (X_i) - \mtrue (X_i) \} \mid \curlF, \curlS_k, \piBhat_k, \mhat_k, M^{(.)}_k \right]
  \nonumber \\
  && =
  \Var_{\curlG} \left( \frac{1}{n_k} \sum_{i \in \curlS_k} \left[ R^B_i Y_i \left\{ \frac{1}{ \piBhat_k (X_i) } - \frac{1}{ \piBtrue (X_i) } \right\}
    - R^B_i \left\{ \frac{ \mhat_k (X_i) }{ \piBhat_k (X_i) } - \frac{ \mtrue (X_i) }{ \piBtrue (X_i) } \right\} \right]
  \right.
  \nonumber \\
  && \hspace{2.5cm}
  \left. \mid \curlF, \curlS_k, \piBhat_k, \mhat_k \right)
  \nonumber \\
  && \hspace{0.4cm} +
  \Var_{\curlG} \left[ \frac{1}{n_k} \sum_{i \in \curlS_k} \frac{ R^A_i }{ \pi^A_i } \{ \mhat_k (X_i) - \mtrue (X_i) \}  \mid \curlF, \curlS_k, \piBhat_k, \mhat_k, M^{(.)}_k \right]
  \label{eq:sum.of.2vars}
\end{eqnarray}

The first of the two variance terms in expression~(\ref{eq:sum.of.2vars}) was shown earlier (see expression~(\ref{eq:first.variance})) to be $O_p(M^{-1})$ and moreover to be $o_p(M^{-1})$ if $\piBhat_k \xrightarrow{p} \piBtrue$ and $\mhat_k \xrightarrow{p} \mtrue$.

Consider the second of the two terms in expression~(\ref{eq:sum.of.2vars}).
We have
\begin{eqnarray}
  &&
  \Var_{\curlG} \left[ \frac{1}{n_k} \sum_{i \in \curlS_k} \frac{ R^A_i }{ \pi^A_i } \{ \mhat_k (X_i) - \mtrue (X_i) \}  \mid \curlF, \curlS_k, \piBhat_k, \mhat_k, M^{(.)}_k \right]
  \nonumber \\
  && =
  \Var_{\curlG} \left[ \frac{1}{n_k} \sum_{i \in \curlS_k} \sum_{l=1}^L I (\pi^C_i = \pi^{C(l)}) \frac{ R^A_i }{ \pi^A_i M^{(l)}_k / (J^{(l)}_k \pi^{C(l)}) } \; \frac{ M^{(l)}_k } { J^{(l)}_k \pi^{C(l)} }
    \right.
    \nonumber \\
    && \hspace{2cm} \times
    \left.
    \{ \mhat_k (X_i) - \mtrue (X_i) \}  \mid \curlF, \curlS_k, \piBhat_k, \mhat_k, M^{(.)}_k \right].
  \label{eq:HT.var.unequal}
\end{eqnarray}
By a very similar argument to that used in Section~\ref{sect:equal.probs}, we see that, because of the way that fold $S_k$ has been chosen, the distribution of $\{ R_i^A: i \in \curlS_k \}$ given $\curlF, \curlS_k, \piBhat_k, \mhat_k, M^{(.)}_k$ corresponds to stratified SRSWOR of $M^{(l)}_k$ clusters from the stratum of $J^{(l)}_k$ clusters with the same value of $\pi^{C(l)}$ followed by the original second-stage sampling of individuals within clusters.
The probability of sampling individual $i$ is the original probability $\pi^A_i$ multiplied by $M^{(l)}_k / (J^{(l)}_k \pi^{C(l)})$, because the fraction of clusters sampled from this stratum is $M^{(l)}_k / J^{(l)}_k$, rather than $\pi^{C(l)}$.
Hence, for any fixed function $\mhat_k$, expression~(\ref{eq:HT.var.unequal}) is the variance of the Horwitz-Thompson estimator of the population mean in the $k$th fold of $\sum_{l=1}^L I (\pi^C_i = \pi^{C(l)}) \{ M^{(l)}_k / (J^{(l)}_k \pi^{C(l)}) \} \{ \mhat_k (X_i) - \mtrue (X_i) \}$ when clusters are sampled in this stratified way.
Subject to some regularity conditions, this variance is $O_p(M^{-1})$ (Proposition 2 of \citet{ChauvetVallee2020}).
Consequently, if $\mhat_k \xrightarrow{p} \mtrue$ as $M \rightarrow \infty$, then this variance is $o_p(M^{-1})$.
Hence, the conditional expectation given $\curlF, S_k, \piBhat_k, \mhat_k$ of this variance is also $o_p(M^{-1})$.
Therefore, the first of the two terms in expression~(\ref{eq:towerformulaM}) is $o_p(M^{-1})$, as required.

Now consider the second of the two terms in expression~(\ref{eq:towerformulaM}).
We have
\begin{eqnarray*}
  &&
  E_{\curlG} \left[ \frac{1}{n_k} \sum_{i \in \curlS_k} \{ U_i (\piBhat_k, \mhat_k) - U_i (\piBtrue, \mtrue) \} \mid \curlF, \curlS_k, \piBhat_k, \mhat_k, M^{(.)}_k \right]
  \nonumber \\
  && \hspace{.2cm} =
  \frac{1}{n_k} \sum_{i \in \curlS_k}
  \left\{ \frac{ E(R^A_i \mid \curlF, \curlS_k, \piBhat_k, \mhat_k, M^{(.)}_k) }{ \pi^A_i } - \frac{ \piBtrue (X_i) }{ \piBhat_k (X_i) } \right\}
  \{ \mhat_k (X_i) - \mtrue (X_i) \}
     \nonumber \\
  \nonumber \\
  && \hspace{.2cm} =
  \frac{1}{n_k} \sum_{i \in \curlS_k}
  \left\{ \sum_{l=1}^L I(\pi^C_i = \pi^{C(l)}) \frac{ M^{(l)}_k }{ \pi^{C(l)} J^{(l)}_k } - \frac{ \piBtrue (X_i) }{ \piBhat_k (X_i) } \right\}
  \{ \mhat_k (X_i) - \mtrue (X_i) \}
     \nonumber
\end{eqnarray*}
So,
\begin{eqnarray}
  &&
  \Var \left( E_{\curlG} \left[ \frac{1}{n_k} \sum_{i \in \curlS_k} \{ U_i (\piBhat_k, \mhat_k) - U_i (\piBtrue, \mtrue) \} \mid \curlF, \curlS_k, \piBhat_k, \mhat_k, M^{(.)}_k \right] \mid \curlF, \curlS_k, \piBhat_k, \mhat_k \right)
  \nonumber \\
  && =
  \Var \left[
  \sum_{l=1}^L \frac{ M^{(l)}_k }{ \pi^{C(l)} J^{(l)}_k }
  \; \frac{1}{n_k} \sum_{i \in \curlS_k} I(\pi^C_i = \pi^{C(l)}) \{ \mhat_k (X_i) - \mtrue (X_i) \}
   \mid \curlF, \curlS_k, \piBhat_k, \mhat_k \right]
  \nonumber \\
  && =
  \Var \left[
  \sum_{l=1}^L \frac{ M^{(l)} }{ \pi^{C(l)} J^{(l)} }
  \; \frac{1}{n_k} \sum_{i \in \curlS_k} I(\pi^C_i = \pi^{C(l)}) \{ \mhat_k (X_i) - \mtrue (X_i) \}
  \right.
  \nonumber \\
  && \hspace{1.4cm} + \left.
    \sum_{l=1}^L \frac{ 1 }{ \pi^{C(l)} }
      \left(  \frac{ M^{(l)}_k }{ J^{(l)}_k } - \frac{ M^{(l)} }{ J^{(l)} } \right)
      \right.
  \nonumber \\
  && \hspace{2.5cm} \times \left.
  \frac{1}{n_k} \sum_{i \in \curlS_k} I(\pi^C_i = \pi^{C(l)}) \{ \mhat_k (X_i) - \mtrue (X_i) \} \mid \curlF, \curlS_k, \piBhat_k, \mhat_k \right]
  \nonumber \\
  && =
  \Var \left[
  \sum_{l=1}^L \frac{ M^{(l)}_k }{ \pi^{C(l)} J^{(l)}_k }
  \; \frac{1}{n_k} \sum_{i \in \curlS_k} I(\pi^C_i = \pi^{C(l)}) \{ \mhat_k (X_i) - \mtrue (X_i) \}
   \mid \curlF, \curlS_k, \piBhat_k, \mhat_k \right]
  \nonumber \\
  && \hspace{0.6cm} + D
  \label{eq:beforeD}
\end{eqnarray}
where
\begin{eqnarray}
  D & = &
  \Var \left[
    \sum_{l=1}^L \frac{ 1 }{ \pi^{C(l)} }
      \left(  \frac{ M^{(l)}_k }{ J^{(l)}_k } - \frac{ M^{(l)} }{ J^{(l)} } \right)
      \right.
  \nonumber \\
  && \hspace{1.5cm} \times \left.
  \frac{1}{n_k} \sum_{i \in \curlS_k} I(\pi^C_i = \pi^{C(l)}) \{ \mhat_k (X_i) - \mtrue (X_i) \} \mid \curlF, \curlS_k, \piBhat_k, \mhat_k \right]
  \nonumber \\
  && + \;
    2 \; \Cov \left[
    \sum_{l=1}^L \frac{ 1 }{ \pi^{C(l)} }
      \left(  \frac{ M^{(l)}_k }{ J^{(l)}_k } - \frac{ M^{(l)} }{ J^{(l)} } \right)
      \right.
  \frac{1}{n_k} \sum_{i \in \curlS_k} I(\pi^C_i = \pi^{C(l)}) \{ \mhat_k (X_i) - \mtrue (X_i) \}
  \; ,
  \nonumber \\
  && \hspace{1cm} \left.
  \sum_{l=1}^L \frac{ 1 }{ \pi^{C(l)} }
  \frac{ M^{(l)} }{ J^{(l)} }
  \frac{1}{n_k} \sum_{i \in \curlS_k} I(\pi^C_i = \pi^{C(l)}) \{ \mhat_k (X_i) - \mtrue (X_i) \}
  \mid \curlF, \curlS_k, \piBhat_k, \mhat_k \right]
    \nonumber
\end{eqnarray}

This $D$ term is a term accounts for the possible difference between $M^{(l)} / J^{(l)}$, the overall proportion of sampled clusters with $\pi^C_j = \pi^{C(l)}$, and $M^{(l)}_k / J^{(l)}_k$, the proportion in fold $k$.
Given the way that clusters are assigned to folds, this difference between proportions will become smaller as $M \rightarrow \infty$.
In Appendix~\ref{sect:MJratios} we show that
\begin{equation}
M^{(l)}_k / J^{(l)}_k - M^{(l)} / J^{(l)} = O_p (M^{-1})
\label{eq:bound.MJdiff}
\end{equation}
and in Appendix~\ref{sect:Dnegligible}, we show that this implies that $D$ is $o_p(M^{-1})$.

Now consider the first term in expression~(\ref{eq:beforeD}).
Recall that we aim to show that the conditional expectation given $\curlF$ of this is $o_p(M^{-1})$.
We have
\begin{eqnarray}
  &&
  \Var \left[
  \sum_{l=1}^L \frac{ M^{(l)} }{ \pi^{C(l)} J^{(l)} }
  \; \frac{1}{n_k} \sum_{i \in \curlS_k} I(\pi^C_i = \pi^{C(l)}) \{ \mhat_k (X_i) - \mtrue (X_i) \}
  \mid \curlF, \curlS_k, \piBhat_k, \mhat_k \right]
  \nonumber \\
  && =
  \sum_{l=1}^L \sum_{l'=l}^L \Cov \left[
    \frac{ M^{(l)} }{ \pi^{C(l)} J^{(l)} }
    \; \frac{1}{n_k} \sum_{i \in \curlS_k} I(\pi^C_i = \pi^{C(l)}) \{ \mhat_k (X_i) - \mtrue (X_i) \}
    \; ,
    \right.
  \nonumber \\
  && \hspace{2cm} \left.
    \frac{ M^{(l')} }{ \pi^{C(l')} J^{(l')} }
    \; \frac{1}{n_k} \sum_{i \in \curlS_k} I(\pi^C_i = \pi^{C(l')}) \{ \mhat_k (X_i) - \mtrue (X_i) \}
    \mid \curlF, \curlS_k, \piBhat_k, \mhat_k \right].
  \nonumber \\
  &&
  \label{eq:vars.and.covars}
\end{eqnarray}

Using the Covariance Inequality and Cauchy-Schwartz Inequality (see Appendix~\ref{sect:covariance.inequality}), we obtain
\begin{eqnarray*}
  &&
  \left| E 
  \left( \Cov \left[
    \frac{ M^{(l)} }{ \pi^{C(l)} J^{(l)} }
    \; \frac{1}{n_k} \sum_{i \in \curlS_k} I(\pi^C_i = \pi^{C(l)}) \{ \mhat_k (X_i) - \mtrue (X_i) \}
    \; ,
    \right. \right. \right.
    \nonumber \\
   && \hspace{0.5cm} \left. \left. \left.
     \frac{ M^{(l')} }{ \pi^{C(l')} J^{(l')} }
     \; \frac{1}{n_k} \sum_{i \in \curlS_k} I(\pi^C_i = \pi^{C(l')}) \{ \mhat_k (X_i) - \mtrue (X_i) \}
     \mid \curlF, \curlS_k, \piBhat_k, \mhat_k \right]
  \mid \curlF \right)
  \right|
  \nonumber \\
  && \leq
  \left\{ E \left( \Var \left[
    \frac{ M^{(l)} }{ \pi^{C(l)} J^{(l)} }
    \; \frac{1}{n_k} \sum_{i \in \curlS_k} I(\pi^C_i = \pi^{C(l)}) \{ \mhat_k (X_i) - \mtrue (X_i) \}
    \right. \right. \right.
    \nonumber \\
    && \hspace{3.5cm}
    \mid \curlF, \curlS_k, \piBhat_k, \mhat_k \bigg]
  \mid \curlF \bigg) \bigg\}^{1/2}
  \nonumber \\
  && \hspace{.4cm} \times
  \left\{ E \left( \Var \left[
    \frac{ M^{(l')} }{ \pi^{C(l')} J^{(l')} }
    \; \frac{1}{n_k} \sum_{i \in \curlS_k} I(\pi^C_i = \pi^{C(l')}) \{ \mhat_k (X_i) - \mtrue (X_i) \}
    \right. \right. \right.
    \nonumber \\
    && \hspace{3.5cm}
    \mid \curlF, \curlS_k, \piBhat_k, \mhat_k \bigg]
  \mid \curlF \bigg) \bigg\}^{1/2}.
\end{eqnarray*}

Hence, if we can show that
\begin{eqnarray}
  &&
E \left( \Var \left[
    \frac{ M^{(l)} }{ \pi^{C(l)} J^{(l)} }
    \; \frac{1}{n_k} \sum_{i \in \curlS_k} I(\pi^C_i = \pi^{C(l)}) \{ \mhat_k (X_i) - \mtrue (X_i) \}
    \right. \right.
    \nonumber \\
    && \hspace{3.5cm}
    \mid \curlF, \curlS_k, \piBhat_k, \mhat_k \bigg]
\mid \curlF \bigg)
\; = \;
O_p (M^{-1}),
\label{eq:expect.var.equalOM}
\end{eqnarray}
then the conditional expectation given $\curlF$ of expression~(\ref{eq:vars.and.covars}) is also $O_p (M^{-1})$.
From that, it follows that if $\mhat_k \xrightarrow{p} \mtrue$, then this conditional expectation will be $o_p (M^{-1})$, as required.

So, it only remains to show that equation~(\ref{eq:expect.var.equalOM}) holds.
Now,
\begin{eqnarray*}
  &&
E \left( \Var \left[
    \frac{ M^{(l)} }{ \pi^{C(l)} J^{(l)} }
    \; \frac{1}{n_k} \sum_{i \in \curlS_k} I(\pi^C_i = \pi^{C(l)}) \{ \mhat_k (X_i) - \mtrue (X_i) \}
    \right. \right.
    \nonumber \\
    && \hspace{3.5cm}
    \mid \curlF, \curlS_k, \piBhat_k, \mhat_k \bigg]
\mid \curlF, \curlS_k, \mhat_k \bigg)
\nonumber \\
&& =
\left( \frac{J}{ J^{(l)} } \right)^2 \left[ \frac{1}{n_k} \sum_{i \in \curlS_k} I(\pi^C_i = \pi^{C(l)}) \{ \mhat_k (X_i) - \mtrue (X_i) \} \right]^2
\nonumber \\
&& \hspace{0.5cm} \times
E \left\{ \Var \left(
    \frac{ M^{(l)} }{ \pi^{C(l)} J }
    \mid \curlF, \curlS_k, \piBhat_k, \mhat_k \right)
    \mid \curlF, \curlS_k, \mhat_k \right\}
\end{eqnarray*}

Also,
\begin{eqnarray}
  &&
E \left\{ \Var \left(
    \frac{ M^{(l)} }{ \pi^{C(l)} J }
    \mid \curlF, \curlS_k, \piBhat_k, \mhat_k \right)
\mid \curlF, \curlS_k, \mhat_k \right\}
\nonumber \\
&& =
\Var \left(
    \frac{ M^{(l)} }{ \pi^{C(l)} J }
    \mid \curlF, \curlS_k, \mhat_k \right)
\nonumber \\
&& \hspace{0.5cm} -
\Var \left\{ E \left(
    \frac{ M^{(l)} }{ \pi^{C(l)} J }
    \mid \curlF, \curlS_k, \piBhat_k, \mhat_k \right)
\mid \curlF, \curlS_k, \mhat_k \right\}
\nonumber \\
&& =
\Var \left(
    \frac{ M^{(l)} }{ \pi^{C(l)} J }
    \mid \curlF \right)
\nonumber \\
&& \hspace{0.5cm} -
\Var \left\{ E \left(
    \frac{ M^{(l)} }{ \pi^{C(l)} J }
    \mid \curlF, \curlS_k, \piBhat_k, \mhat_k \right)
    \mid \curlF, \curlS_k, \mhat_k \right\}
    \nonumber \\
    && =
    \Var \left( \frac{1}{J} \sum_{j=1}^J \frac{ R^C_j }{ \pi^C_j } I(\pi^C_j = \pi^{C(l)}) \mid \curlF \right)
\nonumber \\
&& \hspace{0.5cm} -
\Var \left\{ E \left(
    \frac{1}{J} \sum_{j=1}^J \frac{ R^C_j }{ \pi^C_j } I(\pi^C_j = \pi^{C(l)})
    \mid \curlF, \curlS_k, \piBhat_k, \mhat_k \right)
    \mid \curlF, \curlS_k, \mhat_k \right\}
    \nonumber \\
    &&
    \label{eq:EV.V.VE}
\end{eqnarray}
The first term in expression~(\ref{eq:EV.V.VE}) is $O_p(M^{-1})$.
Consider the expectation inside the second term of expression~(\ref{eq:EV.V.VE}).
\begin{eqnarray*}
&&
E \left(
    \frac{1}{J} \sum_{j=1}^J \frac{ R^C_j }{ \pi^C_j } I(\pi^C_j = \pi^{C(l)})
    \mid \curlF, \curlS_k, \piBhat_k, \mhat_k \right)
    \nonumber \\
    && =
    E \left(
    \frac{1}{J} \sum_{j=1}^J \frac{ R^C_j }{ \pi^C_j } I(\pi^C_j = \pi^{C(l)})
    \mid \curlF \right)
    \nonumber \\
    && \hspace{0.5cm} +
    E \left(
    \frac{1}{J} \sum_{j=1}^J \frac{ R^C_j }{ \pi^C_j } I(\pi^C_j = \pi^{C(l)})
    \mid \curlF, \curlS_k, \piBhat_k, \mhat_k \right)    
    \nonumber \\
    && \hspace{0.5cm} -
    E \left(
    \frac{1}{J} \sum_{j=1}^J \frac{ R^C_j }{ \pi^C_j } I(\pi^C_j = \pi^{C(l)})
    \mid \curlF \right)
    \nonumber \\
    && =
    \frac{ J^{(l)} }{ J }
    + E \left(
    \frac{1}{J} \sum_{j=1}^J \frac{ R^C_j }{ \pi^C_j } I(\pi^C_j = \pi^{C(l)})
    \mid \curlF, \curlS_k, \piBhat_k, \mhat_k \right)    
    \nonumber \\
    && \hspace{0.5cm} -
    E \left(
    \frac{1}{J} \sum_{j=1}^J \frac{ R^C_j }{ \pi^C_j } I(\pi^C_j = \pi^{C(l)})
    \mid \curlF \right)
\end{eqnarray*}
Hence, the second term in expression~(\ref{eq:EV.V.VE}) is $O_p(M^{-1})$ if
\begin{eqnarray}
&&
E \left(
    \frac{1}{J} \sum_{j=1}^J \frac{ R^C_j }{ \pi^C_j } I(\pi^C_j = \pi^{C(l)})
    \mid \curlF, \curlS_k, \piBhat_k, \mhat_k \right)    
    \nonumber \\
    && \hspace{0.5cm} -
    E \left(
    \frac{1}{J} \sum_{j=1}^J \frac{ R^C_j }{ \pi^C_j } I(\pi^C_j = \pi^{C(l)})
    \mid \curlF \right)
    = O_p (M^{-1/2}).
    \label{eq:noninformative.expect}
\end{eqnarray}
In Appendix~\ref{sect:noninformative.expect}, we show that equation~(\ref{eq:noninformative.expect}) holds.
Thus, equation~(\ref{eq:expect.var.equalOM}) does hold, as required.

\subsection{Remainder term}

Consider the remainder term for fold $k$, i.e.\
\begin{equation*}
  \frac{1}{n_k} \sum_{i \in \curlS_k}
   E_{\curlG} \left\{ U_i (\piBhat_k, \mhat_k) - U_i (\piBtrue, \mtrue) \mid \curlF, \curlS_k, \piBhat_k, \mhat_k \right\}.
\end{equation*}
Equation~(\ref{eq:remainder}) still applies when clusters are selected using SWOR with unequal probabilities.
As before, because of the way that the folds have been chosen, and using equation~(\ref{eq:bound.MJdiff}), we have
\begin{eqnarray}
  &&
  \frac{ E(R^A_i \mid \curlF, \curlS_k, \piBhat_k, \mhat_k) }{ \pi^A_i }
  \nonumber \\
  && \hspace{0.3cm} =
  \frac{ E \{ E(R^A_i \mid \curlF, \curlS_k, \piBhat_k, \mhat_k, M^{(.)}) 
  \mid \curlF, \curlS_k, \piBhat_k, \mhat_k \} }{ \pi^A_i }
  \nonumber \\
  && \hspace{0.3cm} =
  E \left\{ \sum_{l=1}^L I(\pi^C_i = \pi^{C(l)}) \frac{ M^{(l)}_k }{ \pi^{C(l)} J^{(l)}_k }
  \mid \curlF, \curlS_k, \piBhat_k, \mhat_k \right\}
  \nonumber \\
  && \hspace{0.3cm} =
  \sum_{l=1}^L I(\pi^C_i = \pi^{C(l)}) \frac{ 1 }{ \pi^{C(l)} }
  \; E \left( \frac{ M^{(l)}_k }{ J^{(l)}_k } \mid \curlF, \curlS_k, \piBhat_k, \mhat_k \right)
  \nonumber \\
  && \hspace{0.3cm} =
  \sum_{l=1}^L I(\pi^C_i = \pi^{C(l)}) \frac{ 1 }{ \pi^{C(l)} }
  \; E \left( \frac{ M^{(l)} }{ J^{(l)} } \mid \curlF, \curlS_k, \piBhat_k, \mhat_k \right)
  \nonumber \\
  && \hspace{0.8cm} +
  \sum_{l=1}^L I(\pi^C_i = \pi^{C(l)}) \frac{ 1 }{ \pi^{C(l)} }
  \; E \left( \frac{ M^{(l)}_k }{ J^{(l)}_k } - \frac{ M^{(l)} }{ J^{(l)} } \mid \curlF, \curlS_k, \piBhat_k, \mhat_k \right)
  \nonumber \\
  && \hspace{0.3cm} =
  \sum_{l=1}^L I(\pi^C_i = \pi^{C(l)}) \frac{ 1 }{ \pi^{C(l)} }
  \; E \left( \frac{ M^{(l)} }{ J^{(l)} } \mid \curlF, \curlS_k, \piBhat_k, \mhat_k \right)
  + O_p (M^{-1})
  \label{eq:whyOp} \\
  && \hspace{0.3cm} =
  \sum_{l=1}^L I(\pi^C_i = \pi^{C(l)})
  \frac{ J }{ J^{(l)} }
  \; E \left( \frac{1}{J} \sum_{j=1}^J \frac{ R^C_j }{ \pi^C_j } I(\pi^C_j = \pi^{C(l)}) \mid \curlF, \curlS_k, \piBhat_k, \mhat_k \right)
  + O_p (M^{-1}).
  \nonumber \\
  &&
  \label{eq:ratio.of.RA.to.piA}
\end{eqnarray}
Note that equation~(\ref{eq:whyOp}) follows from equation~(\ref{eq:bound.MJdiff}).
It follows from equation~(\ref{eq:noninformative.expect}) that equation~(\ref{eq:ratio.of.RA.to.piA}) reduces to
\begin{eqnarray}
  &&
  \frac{ E(R^A_i \mid \curlF, \curlS_k, \piBhat_k, \mhat_k) }{ \pi^A_i }
  \nonumber \\
  && \hspace{0.3cm} =
  \sum_{l=1}^L I(\pi^C_i = \pi^{C(l)})
  \frac{ J }{ J^{(l)} }
   \; E \left( \frac{1}{J} \sum_{j=1}^J \frac{ R^C_j }{ \pi^C_j } I(\pi^C_j = \pi^{C(l)}) \mid \curlF \right)
   + O_p (M^{-1/2})  
   + O_p (M^{-1}) 
   \nonumber \\
   && \hspace{0.3cm} =
   \sum_{l=1}^L I(\pi^C_i = \pi^{C(l)})
   \times 1 + O_p (M^{-1/2}) 
   \nonumber \\
   && \hspace{0.3cm} =
   1 + O_p (M^{-1/2}).
   \nonumber
\end{eqnarray}

Plugging this into equation~(\ref{eq:remainder}), we obtain
\begin{eqnarray}
  &&
  \frac{1}{n_k} \sum_{i \in \curlS_k}
  E_{\curlG} \left\{ U_i (\piBhat_k, \mhat_k) - U_i (\piBtrue, \mtrue) \mid \curlF, \curlS_k, \piBhat_k, \mhat_k \right\}
  \nonumber \\
  && \hspace{0.5cm} =
  \frac{1}{n_k} \sum_{i \in \curlS_k} \frac{ \piBtrue(X_i) }{ \piBhat_k (X_i) } \mtrue(X_i)
  + \left\{ 1 - \frac{ \piBtrue(X_i) }{ \piBhat_k (X_i) } \right\} \mhat_k (X_i)
  - \frac{1}{n_k} \sum_{i \in \curlS_k} \mtrue (X_i)
  \nonumber \\
  && \hspace{1cm} +
  \frac{1}{n_k} \sum_{i \in \curlS_k} O_p (M^{-1/2}) \{ \mhat_k (X_i) - \mtrue(X_i) \}.
  \label{eq:finalterm}
\end{eqnarray}
The final term in expression~(\ref{eq:finalterm}) is $O_p(M^{-1/2})$ for fixed $\mhat_k$.
If $\mhat_k \xrightarrow{P} \mtrue$, then this term becomes $o_p(M^{-1/2})$, and so can be ignored.
The rest of expression~(\ref{eq:finalterm}) is the same as expression~(\ref{eq:remainder.reduces.to}).
Hence, we are back to the situation we have where clusters are selected by SRSWOR.
Thus, under the conditions given there, the remainder term is $o_p(M^{-1/2})$, as required.

\section{Proof that $M^{(l)}_k / J^{(l)}_k - M^{(l)} / J^{(l)} = O_p (M^{-1})$}

\label{sect:MJratios}

By the way that the folds are chosen, we have the following bounds on $M^{(l)}_k / J^{(l)}_k - M^{(l)} / J^{(l)}$.
\begin{eqnarray*}
  \frac{ \lfloor M^{(l)} / K \rfloor }{ \lfloor J^{(l)} / K \rfloor + 1 }
  -
  \frac{ \lfloor M^{(l)} / K \rfloor + 1-K^{-1} }{ \lfloor J^{(l)} / K \rfloor }
  \leq
  \frac{ M^{(l)}_k }{ J^{(l)}_k } - \frac{ M^{(l)} }{ J^{(l)} }
  \leq
  \frac{ \lfloor M^{(l)} / K \rfloor + 1 }{ \lfloor J^{(l)} / K \rfloor }
  -
  \frac{ \lfloor M^{(l)} / K \rfloor }{ \lfloor J^{(l)} / K \rfloor + 1-K^{-1} }.
\end{eqnarray*}  
Consider the upper bound.
\begin{eqnarray*}
  \frac{ \lfloor M^{(l)} / K \rfloor + 1 }{ \lfloor J^{(l)} / K \rfloor }
  -
  \frac{ \lfloor M^{(l)} / K \rfloor }{ \lfloor J^{(l)} / K \rfloor + 1-K^{-1} }
  & = &
  \frac{ \lfloor J^{(l)} / K \rfloor + \lfloor M^{(l)} / K \rfloor (1-K^{-1}) + 1-K^{-1} }
  { \left( \lfloor J^{(l)} / K \rfloor \right)^2 + \lfloor J^{(l)} / K \rfloor (1-K^{-1}) }
  \nonumber \\
  & \rightarrow &
  \frac{ \lfloor J^{(l)} / K \rfloor + \lfloor M^{(l)} / K \rfloor (1-K^{-1}) }
       { \left( \lfloor J^{(l)} / K \rfloor \right)^2 }
       \hspace{0.5cm}
       \mbox{as } M \rightarrow \infty
  \nonumber \\
  & = &
  \frac{ 1 }{ \lfloor J^{(l)} / K \rfloor }
  +
  \frac{ \lfloor M^{(l)} / K \rfloor (1-K^{-1}) }
       { \left( \lfloor J^{(l)} / K \rfloor \right)^2 }
  \nonumber \\
  & = &
  O_p (M^{-1}).
\end{eqnarray*}  
Similarly, the lower bound is $O_p (M^{-1})$.
Hence,
\begin{equation*}
M^{(l)}_k / J^{(l)}_k - M^{(l)} / J^{(l)} = O_p (M^{-1}).
\end{equation*}

\section{Proof that $D = o_p(M^{-1})$}

\label{sect:Dnegligible}

We can write
\begin{eqnarray}
  D & = &
    \sum_{l=1}^L \sum_{l'=1}^L
  \left[ \frac{1}{n_k} \sum_{i \in \curlS_k} I(\pi^C_i = \pi^{C(l)}) \{ \mhat_k (X_i) - \mtrue (X_i) \} \right]
  \nonumber \\
  && \hspace{1.5cm} \times \;
  \left[ \frac{1}{n_k} \sum_{i \in \curlS_k} I(\pi^C_i = \pi^{C(l')}) \{ \mhat_k (X_i) - \mtrue (X_i) \} \right]
  \times
  \frac{ 1 }{ \pi^{C(l)} \pi^{C(l')} }
  \nonumber \\
  && \hspace{1.5cm} \times \;
  \Cov \left(  \frac{ M^{(l)}_k }{ J^{(l)}_k } - \frac{ M^{(l)} }{ J^{(l)} }
  \; , \;
  \frac{ M^{(l)}_k }{ J^{(l)}_k } - \frac{ M^{(l)} }{ J^{(l)} }
  \mid \curlF, \curlS_k, \piBhat_k, \mhat_k \right)
  \nonumber \\
  && + \;
  2 \; \sum_{l=1}^L \sum_{l'=1}^L
  \left[ \frac{1}{n_k} \sum_{i \in \curlS_k} I(\pi^C_i = \pi^{C(l)}) \{ \mhat_k (X_i) - \mtrue (X_i) \} \right]
    \nonumber \\
    && \hspace{2cm} \times    
    \left[ \frac{1}{n_k} \sum_{i \in \curlS_k} I(\pi^C_i = \pi^{C(l')}) \{ \mhat_k (X_i) - \mtrue (X_i) \} \right]
    \nonumber \\
    && \hspace{2cm} \times
    \frac{ 1 }{ \pi^{C(l)} \pi^{C(l')} }
    \times
    \Cov
      \left(
      \frac{ M^{(l)}_k }{ J^{(l)}_k } - \frac{ M^{(l)} }{ J^{(l)} }
      \; , \;
      \frac{ M^{(l')} }{ J^{(l')} }
      \mid \curlF, \curlS_k, \piBhat_k, \mhat_k \right).
      \nonumber \\
      &&
  \label{eq:defineD}
\end{eqnarray}

Consider the first of the two summands in expression~(\ref{eq:defineD}).
Using expression~(\ref{eq:bound.MJdiff}), we have
\begin{eqnarray*}
  &&
  \Cov_{M^{(.)}} \left( \frac{ M^{(l)}_k }{ J^{(l)}_k } - \frac{ M^{(l)} }{ J^{(l)} }
  \; , \;
  \frac{ M^{(l')}_k }{ J^{(l')}_k } - \frac{ M^{(l')} }{ J^{(l')} }
  \mid \curlF, \curlS_k, \piBhat_k, \mhat_k \right)
  \nonumber \\
  && \hspace{0.3cm} =
  \Cov_{M^{(.)}} \left\{ O_p(M^{-1}), O_p(M^{-1})
    \mid \curlF, \curlS_k, \piBhat_k, \mhat_k \right\}
    \nonumber \\
    && \hspace{0.3cm} =
    O_p (M^{-2})
    \nonumber \\
    && \hspace{0.3cm} =
    o_p (M^{-1}).
\end{eqnarray*}

Consider the second of the two summands in expression~(\ref{eq:defineD}).
We have
\begin{eqnarray}
  &&
    \left| \Cov \left(
      \frac{ M^{(l)}_k }{ J^{(l)}_k } - \frac{ M^{(l)} }{ J^{(l)} }
      \; , \;
      \frac{ M^{(l')} }{ J^{(l')} }
      \mid \curlF, \curlS_k, \piBhat_k, \mhat_k \right)
      \right|
      \nonumber \\
      && \leq
      \sqrt{ \Var \left(
      \frac{ M^{(l)}_k }{ J^{(l)}_k } - \frac{ M^{(l)} }{ J^{(l)} }
      \mid \curlF, \curlS_k, \piBhat_k, \mhat_k \right) } 
      \times
      \sqrt{ \Var \left(
      \frac{ M^{(l')} }{ J^{(l')} }
      \mid \curlF, \curlS_k, \piBhat_k, \mhat_k \right) }
      \nonumber \\
      && =
      O_p(M^{-1}) \times
      \sqrt{ \Var \left(
      \frac{ M^{(l')} }{ J^{(l')} }
      \mid \curlF, \curlS_k, \piBhat_k, \mhat_k \right) }.
      \label{eq:bound.abs.covar}
\end{eqnarray}
Note that equation~(\ref{eq:bound.abs.covar}) follows from expression~(\ref{eq:bound.MJdiff}).

The variance of a proportion cannot be greater than 1, and hence expression~(\ref{eq:bound.abs.covar}) is $O_p(M^{-1})$.
Hence, if $\mhat_k \xrightarrow{p} \mtrue$, the second of the two summands in expression~(\ref{eq:defineD}) is $o_p(M^{-1})$, as required.

Since both of the summands in expression~(\ref{eq:defineD}) are $o_p(M^{-1})$, $D$ is itself $o_p(M^{-1})$.

\section{Application of covariance inequality and Cauchy-Schwartz inequality}

\label{sect:covariance.inequality}

For any three random variables $A$, $B$ and $C$, the covariance inequality is that $\{ \Cov(A, B \mid C) \}^2 \leq \Var (A \mid C) \times \Var(B \mid C)$, and hence
\[
| \Cov(A, B \mid C) | \leq \sqrt{ \Var (A \mid C) } \times \sqrt{ \Var(B \mid C) }.
\]
Also, the Cauchy-Schwarz inequality says that $[ E_C \{ \sqrt{ \Var (A \mid C) } \times \sqrt{ \Var(B \mid C) } \} ]^2 \leq E_C [ \{ \sqrt{ \Var (A \mid C) } \}^2 ] \times E_C [ \{ \sqrt{ \Var (B \mid C) } \}^2 ] = E_C \{ \Var (A \mid C) \} \times E_C \{ \Var (B \mid C) \}$, and hence
\[
E_C \{ \sqrt{ \Var (A \mid C) } \times \sqrt{ \Var(B \mid C) } \} ]
 \leq
\sqrt{ E_C \{ \Var (A \mid C) \} \times E_C \{ \Var (B \mid C) \} }.
\]
Also,
\[
| E_C \{ \Cov(A, B \mid C) \} | \leq E_C \{ | \Cov(A, B \mid C) | \}.
\]
Putting this together, we obtain
\begin{equation}
| E_C \{ \Cov(A, B \mid C) \} |
\leq
\sqrt{ E_C \{ \Var (A \mid C) \} \times E_C \{ \Var (B \mid C) \} }.
\label{eq:covariance.inequality.result}
\end{equation}

Interpret the random variables $A$, $B$ and $C$ as follows.
\begin{eqnarray*}
  A
  & = &
  \frac{ M^{(l)} }{ \pi^{C(l)} J^{(l)} }
  \; \frac{1}{n_k} \sum_{i \in \curlS_k} I(\pi^C_i = \pi^{C(l)}) \{ \mhat_k (X_i) - \mtrue (X_i) \}
  \\
  B
  & = &
  \frac{ M^{(l')} }{ \pi^{C(l')} J^{(l')} }
  \; \frac{1}{n_k} \sum_{i \in \curlS_k} I(\pi^C_i = \pi^{C(l')}) \{ \mhat_k (X_i) - \mtrue (X_i) \}
  \\
  C
  & = &
  \{ \curlF, \curlS_k, \piBhat_k, \mhat_k \}
\end{eqnarray*}
Using inequality~(\ref{eq:covariance.inequality.result}) and conditioning throughout on $\curlF$, we obtain
\begin{eqnarray*}
  &&
  \left| E \left( \Cov \left[
    \frac{ M^{(l)} }{ \pi^{C(l)} J^{(l)} }
    \; \frac{1}{n_k} \sum_{i \in \curlS_k} I(\pi^C_i = \pi^{C(l)}) \{ \mhat_k (X_i) - \mtrue (X_i) \}
    \; ,
    \right. \right. \right.
    \nonumber \\
   && \hspace{1cm} \left. \left. \left.
     \frac{ M^{(l')} }{ \pi^{C(l')} J^{(l')} }
     \; \frac{1}{n_k} \sum_{i \in \curlS_k} I(\pi^C_i = \pi^{C(l')}) \{ \mhat_k (X_i) - \mtrue (X_i) \}
     \mid \curlF, \curlS_k, \piBhat_k, \mhat_k \right]
  \mid \curlF \right)
  \right|
  \nonumber \\
  && \leq
  \left\{ E \left( \Var \left[
    \frac{ M^{(l)} }{ \pi^{C(l)} J^{(l)} }
    \; \frac{1}{n_k} \sum_{i \in \curlS_k} I(\pi^C_i = \pi^{C(l)}) \{ \mhat_k (X_i) - \mtrue (X_i) \}
    \right. \right. \right.
    \nonumber \\
    && \hspace{3.5cm}
    \mid \curlF, \curlS_k, \piBhat_k, \mhat_k \bigg]
  \mid \curlF \bigg) \bigg\}^{1/2}
  \nonumber \\
  && \hspace{.4cm} \times
  \left\{ E \left( \Var \left[
    \frac{ M^{(l')} }{ \pi^{C(l')} J^{(l')} }
    \; \frac{1}{n_k} \sum_{i \in \curlS_k} I(\pi^C_i = \pi^{C(l')}) \{ \mhat_k (X_i) - \mtrue (X_i) \}
    \right. \right. \right.
    \nonumber \\
    && \hspace{3.5cm}
    \mid \curlF, \curlS_k, \piBhat_k, \mhat_k \bigg]
  \mid \curlF \bigg) \bigg\}^{1/2}
\end{eqnarray*}

\section{Proof of equation~(\ref{eq:noninformative.expect})}

\label{sect:noninformative.expect}

By multiplying both sides of equation~(\ref{eq:noninformative.expect}) by $\pi^{C(l)} J / J^{(l)}$ and noting that $E ( M^{(l)} / J^{(l)} \mid \curlF ) = \pi^{C(l)}$, we see that equation~(\ref{eq:noninformative.expect}) is true if and only if
\begin{equation}
  \sqrt{M} \left\{ E \left( \frac{ M^{(l)} }{ J^{(l)} } \mid \curlF, \curlS_k, \piBhat_k, \mhat_k \right)
  - \pi^{C(l)} \right\}
  = O_p (1).
  \label{eq:needtoshow}
\end{equation}
We now prove that this holds.

Define $C^* = 1$ if $C^{(l)} < \lfloor \pi^{C(l)} (1 - \truncM) (J^{(l)} - J^{(l)}_k) \rfloor$ for any $l=1, \ldots, L$, and $C^* = 0$ otherwise.
As explained informally in Section~\ref{sect:unequal.probs}, it is very likely that $C^*=0$ when $M$ is large.

Now,
\begin{eqnarray}
&&
E \left( \frac{ M^{(l)} }{ J^{(l)} } \mid \curlF, \curlS_k, \piBhat_k, \mhat_k, C^*=0 \right)  
  \nonumber \\
  && \hspace{.3cm} =
  \frac{1}{ J^{(l)} }
  \sum_{m^{(l)}=0}^M m^{(l)} \; P(M^{(l)} = m^{(l)} \mid \curlF, \curlS_k, \piBhat_k, \mhat_k, C^*=0)
  \nonumber \\
  && \hspace{.3cm} =
  \frac{1}{ J^{(l)} }
  \sum_{m^{(l)}=0}^M m^{(l)} \; P(M^{(l)} = m^{(l)} \mid \curlF, \curlS_k, C^*=0)
  \nonumber \\
  && \hspace{.6cm} \times
  \frac{ p( \piBhat_k, \mhat_k \mid \curlF, \curlS_k, C^*=0, M^{(l)} = m^{(l)} ) }
       { p( \piBhat_k, \mhat_k \mid \curlF, \curlS_k, C^*=0 ) }
  \nonumber \\
  && \hspace{.3cm} =
  \frac{1}{ J^{(l)} }
  \sum_{m^{(l)}=0}^M m^{(l)} \; P(M^{(l)} = m^{(l)} \mid \curlF, \curlS_k, C^*=0)
  \nonumber \\
  && \hspace{1cm} \times
  \frac{ p( \mhat_k \mid \curlF, \curlS_k, C^*=0, M^{(l)} = m^{(l)} ) }
       { p( \mhat_k \mid \curlF, \curlS_k, C^*=0 ) }
       \times
  \frac{ p( \piBhat_k \mid \curlF, \curlS_k, \mhat_k, C^*=0, M^{(l)} = m^{(l)} ) }
       { p( \piBhat_k \mid \curlF, \curlS_k, \mhat_k, C^*=0 ) }
  \nonumber \\
  && \hspace{.3cm} =
  \frac{1}{ J^{(l)} }
  \sum_{m^{(l)}=0}^M m^{(l)} \; P(M^{(l)} = m^{(l)} \mid \curlF, \curlS_k, C^*=0)
  \nonumber \\
  && \hspace{1cm} \times
  \frac{ p( \mhat_k \mid \curlF, \curlS_k ) }
       { p( \mhat_k \mid \curlF, \curlS_k ) }
       \times
  \frac{ p( \piBhat_k \mid \curlF, \curlS_k, \mhat_k, C^*=0, M^{(l)} = m^{(l)} ) }
       { p( \piBhat_k \mid \curlF, \curlS_k, \mhat_k, C^*=0 ) }
       \nonumber \\
  \nonumber \\
  && \hspace{.3cm} =
  \frac{1}{ J^{(l)} }
  \sum_{m^{(l)}=0}^M m^{(l)} \; P(M^{(l)} = m^{(l)} \mid \curlF, \curlS_k, C^*=0)
  \nonumber \\
  && \hspace{1cm} \times
  \frac{ p( \piBhat_k \mid \curlF, \curlS_k, \mhat_k, C^*=0, M^{(l)} = m^{(l)} ) }
       { p( \piBhat_k \mid \curlF, \curlS_k, \mhat_k, C^*=0 ) }
       \label{ratio.of.pis}
\end{eqnarray}
We shall now show that the ratio of densities in expression~(\ref{ratio.of.pis}) equals 1.

When $C^*=0$, define $R^{C*}_{-k}$ to be the set of indices of the $\sum_{l=1}^L \lfloor \pi^{C(l)} (1 - \epsilon) (J^{(l)} - J^{(l)}_k) \rfloor$ clusters in $\curlS_{-k}$ that are used to estimate $\piBhat_k$.
Let $\mathcal{R}^{C*}_{-k}$ denote the set of possible values of $R^{C*}_{-k}$ that are compatible with $\curlF$, $\curlS_k$ and $C^*=0$.
The size of this set is
\[
\prod_{l=1}^L \left( \begin{array}{c}
  J^{(l)} - J^{(l)}_k \\
  \lfloor \pi^{C(l)} (1 - \epsilon) (J^{(l)} - J^{(l)}_k) \rfloor
\end{array} \right),
\]
i.e.\ the product over $l=1, \ldots, L$ of the number of ways of choosing $\lfloor \pi^{C(l)} (1 - \epsilon) (J^{(l)} - J^{(l)}_k) \rfloor$ elements from a set of $J^{(l)} - J^{(l)}_k$ elements.

Now,
\begin{eqnarray}
  &&
  p( \piBhat_k \mid \curlF, \curlS_k, \mhat_k, C^*=0, M^{(l)} = m^{(l)} )
  \nonumber \\
  && \hspace{.3cm} =
  \sum_{r^{C*}_{-k} \in \mathcal{R}^{C*}_{-k}} p( \piBhat_k \mid \curlF, \curlS_k, \mhat_k, C^*=0, M^{(l)} = m^{(l)}, R^{C*}_{-k} = r^{C*}_{-k} )
    \nonumber \\
    && \hspace{2cm} \times
  P( R^{C*}_{-k} = r^{C*}_{-k} \mid \curlF, \curlS_k, \mhat_k, C^*=0, M^{(l)} = m^{(l)} )
  \nonumber \\
  && \hspace{.3cm} =
  \sum_{r^{C*}_{-k} \in \mathcal{R}^{C*}_{-k}} p( \piBhat_k \mid \curlF, \curlS_k, \mhat_k, R^{C*}_{-k} = r^{C*}_{-k} )
    \nonumber \\
    && \hspace{2cm} \times
  P( R^{C*}_{-k} = r^{C*}_{-k} \mid \curlF, \curlS_k, \mhat_k, C^*=0, M^{(l)} = m^{(l)} )
  \nonumber \\
  && \hspace{.3cm} =
  \sum_{r^{C*}_{-k} \in \mathcal{R}^{C*}_{-k}} p( \piBhat_k \mid \curlF, \curlS_k, \mhat_k, R^{C*}_{-k} = r^{C*}_{-k} )
  \bigg/
    \prod_{l=1}^L \left( \begin{array}{c}
      J^{(l)} - J^{(l)}_k \\
      \lfloor \pi^{C(l)} (1 - \epsilon) (J^{(l)} - J^{(l)}_k) \rfloor
    \end{array} \right).
    \nonumber \\
    &&    
    \label{eq:not.func.of.M}
\end{eqnarray}
Since expression~(\ref{eq:not.func.of.M}) is not a function of $m^{(l)}$, it follows that
\[
  p( \piBhat_k \mid \curlF, \curlS_k, \mhat_k, C^*=0, M^{(l)} = m^{(l)} )
  =
  p( \piBhat_k \mid \curlF, \curlS_k, \mhat_k, C^*=0 ).
\]

Therefore, returning to equation~(\ref{ratio.of.pis}), we have
\begin{eqnarray}
&&
E \left( \frac{ M^{(l)} }{ J^{(l)} } \mid \curlF, \curlS_k, \piBhat_k, \mhat_k, C^*=0 \right)  
\nonumber \\
&& \hspace{.3cm} =
\frac{1}{ J^{(l)} }
\sum_{m^{(l)}=0}^M m^{(l)} \; P(M^{(l)} = m^{(l)} \mid \curlF, \curlS_k, C^*=0)
  \nonumber \\
&& \hspace{.3cm} =
  E \left( \frac{ M^{(l)} }{ J^{(l)} } \mid \curlF, \curlS_k, , C^*=0 \right).  
\label{eq:equality.of.EMJ}
\end{eqnarray}

Using equation~(\ref{eq:equality.of.EMJ}), we have
\begin{eqnarray}
&&
  \sqrt{M} \; \left\{ E \left( \frac{ M^{(l)} }{ J^{(l)} } \mid \curlF, \curlS_k, \piBhat_k, \mhat_k \right) - \pi^{C(l)} \right\}
  \nonumber \\
  && \hspace{0.5cm} =
  \sqrt{M} \; P(C^*=1 \mid \curlF, \curlS_k, \piBhat_k, \mhat_k) \;
  \left\{ E \left( \frac{ M^{(l)} }{ J^{(l)} } \mid \curlF, \curlS_k, \piBhat_k, \mhat_k, C^*=1 \right) - \pi^{C(l)} \right\}
  \nonumber \\
  && \hspace{1cm} +
  \sqrt{M} \; P(C^*=0 \mid \curlF, \curlS_k, \piBhat_k, \mhat_k) \;
  \left\{ E \left( \frac{ M^{(l)} }{ J^{(l)} } \mid \curlF, \curlS_k, \piBhat_k, \mhat_k, C^*=0 \right) - \pi^{C(l)} \right\}
  \nonumber \\
  && \hspace{0.5cm} =
  \sqrt{M} \; P(C^*=1 \mid \curlF, \curlS_k, \piBhat_k, \mhat_k) \;
  \left\{ E \left( \frac{ M^{(l)} }{ J^{(l)} } \mid \curlF, \curlS_k, \piBhat_k, \mhat_k, C^*=1 \right) - \pi^{C(l)} \right\}
  \nonumber \\
  && \hspace{1cm} +
  \sqrt{M} \; P(C^*=0 \mid \curlF, \curlS_k, \piBhat_k, \mhat_k) \;
  \left\{ E \left( \frac{ M^{(l)} }{ J^{(l)} } \mid \curlF, \curlS_k, C^*=0 \right) - \pi^{C(l)} \right\}.
  \nonumber \\
  &&
  \label{eq:separateC1andC0}
\end{eqnarray}

Now,
\begin{eqnarray}
  P(C^* = 1 \mid \curlF, \curlS_k)
  & \leq &
  \sum_{l=1}^L P \left( \frac{1}{J} \sum_{j=1}^J \frac{ R^C_j }{ \pi^C_j } I(\pi^C_j = \pi^{C(l)}) < \frac{ J^{(l)} }{ J } (1 - \epsilon) \mid \curlF, \curlS_k \right)
  \nonumber \\
  & = &
  \sum_{l=1}^L P \left( \frac{1}{J} \sum_{j=1}^J \frac{ R^C_j }{ \pi^C_j } I(\pi^C_j = \pi^{C(l)}) - \frac{ J^{(l)} } { J } < - \frac{ J^{(l)} }{ J } \epsilon \mid \curlF, \curlS_k \right)
  \nonumber \\
  & \leq &
  \sum_{l=1}^L P \left( \left| \frac{1}{J} \sum_{j=1}^J \frac{ R^C_j }{ \pi^C_j } I(\pi^C_j = \pi^{C(l)}) - \frac{ J^{(l)} } { J } \right| > \frac{ J^{(l)} }{ J } \epsilon \mid \curlF, \curlS_k \right)
  \nonumber \\
  & \leq &
  \sum_{l=1}^L \frac{ \Var \left( \frac{1}{J} \sum_{j=1}^J \frac{ R^C_j }{ \pi^C_j } I(\pi^C_j = \pi^{C(l)}) \mid \curlF, \curlS_k \right) }
      { \left( \frac{ J^{(l)} \epsilon }{ J } \right)^2  }
      \label{eq:Chebeshev.again} \\
  & = &
  \sum_{l=1}^L \left( \frac{ J }{ J^{(l)} \epsilon } \right)^2
  \Var \left( \frac{1}{J} \sum_{j=1}^J \frac{ R^C_j }{ \pi^C_j } I(\pi^C_j = \pi^{C(l)}) \mid \curlF \right)
  \nonumber \\
      & = &
      O_p (M^{-1}).
      \label{eq:probCequals1}
\end{eqnarray}
Note that line~(\ref{eq:Chebeshev.again}) uses Chebeshev's Inequality.

It follows from equation~(\ref{eq:probCequals1}) that
\begin{eqnarray}
  0
  & = &
\sqrt{M} \; \left\{ E \left( \frac{ M^{(l)} }{ J^{(l)} } \mid \curlF, \curlS_k \right) - \pi^{C(l)} \right\}
  \nonumber \\
  & = &  
  \sqrt{M} \; P(C^* = 0 \mid \curlF, \curlS_k) \left\{ E \left( \frac{ M^{(l)} }{ J^{(l)} } \mid \curlF, \curlS_k, C^*=0 \right) - \pi^{C(l)} \right\}
  \nonumber \\
  && +  
  \sqrt{M} \; P(C^* = 1 \mid \curlF, \curlS_k) \left\{ E \left( \frac{ M^{(l)} }{ J^{(l)} } \mid \curlF, \curlS_k, C^*=1 \right) - \pi^{C(l)} \right\}
  \nonumber \\
  & = &  
  \sqrt{M} \; \{ 1 - O_p (M^{-1}) \} \left\{ E \left( \frac{ M^{(l)} }{ J^{(l)} } \mid \curlF, \curlS_k, C^*=0 \right) - \pi^{C(l)} \right\}
  \nonumber \\
  && +
  \sqrt{M} \; O_p (M^{-1}) \left\{ E \left( \frac{ M^{(l)} }{ J^{(l)} } \mid \curlF, \curlS_k, C^*=1 \right) - \pi^{C(l)} \right\}
  \nonumber \\
  & = &  
  \sqrt{M} \; \left\{ E \left( \frac{ M^{(l)} }{ J^{(l)} } \mid \curlF, \curlS_k, C^*=0 \right) - \pi^{C(l)} \right\}
  + o_p(1).
  \nonumber
\end{eqnarray}

Hence
\begin{equation}
\sqrt{M} \left\{ E \left( \frac{ M^{(l)} }{ J^{(l)} } \mid \curlF, \curlS_k, C^*=0 \right) - \pi^{C(l)} \right\}
= o_p (1).
\label{eq:expectCequals0isop0}
\end{equation}

Now, returning to equation~(\ref{eq:separateC1andC0}) and using equation~(\ref{eq:expectCequals0isop0}), we have
\begin{eqnarray}
&&
  \sqrt{M} \; \left\{ E \left( \frac{ M^{(l)} }{ J^{(l)} } \mid \curlF, \curlS_k, \piBhat_k, \mhat_k \right) - \pi^{C(l)} \right\}
  \nonumber \\
  && \hspace{0.5cm} =
  \sqrt{M} \; P(C^*=1 \mid \curlF, \curlS_k, \piBhat_k, \mhat_k) \;
  \left\{ E \left( \frac{ M^{(l)} }{ J^{(l)} } \mid \curlF, \curlS_k, \piBhat_k, \mhat_k, C^*=1 \right) - \pi^{C(l)} \right\}
  \nonumber \\
  && \hspace{1cm} +
  \sqrt{M} \; P(C^*=0 \mid \curlF, \curlS_k, \piBhat_k, \mhat_k) \;
  \left\{ E \left( \frac{ M^{(l)} }{ J^{(l)} } \mid \curlF, \curlS_k, C^*=0 \right) - \pi^{C(l)} \right\}
  \nonumber \\
  && \hspace{0.5cm} =
  \sqrt{M} \; P(C^*=1 \mid \curlF, \curlS_k, \piBhat_k, \mhat_k) \;
  \left\{ E \left( \frac{ M^{(l)} }{ J^{(l)} } \mid \curlF, \curlS_k, \piBhat_k, \mhat_k, C^*=1 \right) - \pi^{C(l)} \right\}
  \nonumber \\
  && \hspace{1cm} +
  o_p(1).
  \label{eq:C0termvanishes}
\end{eqnarray}

Since the expectation of a proportion is bounded by zero and one, it follows from equation~(\ref{eq:C0termvanishes}) that equation~(\ref{eq:needtoshow}) holds if
\begin{equation}
  \sqrt{M} \; P(C^*=1 \mid \curlF, \curlS_k, \piBhat_k, \mhat_k) = O_p(1).
\label{eq:MP.Oone}
\end{equation}
We shall now prove that equation~(\ref{eq:MP.Oone}) does hold.

Making use of equation~(\ref{eq:probCequals1}), we have
\begin{eqnarray}
  &&
  \sqrt{M} \; P(C^*=1 \mid \curlF, \curlS_k, \piBhat_k, \mhat_k)
  \nonumber \\
  && \hspace{0.3cm} =
  \sqrt{M} \; P(C^*=1 \mid \curlF, \curlS_k)
  \; \frac{ p(\piBhat_k, \mhat_k \mid \curlF, \curlS_k, C^*=1) }
  { p(\piBhat_k, \mhat_k \mid \curlF, \curlS_k) }
  \nonumber \\
  && \hspace{0.3cm} =
  O_p(M^{-1/2}) \;
  \frac{ p(\piBhat_k, \mhat_k \mid \curlF, \curlS_k, C^*=1) }
  { p(\piBhat_k, \mhat_k \mid \curlF, \curlS_k) }
  \nonumber \\
  && \hspace{0.3cm} =
  O_p(M^{-1/2}) \;
  \frac{ p(\mhat_k \mid \curlF, \curlS_k, C^*=1) }
  { p(\mhat_k \mid \curlF, \curlS_k) }
  \; \frac{ p(\piBhat_k \mid \curlF, \curlS_k, \mhat_k, C^*=1) }
  { p(\piBhat_k \mid \curlF, \curlS_k, \mhat_k) }
  \nonumber \\
  && \hspace{0.3cm} =
  O_p(M^{-1/2}) \;
  \frac{ p(\mhat_k \mid \curlF, \curlS_k) }
  { p(\mhat_k \mid \curlF, \curlS_k) }
  \; \frac{ p(\piBhat_k \mid \curlF, \curlS_k, \mhat_k, C^*=1) }
  { p(\piBhat_k \mid \curlF, \curlS_k, \mhat_k) }
  \label{eq:becauseMdoesnotdepend} \\
  && \hspace{0.3cm} =
  O_p(1) \;
  \frac{ p(\piBhat_k \mid \curlF, \curlS_k, \mhat_k, C^*=1) }
  { \sqrt{M} \; p(\piBhat_k \mid \curlF, \curlS_k, \mhat_k) }.
  \label{eq:OpM.ratio}
\end{eqnarray}
Note that line~(\ref{eq:becauseMdoesnotdepend}) follows because $\mhat_k$ depends only on $\{ (X_i, Y_i): i \in \curlS_{-k} \mbox{ and } R^B_i=1 \}$, which are conditionally independent of the $R^C_j$'s, and hence of $C^*$, given $\curlF$ and $\curlS_k$.
 
Hence, equation~(\ref{eq:MP.Oone}) holds if
\[
  \frac{ p(\piBhat_k \mid \curlF, \curlS_k, \mhat_k, C^*=1) }
       { \sqrt{M} \; p(\piBhat_k \mid \curlF, \curlS_k, \mhat_k) }
      = O_p(1).
\]

Now,
\begin{eqnarray}
  &&
E \left[ \frac{ p(\piBhat_k \mid \curlF, \curlS_k, \mhat_k, C^*=1) }
  { \sqrt{M} \; p(\piBhat_k \mid \curlF, \curlS_k, \mhat_k) }
  \mid \curlF, \curlS_k, \mhat_k \right]
\nonumber \\
&& =
\int \frac{ p(\piBhat_k \mid \curlF, \curlS_k, \mhat_k, C^*=1) }
{ \sqrt{M} \; p(\piBhat_k \mid \curlF, \curlS_k, \mhat_k) }
\; p(\piBhat_k \mid \curlF, \curlS_k, \mhat_k) \; d \piBhat_k
\nonumber \\
&& =
\int \frac{ p(\piBhat_k \mid \curlF, \curlS_k, \mhat_k, C^*=1) }
{ \sqrt{M} } \; d \piBhat_k
\nonumber \\
&& =
\frac{1}{ \sqrt{M} }.
\nonumber
\end{eqnarray}

Thus, using Markov's inequality, we have that for any $c>0$,
\begin{eqnarray}
  &&
P \left[ \frac{ p(\piBhat_k \mid \curlF, \curlS_k, \mhat_k, C^*=1) }
  { \sqrt{M} \; p(\piBhat_k \mid \curlF, \curlS_k, \mhat_k) }
  \geq c
  \mid \curlF, \curlS_k, \mhat_k \right]
\nonumber \\
&& \leq
\frac{1}{c} \; E \left[ \frac{ p(\piBhat_k \mid \curlF, \curlS_k, \mhat_k, C^*=1) }
  { \sqrt{M} \; p(\piBhat_k \mid \curlF, \curlS_k, \mhat_k) }
  \mid \curlF, \curlS_k, \mhat_k \right]
\nonumber \\
&& =
\frac{1}{ c \; \sqrt{M} }.
\end{eqnarray}
Hence,
\[
\frac{ p(\piBhat_k \mid \curlF, \curlS_k, \mhat_k, C^*=1) }
     { \sqrt{M} \; p(\piBhat_k \mid \curlF, \curlS_k, \mhat_k) }
     = O_p(M^{-1/2}).
\]
which is a stronger result than we needed.
It now follows from equation~(\ref{eq:OpM.ratio}) that
\[
\sqrt{M} \; P(C^*=1 \mid \curlF, \curlS_k, \piBhat_k, \mhat_k) = O_p(M^{-1/2}).
\]

Therefore, equation~(\ref{eq:needtoshow}) holds.
In fact, it also holds with $O_p(1)$ replaced by $O_p(M^{-1/2})$.

\section{Proof of equation~(\ref{eq:DR3.expansion}) and $\thetahatthr = \thetahattwo + o_p (M^{-1/2})$}

\label{sect:appendix.DR2}

Write $\thetahatthr$ as $\thetahatthr = \thetahatthr (\underpiBhat, \undermhat, \tauhat)$, where
\[
\thetahatthr (\underpiB, \underm, \tau)
=
\frac{1}{n} \sum_{k=1}^K \sum_{i \in \curlS_k} \Uthr_i (\pi^B_k, m_k, \tau),
\]
and
\[
\tauhat = \frac{1}{n} \sum_{i=1}^n \frac{ R^A_i }{ \pi^A_i },
\]
with
\[
\Uthr (\pi^B_k, m_k, \tau)
= \frac{1}{\tau} \left[
  \frac{ R^A }{ \pi^A } m_k (X) + \frac{ R^B }{ \pi^B_k(X) } \{ Y - m_k (X) \}
  \right].
\]
Also define $\bar{m}_0 = \frac{1}{n} \sum_{i=1}^n \mtrue (X_i)$.

Now, by a Taylor-series expansion,
\begin{eqnarray}
  &&
\sqrt{M} \; \{ \thetahatthr (\underpiBhat, \undermhat, \tauhat) - \Ybar \}
\nonumber \\
&& \hspace{.5cm} =
\sqrt{M} \; \frac{ 1 }{ n } \sum_{k=1}^K \sum_{i \in \curlS_k} \Uthr_i (\piBhat_k, \mhat_k, 1)
+
\sqrt{M} \; \frac{ 1 }{ n } \sum_{k=1}^K \sum_{i \in \curlS_k} \left. \frac{ \partial \Uthr_i }{ \partial \tau } (\piBhat_k, \mhat_k, \tau) \right|_{\tau = 1}  \; (\tauhat - 1)
\nonumber \\
&& \hspace{.9cm} -
\sqrt{M} \; \Ybar + o_p(1)
\nonumber \\
&& \hspace{.5cm} =
\sqrt{M} \; \frac{ 1 }{ n } \sum_{k=1}^K \sum_{i \in \curlS_k} \Uthr_i (\piBhat_k, \mhat_k, 1)
-
\sqrt{M} \; \frac{ 1 }{ n } \sum_{k=1}^K \sum_{i \in \curlS_k} \Uthr_i (\piBhat_k, \mhat_k, 1) \times (\tauhat - 1)
\nonumber \\
&& \hspace{.9cm} -
\sqrt{M} \; \Ybar + o_p(1)
\nonumber \\
&& \hspace{.5cm} =
\sqrt{M} \; \frac{ 1 }{ n } \sum_{k=1}^K \sum_{i \in \curlS_k} U_i (\piBhat_k, \mhat_k)
-
\sqrt{M} \; \frac{ 1 }{ n } \sum_{k=1}^K \sum_{i \in \curlS_k} U_i (\piBhat_k, \mhat_k) \times (\tauhat - 1)
\nonumber \\
&& \hspace{.9cm} -
\sqrt{M} \; \Ybar + o_p(1)
\nonumber \\
&& \hspace{.5cm} =
\sqrt{M} \; \left[ \frac{ 1 }{ n } \sum_{i=1}^n U_i (\piBtrue, \mtrue)
  + o_p(M^{-1/2}) \right]
\nonumber \\
&& \hspace{.9cm} -
\sqrt{M} \; \left[ \frac{ 1 }{ n } \sum_{i=1}^n U_i (\piBtrue, \mtrue) + o_p(M^{-1/2}) \right] \times (\tauhat - 1)
\nonumber \\
&& \hspace{.9cm} -
\sqrt{M} \; \Ybar + o_p(1)
\label{eq:use.DR1.valid} \\
&& \hspace{.5cm} =
\sqrt{M} \; \frac{ 1 }{ n } \sum_{i=1}^n U_i (\piBtrue, \mtrue)
-
\sqrt{M} \; \frac{ 1 }{ n } \sum_{i=1}^n U_i (\piBtrue, \mtrue) \times (\tauhat - 1)
-
\sqrt{M} \; \Ybar + o_p(1)
\nonumber \\
&& \hspace{.5cm} =
\sqrt{M} \; \frac{ 1 }{ n } \sum_{i=1}^n
\frac{ R^A_i }{ \pi^A_i } \mtrue (X_i)
+
\sqrt{M} \; \frac{ 1 }{ n } \sum_{i=1}^n
\frac{ R^B_i }{ \piBtrue (X_i) } \{ Y_i - \mtrue (X_i) \}
\nonumber \\
&& \hspace{.9cm} -
\sqrt{M} \; \{ \bar{m}_0 + o_p(1) \}
\times ( \tauhat - 1 )
- \sqrt{M} \; \Ybar + o_p(1)
\nonumber \\
&& \hspace{.5cm} =
\sqrt{M} \; \frac{ 1 }{ n } \sum_{i=1}^n
\frac{ R^A_i }{ \pi^A_i } \mtrue (X_i)
+
\sqrt{M} \; \frac{ 1 }{ n } \sum_{i=1}^n
\frac{ R^B_i }{ \piBtrue (X_i) } \{ Y_i - \mtrue (X_i) \}
\nonumber \\
&& \hspace{.9cm} -
\{ \bar{m}_0 + o_p(1) \} \;
\sqrt{M} \; \frac{1}{n} \sum_{i=1}^n
\left( \frac{ R^A_i }{ \pi^A_i } - 1 \right)
- \sqrt{M} \; \Ybar + o_p(1)
\nonumber \\
&& \hspace{.5cm} =
\sqrt{M} \; \frac{ 1 }{ n } \sum_{i=1}^n
\frac{ R^A_i }{ \pi^A_i } \{ \mtrue (X_i) - \bar{m}_0 \}
+
\sqrt{M} \; \frac{ 1 }{ n } \sum_{i=1}^n
\frac{ R^B_i }{ \piBtrue (X_i) } \{ Y_i - \mtrue (X_i) \}
\nonumber \\
&& \hspace{.9cm} +
\sqrt{M} ( \bar{m}_0 - \Ybar ) + o_p(1)
\label{eq:DR3.expansion.appendix}
\end{eqnarray}

This is the same as equation~(\ref{eq:DR3.expansion}).
Note that line~(\ref{eq:use.DR1.valid}) uses the result that $\thetahat (\underpiBhat, \undermhat) = \thetahat(\piBtrue, \mtrue) + o_p (M^{-1/2})$.

\citet{Seaman2025} show that
\begin{eqnarray}
  &&
\sqrt{M} \; \{ \thetahattwo (\underpiBhat, \undermhat) - \Ybar \}
\nonumber \\
&& \hspace{.5cm} =
\sqrt{M} \; \frac{ 1 }{ n } \sum_{i=1}^n
\frac{ R^A_i }{ \pi^A_i } \{ \mtrue (X_i) - \bar{m}_0 \}
+
\sqrt{M} \; \frac{ 1 }{ n } \sum_{i=1}^n
\frac{ R^B_i }{ \piBtrue (X_i) } \{ Y_i - \mtrue (X_i)  \}
\nonumber \\
&& \hspace{.9cm} +
\sqrt{M} \; \frac{ 1 }{ n } \sum_{i=1}^n
\frac{ R^B_i }{ \piBtrue (X_i) } ( \bar{m}_0 - \Ybar )
+ o_p(1)
\label{eq:CLW2.expansion.appendix}
\end{eqnarray}

Comparing equations~(\ref{eq:DR3.expansion.appendix}) and~(\ref{eq:CLW2.expansion.appendix}), we see that
\begin{eqnarray*}
  \sqrt{M} \; \{ \thetahatthr (\underpiBhat, \undermhat, \tauhat) - \Ybar \}
  & = &
  \sqrt{M} \; \{ \thetahattwo (\underpiBhat, \undermhat) - \Ybar \}
\nonumber \\
&& -
\sqrt{M}
\frac{ 1 }{ n } \sum_{i=1}^n \left\{ \frac{ R^B_i }{ \piBtrue (X_i) } - 1 \right\} ( \bar{m}_0 - \Ybar )
+ o_p(1)
\nonumber
\end{eqnarray*}
If the population is large compared to Sample B, i.e.\ if $n$ is large compared to $\sum_{i=1}^n R^B_i$, then $\bar{m}_0 - \Ybar$ is practically negligible, and thus $\sqrt{M} (\thetahattwo - \Ybar) \approx \sqrt{M} (\thetahatthr - \Ybar) + o_p (1)$.

\section{Proof that $\tmleone = \thetahat + o_p (M^{1/2})$}

\label{sect:appendix.tmle}

Using a Taylor series expansion, we obtain
\begin{eqnarray*}
  &&
  \frac{1}{n}
  \sum_{i \in \curlS_k} U_i ( \piBhat_k, \mhatstar_k )
  \nonumber \\
  && =
    \frac{1}{n}
    \sum_{i \in \curlS_k} U_i ( \piBhat_k, \mhat_k )
    +
    \left. \frac{ d }{ d \epsilon_k }
    \left[
      \frac{1}{n}
      \sum_{i \in \curlS_k} U_i \{ \piBhat_k, \mhat_k (\epsilon_k) \}
      \right] \right|_{\epsilon_k = 0}
    \epsilonhat_k
    +
    o_p (\epsilonhat_k^2).
    \nonumber
\end{eqnarray*}

Now, if $\mhat_k (x; \epsilon_k)$ is defined by equation~(\ref{eq:TMLE.linear}), then
\begin{eqnarray}
    \left. \frac{ d }{ d \epsilon_k }
    \left[
      \frac{1}{n}
      \sum_{i \in \curlS_k} U_i \{ \piBhat_k, \mhat_k (\epsilon_k) \}
      \right] \right|_{\epsilon_k = 0}
    & = &
    \frac{1}{n} \sum_{i \in \curlS_k}
  \left\{ \frac{ R^A_i }{ \pi^A_i } - \frac{ R^B_i }{ \piBhat (X_i) } \right\}
  \frac{ 1 }{ \piBhat (X_i) }
  \nonumber \\
    & = &
    \frac{1}{n} \sum_{i \in \curlS_k}
    \left\{ 1 - \frac{ R^B_i }{ \piBhat (X_i) } \right\}
  \frac{ 1 }{ \piBhat (X_i) }
  \nonumber \\
  && +
  \frac{1}{n} \sum_{i \in \curlS_k}
  \left( \frac{ R^A_i }{ \pi^A_i } - 1 \right)
  \frac{ 1 }{ \piBhat (X_i) }
  \label{eq:tmle.rA}
\end{eqnarray}

Now, for any fixed $\piBhat_k$ and $\mhat_k$,
\begin{eqnarray}
  &&
  \frac{1}{n} \sum_{i \in \curlS_k}
    \left\{ 1 - \frac{ R^B_i }{ \piBhat (X_i) } \right\}
    \frac{ 1 }{ \piBhat (X_i) }
  \nonumber \\
  && \hspace{.3cm} =
  \frac{1}{n} \sum_{i \in \curlS_k}
    \left\{ 1 - \frac{ \pi^B (X_i) }{ \piBhat (X_i) } \right\}
    \frac{ 1 }{ \piBhat (X_i) }
  \nonumber \\
  && \hspace{.8cm} -
    \frac{1}{n} \sum_{i \in \curlS_k}
    \{ R^B_i - \pi^B (X_i) \} \;
    \frac{ 1 }{ \piBhat (X_i)^2 }
  \nonumber \\
  && \hspace{.3cm} =
  E_X \left[
    \left\{ 1 - \frac{ \pi^B (X) }{ \piBhat (X) } \right\}
    \frac{ 1 }{ \piBhat (X) } \right]
    + O_p (M^{-1/2})
  \nonumber \\
  && \hspace{.8cm} -
  E_{X, R^B} \left[
    \frac{1}{n} \sum_{i \in \curlS}
    \{ R^B - \pi^B (X) \} \;
    \frac{ 1 }{ \piBhat (X_i) }
    \right]
  + O_p (M^{-1/2})
    \nonumber \\
    && \hspace{.3cm} =
  E_X \left[
    \left\{ 1 - \frac{ \pi^B (X) }{ \piBhat (X) } \right\}
    \frac{ 1 }{ \piBhat (X) } \right]
    + O_p (M^{-1/2})
  \nonumber \\
  && \hspace{.3cm} \leq
  \sqrt{
    E_X \left[ \left\{ 1 - \frac{ \pi^B (X) }{ \piBhat (X) } \right\}^2 \right] \;
    E_X \left[ \frac{ 1 }{ \piBhat (X)^2 } \right]
  }
  + O_p (M^{-1/2})
  \label{eq:cauchy.schwartz} \\
  && \hspace{.3cm} =
  \sqrt{ o_p (M^{- \piBrate }) } + O_p (M^{-1/2})
  \nonumber \\
  && \hspace{.3cm} =
  o_p (M^{- \piBrate /2}).
\end{eqnarray}
Note that line~(\ref{eq:cauchy.schwartz}) uses the Cauchy-Schwartz inequality.

If $\mhat (x; \epsilon_k)$ is instead defined by equation~(\ref{eq:TMLE.logit}), then
\begin{eqnarray*}
    \left. \frac{ d }{ d \epsilon_k }
    \left[
      \frac{1}{n}
      \sum_{i \in \curlS_k} U_i \{ \piBhat_k, \mhat_k (\epsilon_k) \}
      \right] \right|_{\epsilon_k = 0}
    & = &
    \frac{1}{n} \sum_{i \in \curlS_k}
  \left\{ \frac{ R^A_i }{ \pi^A_i } - \frac{ R^B_i }{ \piBhat (X_i) } \right\}
  \frac{ 1 }{ \piBhat (X_i) }
  \frac{ \mhat (X_i) }{ \{ 1 + \mhat (X_i) \}^2 },
\end{eqnarray*}
and the same argument shows that this is $o_p (M^{- \piBrate /2})$.

By using a similar argument to that used in Appendices~\ref{sect:appendix.simple} and~\ref{sect:appendix.unequal}, the term
\[
\frac{1}{n} \sum_{i \in \curlS}
\left( \frac{ R^A_i }{ \pi^A_i } - 1 \right)
\frac{ 1 }{ \piBhat (X_i) }
\]
in expression~(\ref{eq:tmle.rA}) can be shown to be $O_p(M^{-1/2})$ given any fixed $\piBhat_k$.

We shall now show that $\epsilonhat_k$ is $o_p(M^{- \mrate /2})$.
First, consider the case where $\mhat_k (x; \epsilon_k)$ is defined by equation~(\ref{eq:TMLE.linear}).
For any fixed $\piBhat$ and $\mhat$, the first iteration of a Newton-Raphson algorithm to find the maximum likelihood estimate of $\epsilon_k$ starting from an initial value of $\epsilonhat_k=0$ would set $\epsilonhat_k$ equal to
\begin{eqnarray*}
  \epsilonhat_k
  & = &
\frac{1}{n_k} \left.
\sum_{i \in \curlS_k} \frac{ R^B_i }{ \piBhat_k (X_i) } \{ Y_i - \mhat_k (X_i) \}
\right/
\frac{1}{n_k} \sum_{i \in \curlS_k} \frac{ R^B_i }{ \piBhat_k (X_i)^2 }
\nonumber \\
  & = &
\left.
E_X \left[ \frac{ \pi^B (X) }{ \piBhat_k (X) } \{ m (X) - \mhat_k (X) \} \right]
\right/
E_X \left\{ \frac{ \pi^B (X) }{ \piBhat_k (X)^2 } \right\}
\; + O_p(M^{-1/2}).
\nonumber \\
\end{eqnarray*}

Now, by the Cauchy-Schwartz inequality,
\begin{eqnarray*}
  E_X \left[ \frac{ \pi^B (X) }{ \piBhat_k (X) } \{ m (X) - \mhat_k (X) \} \right]
  & \leq &
  \sqrt{
    E_X \left\{ \frac{ \pi^B (X)^2 }{ \piBhat_k (X)^2 } \right\}
    E_X [ \{ m (X) - \mhat_k (X) \}^2 ]
  }
    \nonumber \\
  & = &
  \sqrt{
    O_p(1) \times o_p (N^{- \mrate})
  }
  \nonumber \\
  & = &
  o_p (N^{- \mrate/2}).
\end{eqnarray*}

Also,
\begin{equation*}
E_X \left\{ \frac{ \pi^B (X) }{ \piBhat_k (X)^2 } \right\}
=
E_X \left\{ \frac{ 1 }{ \piBhat_k (X) } \right\}
+ o_p(1).
\end{equation*}

Hence,
\begin{eqnarray*}
  \epsilonhat_k
  & = &
  \frac{ o_p (N^{- \mrate /2}) }{ E_X \left\{ \frac{ 1 }{ \piBhat_k (X) } \right\} + o_p(1) } + O_p (M^{-1/2})
  \nonumber \\
  & = &
  o_p (N^{- \mrate /2}).
\end{eqnarray*}
Since the first iteration of the Newton-Raphson algorithm leads to $\epsilonhat_k = o_p(N^{- \mrate /2})$, the value of $\epsilonhat_k$ at convergence of the algorthm will also be $o_p(N^{- \mrate /2})$.

Second, consider the case where $\mhat_k (x; \epsilon_k)$ is defined by equation~(\ref{eq:TMLE.logit}).
Now, for any fixed $\piBhat$ and $\mhat$, the first iteration of a Newton-Raphson algorithm to find the maximum likelihood estimate of $\epsilon_k$ starting from an initial value of $\epsilonhat_k=0$ would set $\epsilonhat_k$ equal to
\begin{eqnarray*}
  \epsilonhat_k
  & = &
\frac{1}{n_k} \left.
\sum_{i \in \curlS_k} \frac{ R^B_i }{ \piBhat_k (X_i) } \{ Y_i - \mhat_k (X_i) \}
\right/
\frac{1}{n_k} \sum_{i \in \curlS_k} \frac{ R^B_i }{ \piBhat_k (X_i)^2 }
\frac{ \mhat_k (X_i) }{ \{ 1 + \mhat_k (X_i) \}^2 }.
\nonumber
\end{eqnarray*}
The same logic that was used in the case of where $\mhat_k (x; \epsilon_k)$ is defined by equation~(\ref{eq:TMLE.linear}) again applies, and hence $\epsilonhat_k = o_p(N^{- \mrate /2})$.

Putting this together, we have
\begin{eqnarray*}
    \frac{1}{n}
    \sum_{i \in \curlS_k} U_i ( \piBhat_k, \mhat_k )
    +
    \left. \frac{ \partial }{ \epsilon_k }
    \left[
      \frac{1}{n}
      \sum_{i \in \curlS_k} U_i \{ \piBhat_k, \mhat_k (\epsilon_k) \}
      \right] \right|_{\epsilon_k = 0}
    \epsilonhat_k
    & = &
    o_p (M^{- \piBrate /2}) \times o_p (N^{- \mrate /2})
    \nonumber \\
    & = &
    o_p (M^{-1/2}).
    \nonumber
\end{eqnarray*}
Hence, $\tmleone = \thetahat + o_p (M^{1/2})$.

\section{Data-generating mechanism for simulation study}

\label{sect:appendix.simstudy.dgm}

Let $H_{jq} \sim \mbox{NegativeBinomial} (\mbox{mean}=100, \mbox{variance}=400)$ be the number of households of $q$ individuals in cluster $j$ ($q=1,2,3$ and $j=1, \ldots, J$).
Thus, the number of individuals in cluster $j$ is $N_j = \sum_{q=1}^3 q H_j$, and the population size is $n = \sum_{j=1}^J N_j$, which has expectation $600 J$ and standard deviation $\sqrt{2400 J}$.
We use $J=1000$, and so this expectation and standard deviation are 600000 and 1549, respectively.

Let $H_j = \sum_{q=1}^3 H_{jq}$ be the total number of households in cluster $j$.
Let $H_j^* = H_j \times J \left/ \sum_{j=1}^J H_j \right. + \mbox{Normal} (0, 0.1^2)$ be the ratio of the number of households in cluster $j$ to the mean number of households per cluster plus a cluster-specific random effect.
Recall that $D_i$ denotes the index of the cluster to which individual $i$ belongs, and let $B_i$ denote the number of individuals in the household to which individual $i$ belongs divided by the maximum household size.
So, $B=1/3$, 2/3 or 1.
Let $\mu_i = -1.3 + 0.5 H_{D_i}^* + 0.5 B_i$. 
Given $\mu_1, \ldots, \mu_n$, the $4n$ random variables $X_{11}, X_{21}, X_{31}, X_{41}, \ldots, X_{1n}, X_{2n}, X_{3n}, X_{4n}$ are independently distributed with
\begin{eqnarray*}
  X_{1i} \mid \mu_i & \sim & \mbox{Normal} (\mu_i, 0.4^2) \\
  X_{2i} \mid \mu_i & \sim & \mbox{Normal} (\mu_i, 0.2^2) \\
  X_{3i} \mid \mu_i & \sim & \mbox{Bernoulli} \big( \mbox{expit} (-1.35 + \mu_i) \big) \\
  X_{4i} \mid \mu_i & \sim & \mbox{Bernoulli} \big( \mbox{expit} (2 \mu_i) \big)
\end{eqnarray*}
Note that marginally (over $\mu_i$) $P(X_{3i} = 1) = 0.2$ and $P(X_{3i} = 1) = 0.5$.

Sampford sampling was used to select the clusters in Sample A in Scenarios 1--24; random systematic sampling was used in Scenarios 1b--24b.

\section{Results for simulation study}

\label{sect:appendix.simstudy.results}

Table~\ref{tab:scenarios} shows the values of $E(n^B)$, $M$ and $n_{\rm house}$ in the 48 scenarios.

\subsection{Scenarios 1--24: continuous $Y$}

Tables~\ref{tab:scenario1}--\ref{tab:scenario24} show the bias, empirical SE, mean estimated SE and coverage of 95\% confidence intervals for each estimator in each of Scenarios~1--24.

\begin{figure}
\begin{center}
\includegraphics[width=1\textwidth,page=2]{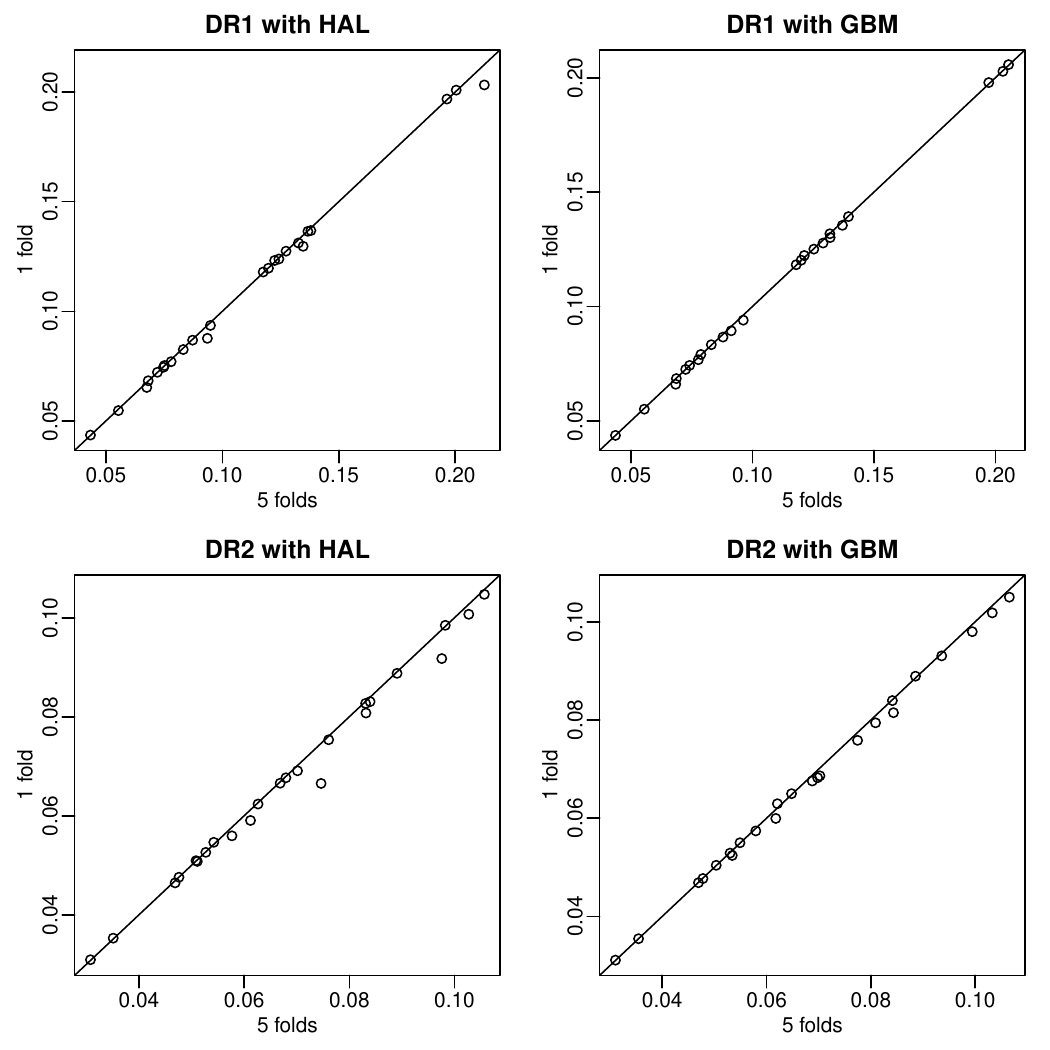}
\hspace*{-5mm}
\end{center}
\caption{Empirical standard error of estimator of $\Ybar$ when $\thetahat$ (top) or $\thetahatthr$ (bottom) is used with HAL (left) or XGBoost (right).  Each plot compares coverage when 5-fold cross-fitting is used versus when no cross-fitting is used.  Each dot corresponds to one of the Scenarios 1--24.}
\label{fig:empSE}
\end{figure}

\begin{figure}
\begin{center}
\includegraphics[width=1\textwidth,page=1]{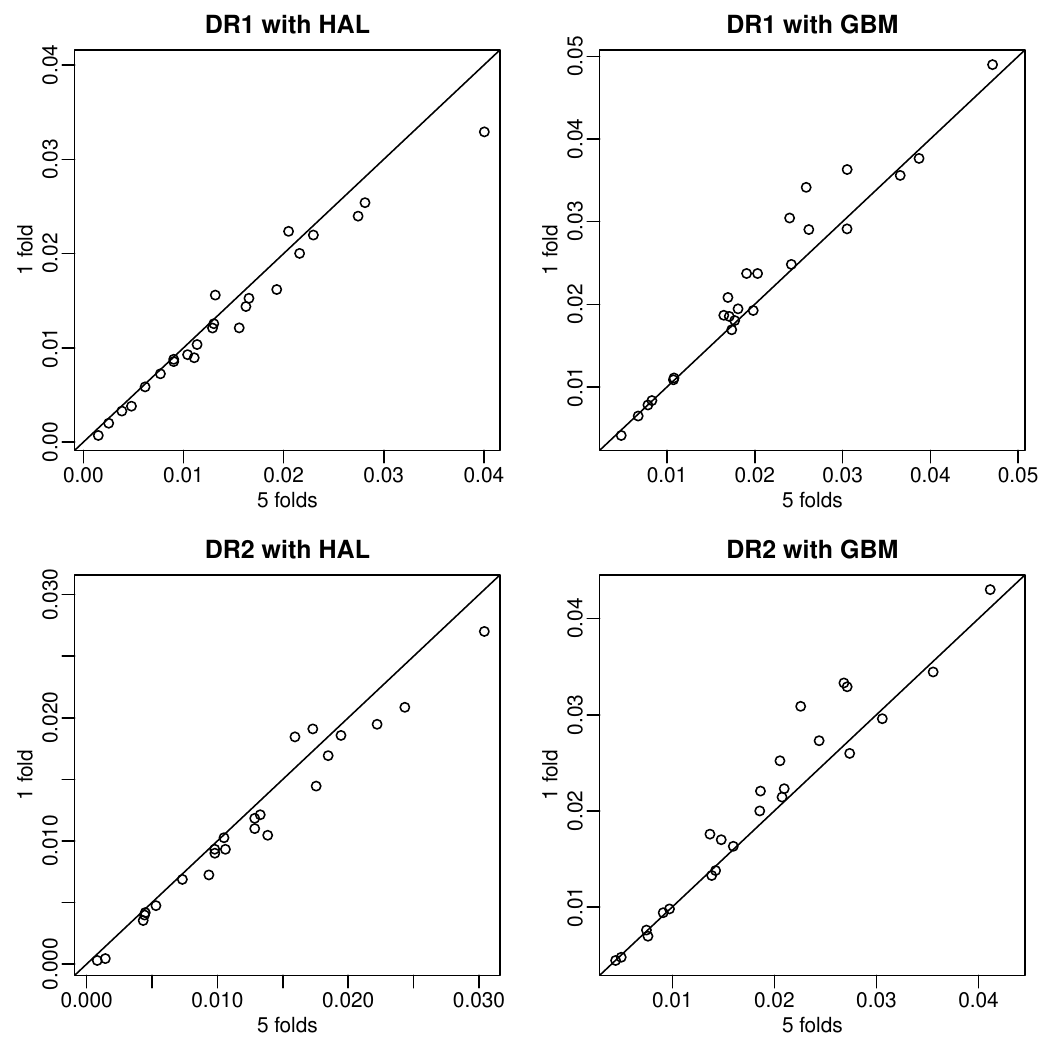}
\hspace*{-5mm}
\end{center}
\caption{Bias of estimator of $\Ybar$ when $\thetahat$ (top) or $\thetahatthr$ (bottom) is used with HAL (left) or XGBoost (right).  Each plot compares coverage when 5-fold cross-fitting is used versus when no cross-fitting is used (1 fold).  Each dot corresponds to one of Scenarios 1--24.}
\label{fig:bias}
\end{figure}

\begin{figure}
\begin{center}
\includegraphics[width=1\textwidth,page=3]{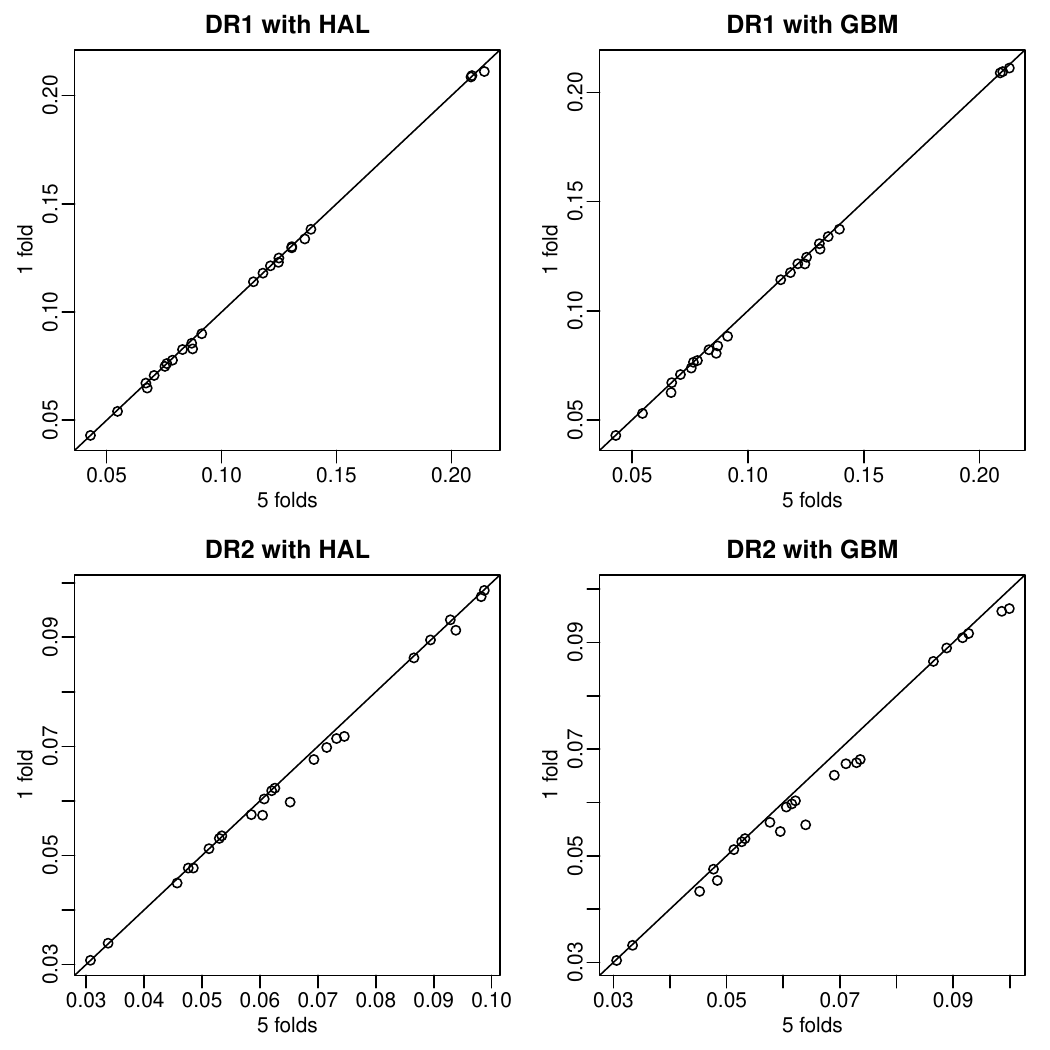}
\hspace*{-5mm}
\end{center}
\caption{Mean estimated standard error of estimator of $\Ybar$ when $\thetahat$ (top) or $\thetahatthr$ (bottom) is used with HAL (left) or XGBoost (right).  Each plot compares coverage when 5-fold cross-fitting is used versus when no cross-fitting is used (1 fold).  Each dot corresponds to one of Scenarios 1--24.}
\label{fig:estimatedSE}
\end{figure}

Figures~\ref{fig:empSE}, \ref{fig:bias} and~\ref{fig:estimatedSE} are analogous to Figure~\ref{fig:coverage}, but show empirical standard error, bias, and mean estimated standard error, respectively, rather than coverage.
Figure~\ref{fig:empSE} shows that cross-fitting has little effect on the empirical standard error.
Figure~\ref{fig:bias} indicates a tendency for the omission of the cross-fitting step to increase bias when XGboost is used.
Figure~\ref{fig:estimatedSE} shows that omitting cross-fitting when using XGboost can decrease the mean estimated standard error.

\subsection{Scenarios 1b--24b: binary $Y$}

\begin{figure}
\begin{center}
\includegraphics[width=1\textwidth,page=1]{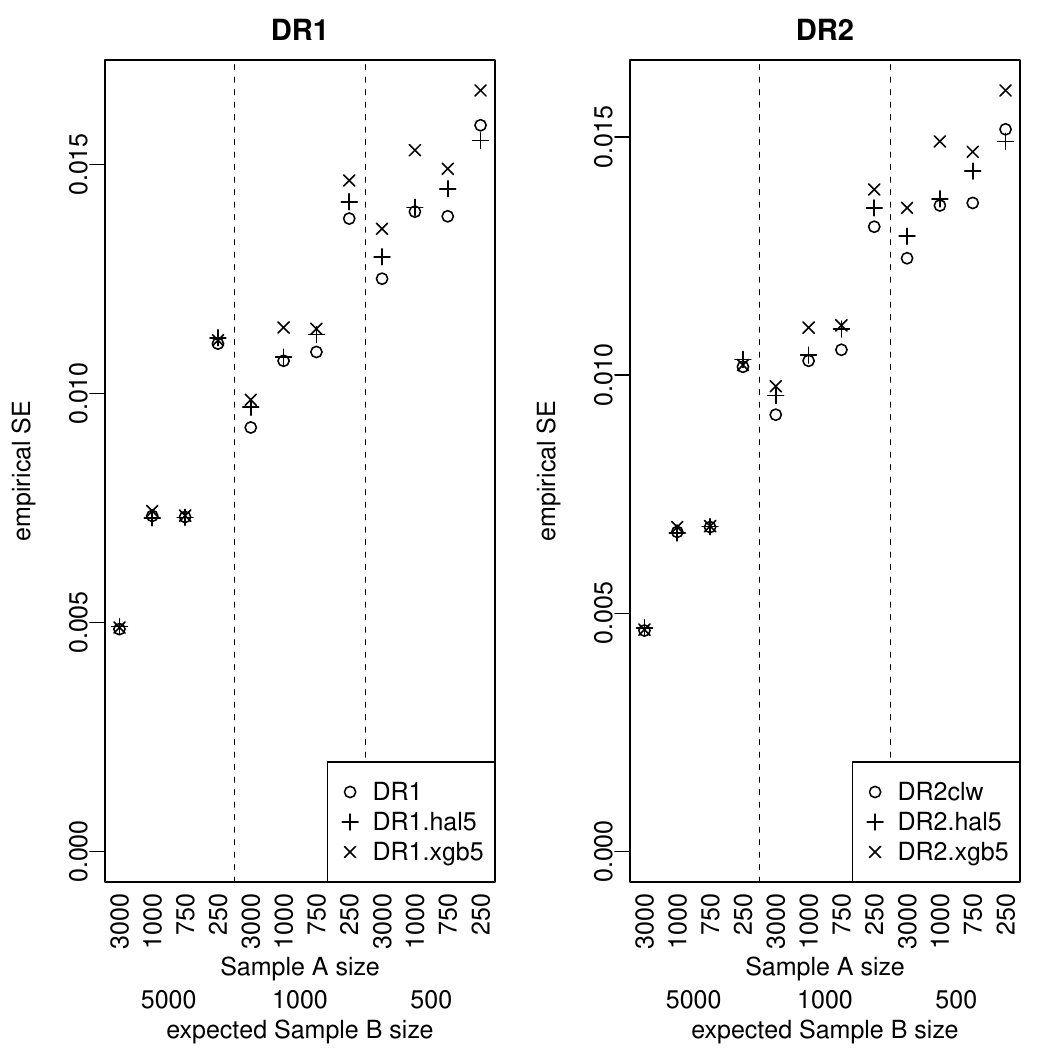}
\hspace*{-5mm}
\end{center}
\caption{Left: Empirical standard error of $\thetahat$ when used with parametric models (DR1), HAL (DR1.hal5) or XGBoost (DR1.xgb5) in Scenarios 1b--12b.
Right: Empirical standard error of $\thetahattwo$ or $\thetahatthr$ when used with parametric models (DR2clw), HAL (DR2.hal5) or XGBoost (DR2.xgb5).}
\label{fig:compareSEs.b}
\end{figure}

\begin{figure}
\begin{center}
\includegraphics[width=1\textwidth,page=2]{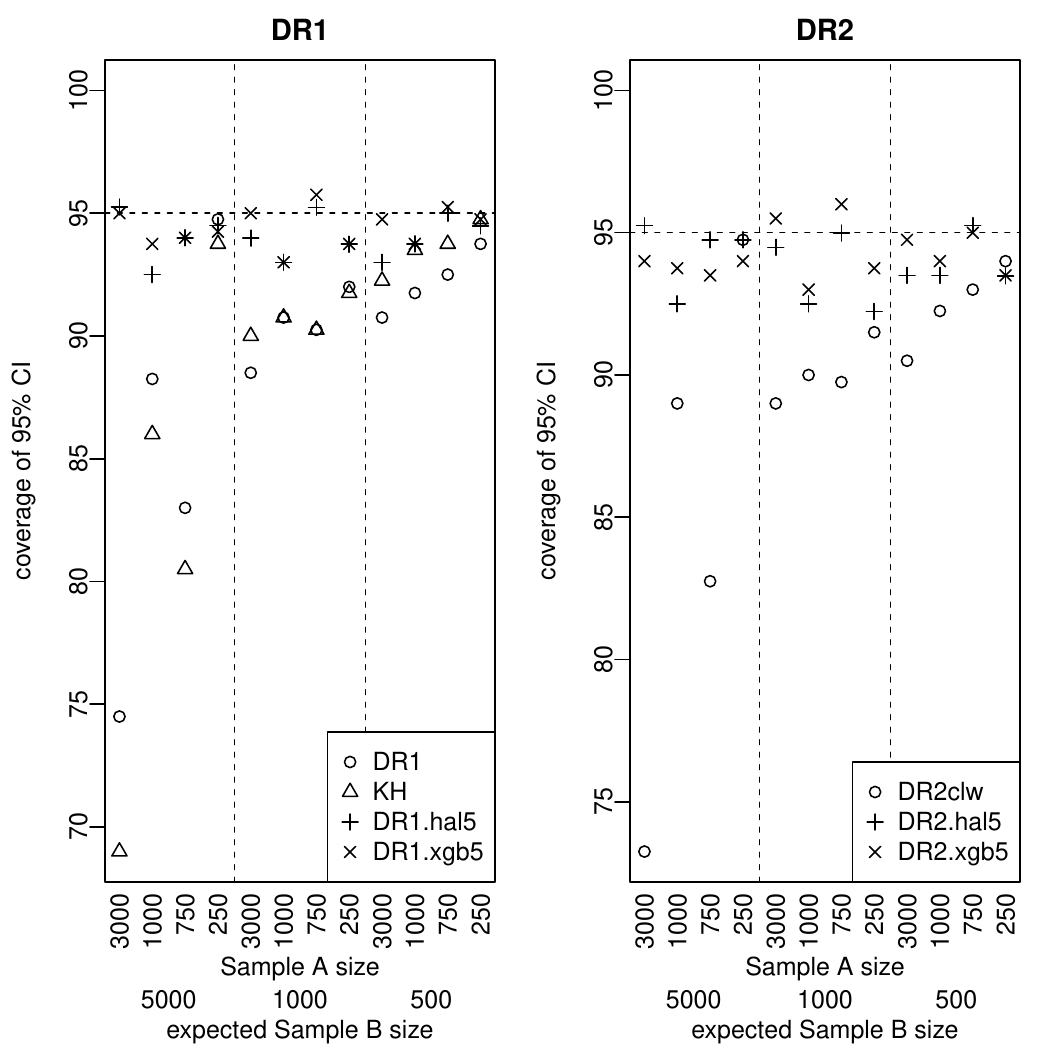}
\hspace*{-5mm}
\end{center}
\caption{Left: Coverage of 95\% confidence interval for $\Ybar$ in Scenarios 13b--24b when $\thetahat$ is used with parametric nuisance models (DR1), Kim and Haziza's method (KH), HAL (DR1.hal5) or XGBoost (DR1.xgb5).  Right: Coverage when $\thetahattwo$ or $\thetahatthr$ is used with parametric nuisance models (DR2clw) or $\thetahatthr$ used with HAL (DR2.hal5) or XGBoost (DR2.xgb5).}
\label{fig:compare.cover.b}
\end{figure}

\begin{figure}
\begin{center}
\includegraphics[width=1\textwidth,page=2]{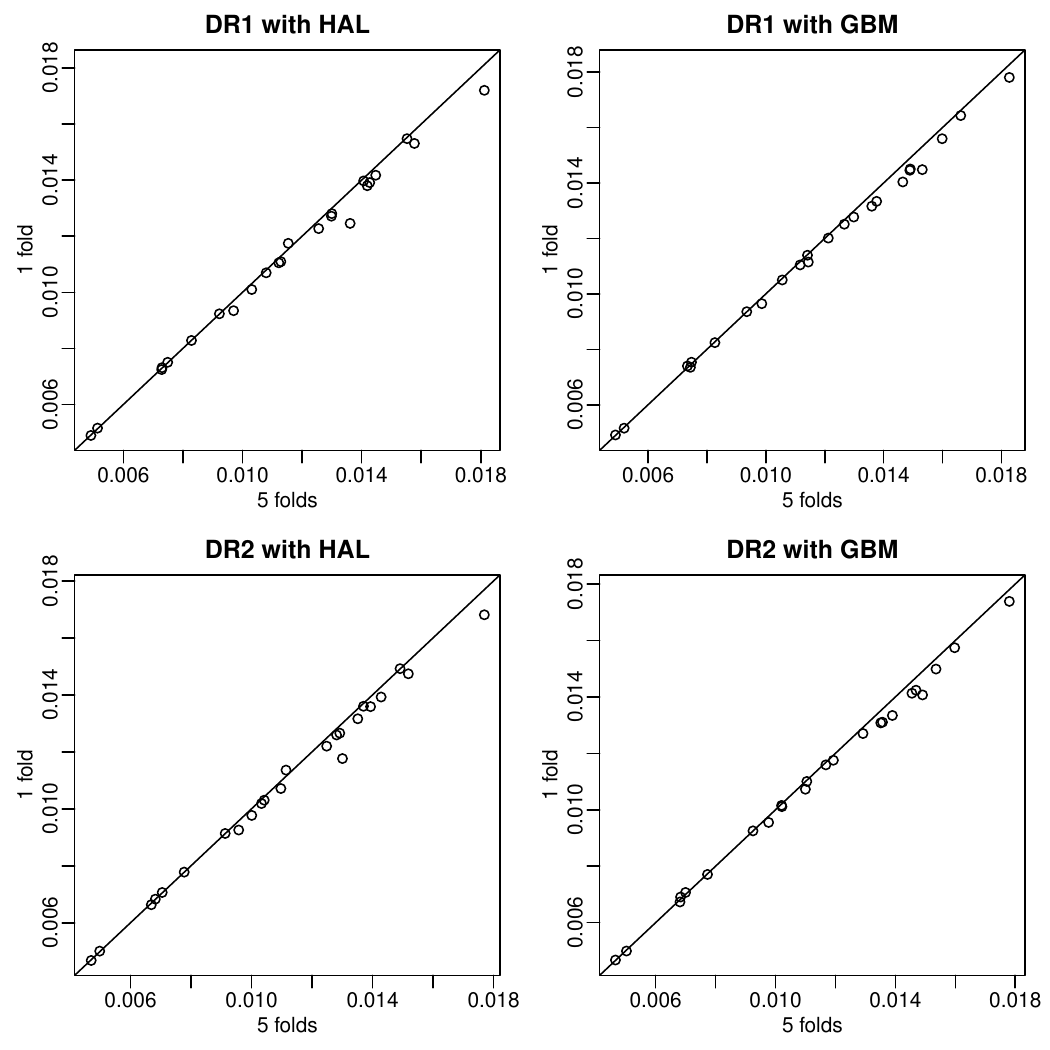}
\hspace*{-5mm}
\end{center}
\caption{Empirical standard error of estimator of $\Ybar$ when $\thetahat$ (top) or $\thetahatthr$ (bottom) is used with HAL (left) or XGBoost (right).  Each plot compares coverage when 5-fold cross-fitting is used versus when no cross-fitting is used (1 fold).  Each dot corresponds to one of Scenarios 1b--24b.}
\label{fig:empSE.b}
\end{figure}

\begin{figure}
\begin{center}
\includegraphics[width=1\textwidth,page=1]{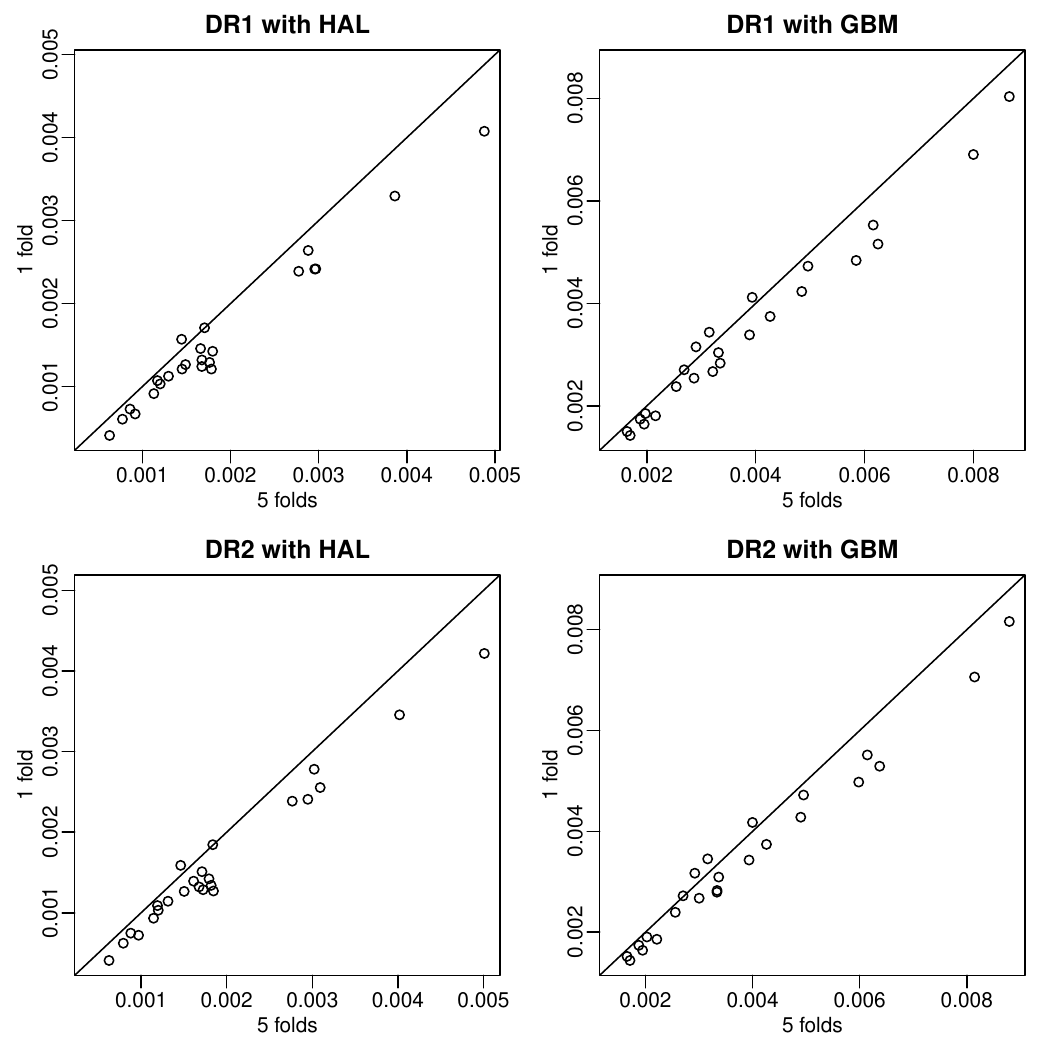}
\hspace*{-5mm}
\end{center}
\caption{Bias of estimator of $\Ybar$ when $\thetahat$ (top) or $\thetahatthr$ (bottom) is used with HAL (left) or XGBoost (right).  Each plot compares coverage when 5-fold cross-fitting is used versus when no cross-fitting is used (1 fold).  Each dot corresponds to one of Scenarios 1b--24b.}
\label{fig:bias.b}
\end{figure}

\begin{figure}
\begin{center}
\includegraphics[width=1\textwidth,page=3]{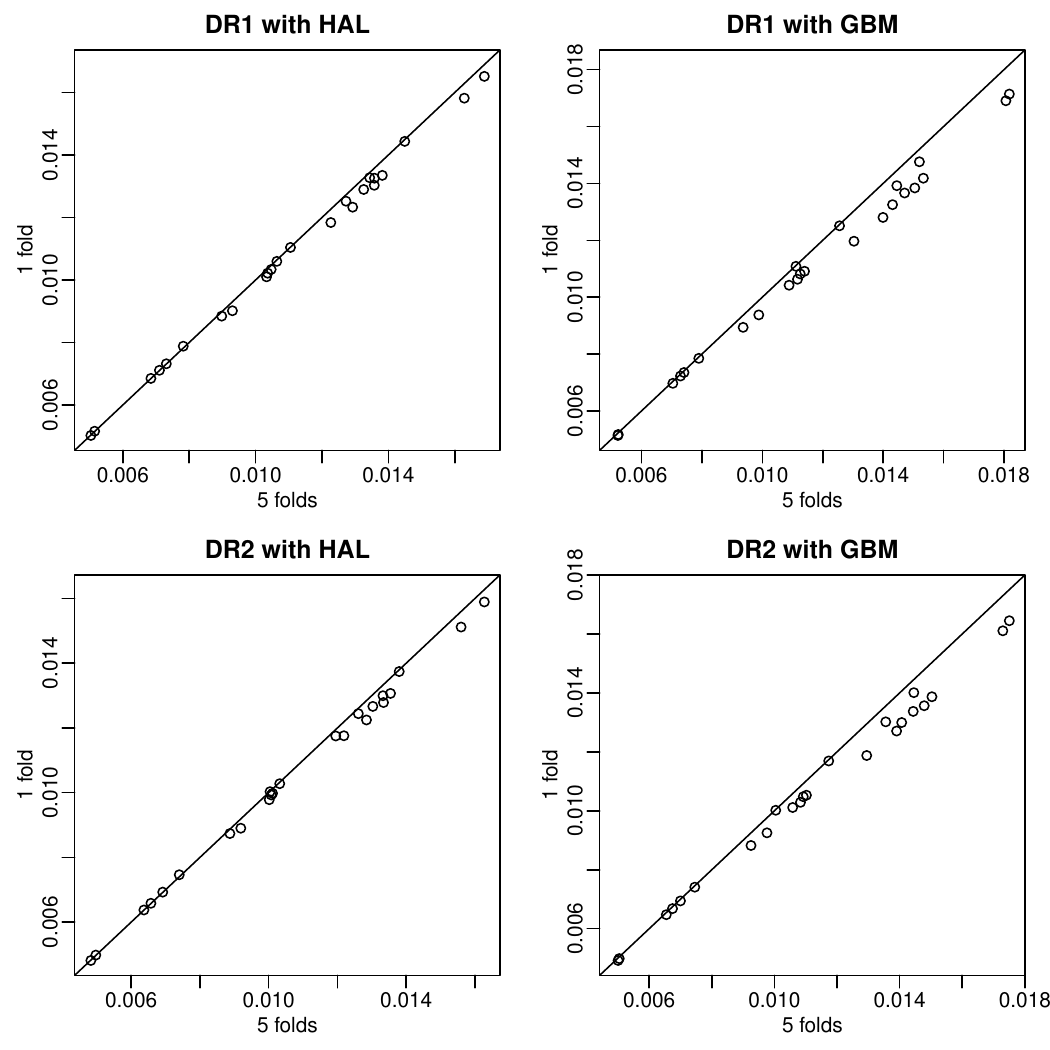}
\hspace*{-5mm}
\end{center}
\caption{Mean estimated standard error of estimator of $\Ybar$ when $\thetahat$ (top) or $\thetahatthr$ (bottom) is used with HAL (left) or XGBoost (right).  Each plot compares coverage when 5-fold cross-fitting is used versus when no cross-fitting is use (1 fold).  Each dot corresponds to one of Scenarios 1b--24b.}
\label{fig:estimatedSE.b}
\end{figure}

\begin{figure}
\begin{center}
\includegraphics[width=1\textwidth]{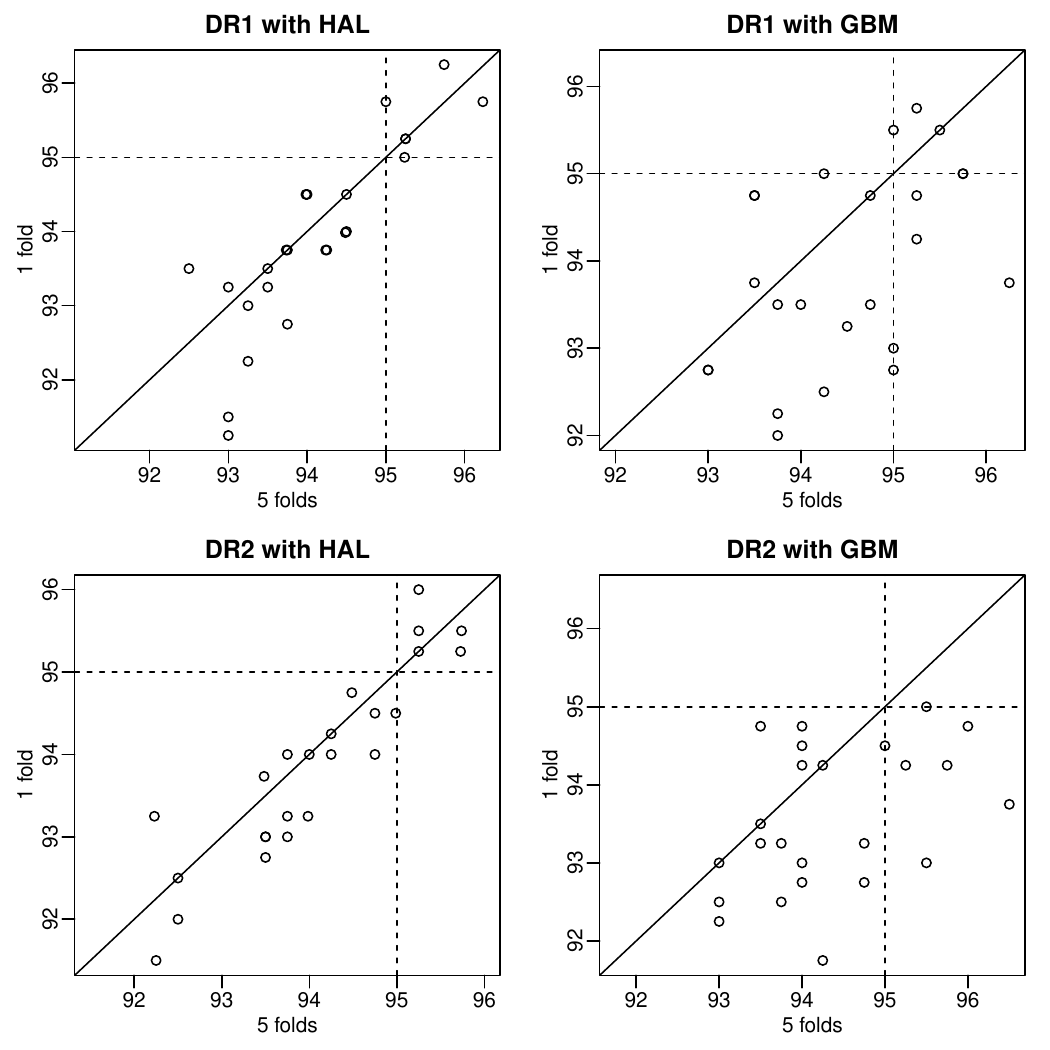}
\hspace*{-5mm}
\end{center}
\caption{Coverage of 95\% confidence interval for $\Ybar$ when $\thetahat$ (top) or $\thetahatthr$ (bottom) is used with HAL (left) or XGBoost (right).  Each plot compares coverage when 5-fold cross-fitting is used versus when no cross-fitting is used (1 fold).  Each dot corresponds to one of Scenarios 1b--24b.  Broken lines indicate perfect 95\% coverage.}
\label{fig:coverage.b}
\end{figure}

Figure~\ref{fig:compareSEs.b} shows that, in Scenarios 1b--12b, where the parametric nuisance models are correctly specified, using data-adaptive estimation of $\piBtrue$ and $\mtrue$ had little effect on the empirical standard errors of the double robust estimators of $\Ybar$, compared to using the parametric nuisance models.
Figure~\ref{fig:compare.cover.b} shows the coverage of 95\% confidence intervals for Scenarios 13b--24b.
Coverage for estimators using the misspecified parametric models was poor in most of the scenarios, whereas coverage was good for the estimators using data-adaptive methods.
Figure~\ref{fig:empSE.b} shows that omitting the cross-fitting step had little effect on the empirical standard errors.
Figure~\ref{fig:bias.b} indicates that cross-fitting causes the bias to increase slightly.
Figure~\ref{fig:estimatedSE.b} shows that omitting cross-fitting when using XGboost decreases the mean estimated standard error in many scenarios and leaves it unchanged in the other scenarios.
Figure~\ref{fig:coverage.b} shows that the net effect of omitting cross-fitting is to worsen the coverage of confidence intervals in most scenarios, at least for XGboost.

Tables~\ref{tab:scenario1b}--\ref{tab:scenario24b} show the bias, empirical SE, mean estimated SE and coverage of 95\% confidence intervals for each estimator in each of Scenarios~1b--24b.

\section{Estimating the mean in a subpopulation}

\label{sect:domain.estimation}

Equation~(\ref{eq:ratio.estimator}) defines $\thetahatthr$ as
\begin{equation*}
  \thetahatthr
  = \thetahatthr (\underpiBhat, \undermhat)
  = \frac{ 1 }{ \hat{n}^A } \sum_{k=1}^K \sum_{i \in \curlS_k} \left[
  \frac{ R^A_i }{ \pi^A_i } \mhat_k (X_i)
  +
  \frac{ R^B_i }{ \piBhat_k (X_i) } \{ Y_i - \mhat_k (X_i) \}
  \right],
\end{equation*}
where $\bar{n}^A = \sum_{i=1}^n R^A_i / \pi^A_i$, and Appendix~\ref{sect:appendix.DR2} shows that
\begin{eqnarray*}
\sqrt{M} \{ \thetahatthr (\underpiBhat, \undermhat) - \Ybar \}
& = &
\sqrt{M} \frac{1}{n} \sum_{k=1}^K \sum_{i \in \curlS_k}
\frac{R^A_i}{\pi^A_i} \left\{ \mtrue (X_i) - \bar{m}_0 \right\}
\nonumber \\
&& +
\sqrt{M} \frac{1}{n} \sum_{k=1}^K \sum_{i \in \curlS_k}
\frac{R^B_i}{\piBtrue (X_i)} \{ Y_i - \mtrue(X_i) \}
\nonumber \\
&& +
\sqrt{M} (\bar{m}_0 - \Ybar) + o_p(1),
\end{eqnarray*}
where $\Ybar = n^{-1} \sum_{i=1}^n Y_i$ and $\bar{m}_0 = n^{-1} \sum_{i=1}^n \mtrue (X_i)$.

This estimator $\thetahatthr$ is a estimator of the mean of $Y$ in the population of size $n$ covered by the probability sample.
If we instead wish to estimate the mean of $Y$ in a subpopulation of the population covered by the probability sample, as is the case in the Natsal-3 application described in Section~\ref{sect:Natsal}, then a small methodological extension is needed.
We now describe this extension.

let $S_i=1$ if individual $i$ belongs to the subpopulation of interest, and $S_i=0$ otherwise.
Define $n_s = \sum_{i=1}^n S_i$ to be the size of the subpopulation, and $\hat{n}^A_s = \sum_{i=1}^n R^A_i S_i / \pi^A_i$ to be the Horwitz-Thompson estimator of the size of this subpopulation.
Define $\Ybar_s = n_s^{-1} \sum_{i=1}^n S_i Y_i$ to be the subpopulation mean of $Y$ and $\bar{m}_{0s} = n_s^{-1} \sum_{i=1}^n S_i \mtrue (X_i)$ to be the subpopulation mean of $E(Y \mid X)$.
Finally, redefine $\thetahatthr$ as
\begin{equation*}
  \thetahatthr
  = \thetahatthr (\underpiBhat, \undermhat)
  = \frac{ 1 }{ \hat{n}^A_s } \sum_{k=1}^K \sum_{i \in \curlS_k} S_i \left[
  \frac{ R^A_i }{ \pi^A_i } \mhat_k (X_i)
  +
  \frac{ R^B_i }{ \piBhat_k (X_i) } \{ Y_i - \mhat_k (X_i) \}
  \right].
\end{equation*}
Note that to use this estimator of $\Ybar_s$, we need to estimate the nuisance functions $\pi^B(X)$ and $m(X)$ only for values of $X$ for which $S=1$.
Also note that this redefined $\thetahatthr$ can be viewed as being just the original $\thetahatthr$ defined in Section~\ref{sect:DR2} but with the outcome $Y$ redefined to be $SY$ and $\hat{n}^A$ replaced by $\hat{n}^A_s$.

A slight modification of the proof in Appendix~\ref{sect:appendix.DR2} shows that
\begin{eqnarray}
\sqrt{M} \{ \thetahatthr (\underpiBhat, \undermhat) - \Ybar_s \}
& = &
\sqrt{M} \frac{1}{n_s} \sum_{k=1}^K \sum_{i \in \curlS_k}
\frac{R^A_i}{\pi^A_i} S_i \left\{ \mtrue (X_i) - \bar{m}_0 \right\}
\nonumber \\
&& +
\sqrt{M} \frac{1}{n_s} \sum_{k=1}^K \sum_{i \in \curlS_k}
\frac{R^B_i}{\piBtrue (X_i)} S_i \{ Y_i - \mtrue(X_i) \}
\nonumber \\
&& +
\sqrt{M} (\bar{m}_{0s} - \Ybar_s) + o_p(1).
\label{eq:subpopulation.expansion}
\end{eqnarray}
\citet{Seaman2025} give formulae for the asymptotic variance of $\thetahatthr - \Ybar_s$ implied by equation~(\ref{eq:subpopulation.expansion}), and an estimator of this variance.

For completeness, we now provide the required modified version of the proof in Appendix~\ref{sect:appendix.DR2}.
This is very similar to the proof in that appendix, but with $n$, $\hat{n}^A$, $\Ybar$ and $\bar{m}_0$ replaced by $n_s$, $\hat{n}^A_s$, $\Ybar_s$ and $\bar{m}_{0s}$ respectively, and the addition of $S_i$ terms.
Note that if $S = 1$ for all individuals in the population, this modified proof reduces to the proof in Appendix~\ref{sect:appendix.DR2}.

Write $\thetahatthr$ as $\thetahatthr = \thetahatthr (\underpiBhat, \undermhat, \tauhat)$, where
\[
\thetahatthr (\underpiB, \underm, \tau)
=
\frac{1}{n_s} \sum_{k=1}^K \sum_{i \in \curlS_k} S_i \; \Uthr_i (\pi^B_k, m_k, \tau),
\]
and
\[
\tauhat = \frac{1}{n_s} \sum_{i=1}^n \frac{ R^A_i }{ \pi^A_i } S_i.
\]
The definitions of $\Uthr_i (\pi^B, m, \tau)$ and $U_i (\pi^B, m)$ are unchanged.

By a Taylor-series expansion,
\begin{eqnarray}
  &&
\sqrt{M} \; \{ \thetahatthr (\underpiBhat, \undermhat, \tauhat) - \Ybar_s \}
\nonumber \\
&& \hspace{.5cm} =
\sqrt{M} \; \frac{ 1 }{ n_s } \sum_{k=1}^K \sum_{i \in \curlS_k} S_i \; \Uthr_i (\piBhat_k, \mhat_k, 1)
+
\sqrt{M} \; \frac{ 1 }{ n_s } \sum_{k=1}^K \sum_{i \in \curlS_k} S_i \; \left. \frac{ \partial \Uthr_i }{ \partial \tau } (\piBhat_k, \mhat_k, \tau) \right|_{\tau = 1}  \; (\tauhat - 1)
\nonumber \\
&& \hspace{.9cm} -
\sqrt{M} \; \Ybar_s + o_p(1)
\nonumber \\
&& \hspace{.5cm} =
\sqrt{M} \; \frac{ 1 }{ n_s } \sum_{k=1}^K \sum_{i \in \curlS_k} S_i \; \Uthr_i (\piBhat_k, \mhat_k, 1)
-
\sqrt{M} \; \frac{ 1 }{ n_s } \sum_{k=1}^K \sum_{i \in \curlS_k} S_i \; \Uthr_i (\piBhat_k, \mhat_k, 1) \times (\tauhat - 1)
\nonumber \\
&& \hspace{.9cm} -
\sqrt{M} \; \Ybar_s + o_p(1)
\nonumber \\
&& \hspace{.5cm} =
\sqrt{M} \; \frac{ 1 }{ n_s } \sum_{k=1}^K \sum_{i \in \curlS_k} S_i \; U_i (\piBhat_k, \mhat_k)
-
\sqrt{M} \; \frac{ 1 }{ n_s } \sum_{k=1}^K \sum_{i \in \curlS_k} S_i \; U_i (\piBhat_k, \mhat_k) \times (\tauhat - 1)
\nonumber \\
&& \hspace{.9cm} -
\sqrt{M} \; \Ybar_s + o_p(1)
\nonumber \\
&& \hspace{.5cm} =
\sqrt{M} \; \left[ \frac{ 1 }{ n_s } \sum_{i=1}^n S_i \; U_i (\piBtrue, \mtrue)
  + o_p(M^{-1/2}) \right]
\nonumber \\
&& \hspace{.9cm} -
\sqrt{M} \; \left[ \frac{ 1 }{ n_s } \sum_{i=1}^n S_i \; U_i (\piBtrue, \mtrue) + o_p(M^{-1/2}) \right] \times (\tauhat - 1)
\nonumber \\
&& \hspace{.9cm} -
\sqrt{M} \; \Ybar_s + o_p(1)
\label{eq:SYreplacesS} \\
&& \hspace{.5cm} =
\sqrt{M} \; \frac{ 1 }{ n_s } \sum_{i=1}^n S_i \; U_i (\piBtrue, \mtrue)
-
\sqrt{M} \; \frac{ 1 }{ n_s } \sum_{i=1}^n S_i \; U_i (\piBtrue, \mtrue) \times (\tauhat - 1)
-
\sqrt{M} \; \Ybar_s + o_p(1)
\nonumber \\
&& \hspace{.5cm} =
\sqrt{M} \; \frac{ 1 }{ n_s } \sum_{i=1}^n
\frac{ R^A_i }{ \pi^A_i } \mtrue (X_i)
+
\sqrt{M} \; \frac{ 1 }{ n } \sum_{i=1}^n
\frac{ R^B_i }{ \piBtrue (X_i) } S_i \; \{ Y_i - \mtrue (X_i) \}
\nonumber \\
&& \hspace{.9cm} -
\sqrt{M} \; \{ \bar{m}_{0s} + o_p(1) \}
\times ( \tauhat - 1 )
- \sqrt{M} \; \Ybar_s + o_p(1)
\nonumber \\
&& \hspace{.5cm} =
\sqrt{M} \; \frac{ 1 }{ n_s } \sum_{i=1}^n
\frac{ R^A_i }{ \pi^A_i } S_i \mtrue (X_i)
+
\sqrt{M} \; \frac{ 1 }{ n_s } \sum_{i=1}^n
\frac{ R^B_i }{ \piBtrue (X_i) } \{ Y_i - \mtrue (X_i) \}
\nonumber \\
&& \hspace{.9cm} -
\{ \bar{m}_{0s} + o_p(1) \} \;
\sqrt{M} \; \frac{1}{n_s} \sum_{i=1}^n
S_i \left( \frac{ R^A_i }{ \pi^A_i } - 1 \right)
- \sqrt{M} \; \Ybar_s + o_p(1)
\nonumber \\
&& \hspace{.5cm} =
\sqrt{M} \; \frac{ 1 }{ n_s } \sum_{i=1}^n
\frac{ R^A_i }{ \pi^A_i } S_i \{ \mtrue (X_i) - \bar{m}_{0s} \}
+
\sqrt{M} \; \frac{ 1 }{ n_s } \sum_{i=1}^n
\frac{ R^B_i }{ \piBtrue (X_i) } S_i \{ Y_i - \mtrue (X_i) \}
\nonumber \\
&& \hspace{.9cm} +
\sqrt{M} ( \bar{m}_{0s} - \Ybar_s ) + o_p(1)
\nonumber
\end{eqnarray}

Note that line (\ref{eq:SYreplacesS}) follows because $S_i U_i (\pi^B, m)$ can be viewed as being $U_i (\pi^B, m)$ with outcome $Y$ redefined to be $SY$.

\section{Conditional Poisson sampling of clusters}

\label{sect:appendix.conditional.poisson}

For simplicity of notation in this section, we shall use $R_j$ to denote $R^C_j$, i.e.\ the binary indicator that cluster $j$ is included in Sample A.

If $M$ clusters are sampled from $J$ clusters using conditional Poisson sampling with binomial sampling probabilities $\pi_1, \ldots, \pi_J$, then
\begin{eqnarray}
  P( R_1 = r_1, \ldots, R_J = r_J \mid \curlF )
  & = &
\frac{ \prod_{j=1}^J \pi_j^{r_j} (1 - \pi_j)^{1 - r_j} }
     { \sum_{ (r_1', \ldots, r_J') } I \left( \sum_{j=1}^J r_j' = M \right) \prod_{j=1}^J \pi_j^{r_j'} (1 - \pi^C_j)^{1 - r_j'} }
     \nonumber \\
     && \label{eq:cond.Pois}
\end{eqnarray}
if $\sum_{j=1}^J r_j = M$, and $P( R_1 = r_1, \ldots, R_J = r_J \mid \curlF ) = 0$ otherwise.
We shall denote this distribution as
\[
(R_1, \ldots, R_J) \mid \curlF \sim \mbox{condPois} \big( (\pi_1, \ldots, \pi_J), \; M \big).
\]

Henceforth, for convenience, we shall omit explicit conditioning on $\curlF$ in expressions for probabilities, but it should be understood that this conditioning is implicit.

It follows from equation~(\ref{eq:cond.Pois}) that, for any $(r_1, \ldots, r_J)$ such that $\sum_{j=1}^J r_j = M$,
\begin{eqnarray*}
  &&
  P( R_1 = r_1, \ldots, R_k = r_k \mid R_{k+1} = r_{k+1}, \ldots, R_J = r_J )
  \\
  && \hspace{0.5cm} =
  \frac{ P( R_1 = r_1, \ldots, R_J = r_J ) }
       { P( R_{k+1} = r_{k+1}, \ldots, R_J = r_J ) }
       \\
       && \hspace{0.5cm} =
\frac{ \prod_{j=1}^J \pi_j^{r_j} (1 - \pi_j)^{1 - r_j} }
     { \sum_{ (r_1', \ldots, r_k') } I \left( \sum_{j=1}^k r_j' = M - \sum_{j=k+1}^J r_j \right)
       \prod_{j=1}^J \pi_j^{r_j'} (1 - \pi_j)^{1 - r_j'} }
       \\
       && \hspace{0.5cm} =
\frac{ \prod_{j=1}^J \pi_j^{r_j} (1 - \pi_j)^{1 - r_j} }
     { \prod_{j=k+1}^J \pi_j^{r_j} (1 - \pi_j)^{1 - r_j}
     \sum_{ (r_1', \ldots, r_k') } I \left( \sum_{j=1}^k r_j' = M - \sum_{j=1}^k r_j \right)
     \prod_{j=1}^k \pi_j^{r_j'} (1 - \pi_j)^{1 - r_j'}
     }
     \\
     && \hspace{0.5cm} =
     \frac{ \prod_{j=1}^k \pi_j^{r_j} (1 - \pi_j)^{1 - r_j} }
          { \sum_{ (r_1', \ldots, r_k') } I \left( \sum_{j=1}^k r_j' = M - \sum_{j=1}^k r_j \right)
            \prod_{j=1}^k \pi_j^{r_j'} (1 - \pi_j)^{1 - r_j'}
          }.
\end{eqnarray*}

We recognise this as
\[
R_1, \ldots, R_k \mid R_{k+1}, \ldots, R_J, \curlF \sim \mbox{condPois} \left( (\pi_1, \ldots, \pi_k), \; M - \sum_{j=k+1}^J R_j \right).
\]

More generally, for any subset $\curlS$ of $\{1, \ldots, J\}$, we have
\[
\{ R_j: j \in \curlS \} \mid \{ R_j: j \notin \curlS \}, \curlF \sim \mbox{condPois} \left( \{ \pi_j: j \in \curlS \}, \; M - \sum_{j \notin \curlS} R_j \right).
\]

Now assume that $M$ is an integer multiple of $K$.
Let $\curlS^C_k$ denote the set of clusters in fold $k$.
Because of the way the folds are chosen, we know that fold $k$ will contain $M/K$ sampled clusters.
So, for any $\{ r_1, \ldots, r_J \}$ such that $\sum_{j=1}^J r_j = M$ and for any set $\curlS$ of $\lfloor J / K \rfloor$ or $\lceil J / K \rceil$ clusters such that $\sum_{j \in \curlS} r_j = M/K$ (and thus $\sum_{j \notin \curlS} r_j = M (K-1)/K$), we can use Bayes' Rule to show
\begin{eqnarray*}
  &&
  P( \{ R_j = r_j: j \in \curlS \} \mid \{ R_j = r_j: j \notin \curlS \}, \curlS^C_k = \curlS )
  \\
  && \hspace{.2cm} =
  \frac{ P( \curlS_k^C = \curlS \mid R_1=r_1, \ldots, R_J=r_j )
    \times P( \{ R_j = r_j: j \in \curlS \} \mid \{ R_j = r_j: j \notin \curlS \} ) }
       { P( \curlS^C_k = \curlS \mid \{ R_j = r_j: j \notin \curlS \} ) }
  \\
  && \hspace{.2cm} =
  \frac{ P( \curlS^C_k = \curlS \mid \sum_{j \in \curlS} R_j = M/K )
    \times P( \{ R_j = r_j: j \in \curlS \} \mid \{ R_j = r_j: j \notin \curlS \} ) }
       { P( \curlS^C_k = \curlS \mid \sum_{j \notin \curlS} R_j = M (K-1)/K) }
       \\
  && \hspace{.2cm} =
  \frac{ P( \curlS^C_k = \curlS \mid \sum_{j \in \curlS} R_j = M/K )
    \times P( \{ R_j = r_j: j \in \curlS \} \mid \{ R_j = r_j: j \notin \curlS \} ) }
       { P( \curlS^C_k = \curlS \mid \sum_{j \in \curlS} R_j = M/K ) }
  \\
  && \hspace{.2cm} =
      P( \{ R_j = r_j: j \in \curlS \} \mid \{ R_j = r_j: j \notin \curlS \} ).
\end{eqnarray*}

Therefore
\[
\{ R_j: j \in \curlS^C_k \} \mid \{ R_j: j \notin \curlS^C_k \}, \curlF, \curlS^C_k \sim \mbox{condPois} \left( \{ \pi_j: j \in \curlS^C_k \}, \; M - \sum_{j \notin \curlS^C_k} R_j \right).
\]

By the way the folds have been chosen, $\sum_{j \notin \curlS^C_k} R_j$ must equal $M (K-1) / K$, and so we have
\[
\{ R_j: j \in \curlS^C_k \} \mid \{ R_j: j \notin \curlS^C_k \}, \curlF, \curlS^C_k \sim \mbox{condPois} \left( \{ \pi_j: j \in \curlS^C_k \}, \; M/K \right).
\]

We see therefore that
$
\{ R_j: j \in \curlS^C_k \} \ci \{ R_j: j \notin \curlS^C_k \} \mid \curlF, \curlS^C_k
$.

This implies that
$
\{ R^A_i: i \in \curlS_k \} \ci \{ R^A_i: i \notin \curlS_k \} \mid \curlF, \curlS_k
$.
Hence, $\piBhat_k$ and $\mhat_k$ (which depend on $\{ R^A_i: i \notin \curlS_k \}$) tell us nothing about $\{ R^A_i: i \in \curlS_k \}$.

\newpage

\begin{table}[ht]
  \centering
\begin{tabular}{rl|rrr}
    \multicolumn{2}{c|}{Scenarios} & $E(n^B)$ & $M$ & $n_{\rm house}$ \\ \hline
    1 &  13 & 5000 & 150 & 20 \\
    2 &  14 & 5000 & 150 & 5 \\
    3 &  15 & 5000 & 50 & 20 \\
    4 &  16 & 5000 & 50 & 5 \\
    5 &  17 & 1000 & 150 & 20 \\
    6 &  18 & 1000 & 150 & 5 \\
    7 &  19 & 1000 & 50 & 20 \\
    8 &  20 & 1000 & 50 & 5 \\
    9 &  21 &  500 & 150 & 20 \\
   10 &  22 &  500 & 150 & 5 \\
   11 &  23 &  500 & 50 & 20 \\
   12 &  24 &  500 & 50 & 5
  \end{tabular}
  \vspace{.8cm}
  \caption{For each of the scenarios in the simulation study, the expected number of individuals in Sample B ($E(n^B)$), and the number of sampled clusters ($M$) and of sampled houses per sampled cluster ($n_{\rm house}$) in Sample A.}
  \label{tab:scenarios}
\end{table}

\begin{table}[ht]
\centering
\begin{tabular}{rrrrr}
  & bias & empSE & SEhat & cover \\ 
  \hline
\hline
HT & 0.005 & 0.069 & 0.068 & 95 \\ 
  Haj & 0.001 & 0.035 & 0.035 & 94 \\ 
  naive & -0.306 & 0.034 &  &  \\ 
   \hline
DR1 & 0.004 & 0.068 & 0.068 & 94 \\ 
  DR2clw & 0.001 & 0.035 & 0.035 & 95 \\ 
  TMLE1 & 0.004 & 0.068 & 0.068 & 94 \\ 
  TMLE2 & 0.001 & 0.035 & 0.035 & 95 \\ 
  KH & 0.004 & 0.068 & 0.068 & 94 \\ 
   \hline
DR1.hal5 & 0.008 & 0.068 & 0.067 & 94 \\ 
  DR2.hal5 & 0.004 & 0.035 & 0.034 & 95 \\ 
  TMLE1.hal5 & 0.007 & 0.068 & 0.067 & 94 \\ 
  TMLE2.hal5 & 0.004 & 0.035 & 0.034 & 95 \\ 
   \hline
DR1.hal1 & 0.007 & 0.068 & 0.067 & 94 \\ 
  DR2.hal1 & 0.004 & 0.035 & 0.034 & 94 \\ 
  TMLE1.hal1 & 0.007 & 0.068 & 0.067 & 94 \\ 
  TMLE2.hal1 & 0.004 & 0.035 & 0.034 & 94 \\ 
   \hline
DR1.xgb5 & 0.011 & 0.069 & 0.067 & 95 \\ 
  DR2.xgb5 & 0.007 & 0.035 & 0.033 & 92 \\ 
  TMLE1.xgb5 & 0.009 & 0.069 & 0.067 & 94 \\ 
  TMLE2.xgb5 & 0.006 & 0.035 & 0.034 & 93 \\ 
   \hline
DR1.xgb1 & 0.011 & 0.068 & 0.067 & 95 \\ 
  DR2.xgb1 & 0.008 & 0.035 & 0.033 & 92 \\ 
  TMLE1.xgb1 & 0.010 & 0.068 & 0.067 & 95 \\ 
  TMLE2.xgb1 & 0.006 & 0.035 & 0.033 & 94 \\ 
    \end{tabular}
  \vspace{.8cm}
\caption{Bias, empirical SE (`empSE'), mean estimated SE (`SEhat'), and CI coverage (`cover') for Scenario 1, where $E(n^B) =5000$, $M=150$, $n_{\rm house}=20$ and parametric nuisance models are correctly specified.} 
\label{tab:scenario1}
\end{table}

\begin{table}[ht]
\centering
\begin{tabular}{rrrrr}
  & bias & empSE & SEhat & cover \\ 
  \hline
\hline
HT & -0.006 & 0.127 & 0.127 & 95 \\ 
  Haj & -0.009 & 0.064 & 0.064 & 96 \\ 
  naive & -0.313 & 0.061 &  &  \\ 
   \hline
DR1 & -0.001 & 0.120 & 0.122 & 94 \\ 
  DR2clw & -0.005 & 0.053 & 0.055 & 94 \\ 
  TMLE1 & -0.001 & 0.120 & 0.122 & 94 \\ 
  TMLE2 & -0.005 & 0.053 & 0.055 & 94 \\ 
  KH & -0.001 & 0.120 & 0.122 & 94 \\ 
   \hline
DR1.hal5 & 0.005 & 0.120 & 0.121 & 94 \\ 
  DR2.hal5 & 0.001 & 0.053 & 0.053 & 94 \\ 
  TMLE1.hal5 & 0.004 & 0.120 & 0.121 & 94 \\ 
  TMLE2.hal5 & 0.001 & 0.053 & 0.053 & 94 \\ 
   \hline
DR1.hal1 & 0.004 & 0.120 & 0.121 & 94 \\ 
  DR2.hal1 & 0.000 & 0.053 & 0.053 & 94 \\ 
  TMLE1.hal1 & 0.003 & 0.120 & 0.121 & 94 \\ 
  TMLE2.hal1 & -0.000 & 0.053 & 0.053 & 94 \\ 
   \hline
DR1.xgb5 & 0.008 & 0.120 & 0.122 & 95 \\ 
  DR2.xgb5 & 0.004 & 0.053 & 0.053 & 93 \\ 
  TMLE1.xgb5 & 0.007 & 0.120 & 0.122 & 95 \\ 
  TMLE2.xgb5 & 0.003 & 0.053 & 0.053 & 94 \\ 
   \hline
DR1.xgb1 & 0.008 & 0.120 & 0.122 & 94 \\ 
  DR2.xgb1 & 0.004 & 0.053 & 0.053 & 93 \\ 
  TMLE1.xgb1 & 0.006 & 0.120 & 0.122 & 94 \\ 
  TMLE2.xgb1 & 0.003 & 0.053 & 0.053 & 93 \\ 
    \end{tabular}
  \vspace{.8cm}
\caption{Bias, empirical SE (`empSE'), mean estimated SE (`SEhat'), and CI coverage (`cover') for Scenario 2, where $E(n^B) =5000$, $M=150$, $n_{\rm house}=5$ and parametric nuisance models are correctly specified.} 
\label{tab:scenario2}
\end{table}

\begin{table}[ht]
\centering
\begin{tabular}{rrrrr}
  & bias & empSE & SEhat & cover \\ 
  \hline
\hline
HT & -0.003 & 0.120 & 0.119 & 93 \\ 
  Haj & -0.000 & 0.060 & 0.062 & 94 \\ 
  naive & -0.308 & 0.058 &  &  \\ 
   \hline
DR1 & -0.004 & 0.118 & 0.115 & 93 \\ 
  DR2clw & -0.001 & 0.055 & 0.055 & 94 \\ 
  TMLE1 & -0.004 & 0.118 & 0.115 & 93 \\ 
  TMLE2 & -0.001 & 0.055 & 0.055 & 94 \\ 
  KH & -0.004 & 0.118 & 0.115 & 93 \\ 
   \hline
DR1.hal5 & 0.001 & 0.118 & 0.114 & 93 \\ 
  DR2.hal5 & 0.004 & 0.054 & 0.053 & 94 \\ 
  TMLE1.hal5 & 0.001 & 0.117 & 0.114 & 93 \\ 
  TMLE2.hal5 & 0.004 & 0.054 & 0.054 & 94 \\ 
   \hline
DR1.hal1 & 0.001 & 0.118 & 0.114 & 93 \\ 
  DR2.hal1 & 0.004 & 0.055 & 0.054 & 94 \\ 
  TMLE1.hal1 & 0.000 & 0.118 & 0.114 & 93 \\ 
  TMLE2.hal1 & 0.003 & 0.055 & 0.054 & 94 \\ 
   \hline
DR1.xgb5 & 0.005 & 0.118 & 0.114 & 92 \\ 
  DR2.xgb5 & 0.008 & 0.055 & 0.053 & 93 \\ 
  TMLE1.xgb5 & 0.004 & 0.118 & 0.114 & 93 \\ 
  TMLE2.xgb5 & 0.007 & 0.055 & 0.054 & 93 \\ 
   \hline
DR1.xgb1 & 0.004 & 0.118 & 0.114 & 93 \\ 
  DR2.xgb1 & 0.007 & 0.055 & 0.053 & 93 \\ 
  TMLE1.xgb1 & 0.003 & 0.118 & 0.114 & 93 \\ 
  TMLE2.xgb1 & 0.005 & 0.055 & 0.053 & 93 \\ 
    \end{tabular}
  \vspace{.8cm}
\caption{Bias, empirical SE (`empSE'), mean estimated SE (`SEhat'), and CI coverage (`cover') for Scenario 3, where $E(n^B) =5000$, $M=50$, $n_{\rm house}=20$ and parametric nuisance models are correctly specified.} 
\label{tab:scenario3}
\end{table}

\begin{table}[ht]
\centering
\begin{tabular}{rrrrr}
  & bias & empSE & SEhat & cover \\ 
  \hline
\hline
HT & 0.010 & 0.208 & 0.220 & 97 \\ 
  Haj & 0.004 & 0.112 & 0.112 & 94 \\ 
  naive & -0.303 & 0.102 &  &  \\ 
   \hline
DR1 & 0.010 & 0.198 & 0.210 & 96 \\ 
  DR2clw & 0.004 & 0.089 & 0.092 & 94 \\ 
  TMLE1 & 0.009 & 0.198 & 0.210 & 96 \\ 
  TMLE2 & 0.003 & 0.089 & 0.092 & 94 \\ 
  KH & 0.009 & 0.198 & 0.210 & 96 \\ 
   \hline
DR1.hal5 & 0.017 & 0.196 & 0.209 & 96 \\ 
  DR2.hal5 & 0.011 & 0.089 & 0.089 & 94 \\ 
  TMLE1.hal5 & 0.016 & 0.196 & 0.208 & 96 \\ 
  TMLE2.hal5 & 0.010 & 0.089 & 0.090 & 94 \\ 
   \hline
DR1.hal1 & 0.015 & 0.197 & 0.209 & 96 \\ 
  DR2.hal1 & 0.009 & 0.089 & 0.090 & 94 \\ 
  TMLE1.hal1 & 0.015 & 0.197 & 0.209 & 96 \\ 
  TMLE2.hal1 & 0.009 & 0.089 & 0.090 & 94 \\ 
   \hline
DR1.xgb5 & 0.020 & 0.197 & 0.209 & 96 \\ 
  DR2.xgb5 & 0.014 & 0.089 & 0.089 & 94 \\ 
  TMLE1.xgb5 & 0.019 & 0.197 & 0.209 & 96 \\ 
  TMLE2.xgb5 & 0.013 & 0.088 & 0.089 & 94 \\ 
   \hline
DR1.xgb1 & 0.019 & 0.198 & 0.209 & 96 \\ 
  DR2.xgb1 & 0.013 & 0.089 & 0.089 & 94 \\ 
  TMLE1.xgb1 & 0.018 & 0.198 & 0.209 & 96 \\ 
  TMLE2.xgb1 & 0.012 & 0.089 & 0.089 & 94 \\ 
    \end{tabular}
  \vspace{.8cm}
\caption{Bias, empirical SE (`empSE'), mean estimated SE (`SEhat'), and CI coverage (`cover') for Scenario 4, where $E(n^B) =5000$, $M=50$, $n_{\rm house}=5$ and parametric nuisance models are correctly specified.} 
\label{tab:scenario4}
\end{table}

\begin{table}[ht]
\centering
\begin{tabular}{rrrrr}
  & bias & empSE & SEhat & cover \\ 
  \hline
\hline
HT & 0.005 & 0.069 & 0.068 & 95 \\ 
  Haj & 0.001 & 0.035 & 0.035 & 94 \\ 
  naive & -0.306 & 0.034 &  &  \\ 
   \hline
DR1 & 0.003 & 0.075 & 0.078 & 96 \\ 
  DR2clw & 0.000 & 0.051 & 0.051 & 94 \\ 
  TMLE1 & 0.003 & 0.075 & 0.078 & 96 \\ 
  TMLE2 & 0.000 & 0.051 & 0.051 & 94 \\ 
  KH & 0.003 & 0.075 & 0.077 & 95 \\ 
   \hline
DR1.hal5 & 0.013 & 0.075 & 0.075 & 96 \\ 
  DR2.hal5 & 0.010 & 0.051 & 0.048 & 94 \\ 
  TMLE1.hal5 & 0.013 & 0.076 & 0.075 & 95 \\ 
  TMLE2.hal5 & 0.010 & 0.051 & 0.048 & 93 \\ 
   \hline
DR1.hal1 & 0.013 & 0.075 & 0.075 & 96 \\ 
  DR2.hal1 & 0.009 & 0.051 & 0.048 & 94 \\ 
  TMLE1.hal1 & 0.011 & 0.075 & 0.075 & 96 \\ 
  TMLE2.hal1 & 0.008 & 0.051 & 0.048 & 94 \\ 
   \hline
DR1.xgb5 & 0.017 & 0.078 & 0.076 & 95 \\ 
  DR2.xgb5 & 0.014 & 0.053 & 0.048 & 92 \\ 
  TMLE1.xgb5 & 0.016 & 0.077 & 0.076 & 94 \\ 
  TMLE2.xgb5 & 0.013 & 0.053 & 0.048 & 91 \\ 
   \hline
DR1.xgb1 & 0.021 & 0.077 & 0.074 & 94 \\ 
  DR2.xgb1 & 0.018 & 0.052 & 0.045 & 89 \\ 
  TMLE1.xgb1 & 0.018 & 0.077 & 0.074 & 95 \\ 
  TMLE2.xgb1 & 0.014 & 0.053 & 0.045 & 89 \\ 
    \end{tabular}
  \vspace{.8cm}
\caption{Bias, empirical SE (`empSE'), mean estimated SE (`SEhat'), and CI coverage (`cover') for Scenario 5, where $E(n^B) =1000$, $M=150$, $n_{\rm house}=20$ and parametric nuisance models are correctly specified.} 
\label{tab:scenario5}
\end{table}

\begin{table}[ht]
\centering
\begin{tabular}{rrrrr}
  & bias & empSE & SEhat & cover \\ 
  \hline
\hline
HT & -0.006 & 0.127 & 0.127 & 95 \\ 
  Haj & -0.009 & 0.064 & 0.064 & 96 \\ 
  naive & -0.313 & 0.061 &  &  \\ 
   \hline
DR1 & 0.002 & 0.130 & 0.128 & 94 \\ 
  DR2clw & -0.002 & 0.069 & 0.067 & 93 \\ 
  TMLE1 & 0.002 & 0.129 & 0.128 & 94 \\ 
  TMLE2 & -0.002 & 0.069 & 0.067 & 93 \\ 
  KH & 0.002 & 0.129 & 0.128 & 94 \\ 
   \hline
DR1.hal5 & 0.016 & 0.127 & 0.125 & 94 \\ 
  DR2.hal5 & 0.013 & 0.068 & 0.062 & 92 \\ 
  TMLE1.hal5 & 0.016 & 0.128 & 0.125 & 94 \\ 
  TMLE2.hal5 & 0.012 & 0.069 & 0.062 & 91 \\ 
   \hline
DR1.hal1 & 0.014 & 0.127 & 0.125 & 94 \\ 
  DR2.hal1 & 0.011 & 0.068 & 0.062 & 91 \\ 
  TMLE1.hal1 & 0.013 & 0.128 & 0.125 & 94 \\ 
  TMLE2.hal1 & 0.009 & 0.068 & 0.062 & 90 \\ 
   \hline
DR1.xgb5 & 0.024 & 0.129 & 0.125 & 93 \\ 
  DR2.xgb5 & 0.021 & 0.069 & 0.062 & 88 \\ 
  TMLE1.xgb5 & 0.023 & 0.129 & 0.125 & 94 \\ 
  TMLE2.xgb5 & 0.020 & 0.069 & 0.062 & 90 \\ 
   \hline
DR1.xgb1 & 0.025 & 0.128 & 0.125 & 93 \\ 
  DR2.xgb1 & 0.021 & 0.068 & 0.060 & 89 \\ 
  TMLE1.xgb1 & 0.022 & 0.128 & 0.124 & 93 \\ 
  TMLE2.xgb1 & 0.018 & 0.068 & 0.060 & 90 \\ 
    \end{tabular}
  \vspace{.8cm}
\caption{Bias, empirical SE (`empSE'), mean estimated SE (`SEhat'), and CI coverage (`cover') for Scenario 6, where $E(n^B) =1000$, $M=150$, $n_{\rm house}=5$ and parametric nuisance models are correctly specified.} 
\label{tab:scenario6}
\end{table}

\begin{table}[ht]
\centering
\begin{tabular}{rrrrr}
  & bias & empSE & SEhat & cover \\ 
  \hline
\hline
HT & -0.003 & 0.120 & 0.119 & 93 \\ 
  Haj & -0.000 & 0.060 & 0.062 & 94 \\ 
  naive & -0.308 & 0.058 &  &  \\ 
   \hline
DR1 & -0.003 & 0.123 & 0.122 & 93 \\ 
  DR2clw & -0.000 & 0.067 & 0.068 & 96 \\ 
  TMLE1 & -0.003 & 0.123 & 0.121 & 93 \\ 
  TMLE2 & -0.000 & 0.067 & 0.067 & 96 \\ 
  KH & -0.003 & 0.123 & 0.121 & 93 \\ 
   \hline
DR1.hal5 & 0.010 & 0.122 & 0.118 & 93 \\ 
  DR2.hal5 & 0.013 & 0.067 & 0.063 & 91 \\ 
  TMLE1.hal5 & 0.010 & 0.122 & 0.119 & 93 \\ 
  TMLE2.hal5 & 0.013 & 0.067 & 0.064 & 93 \\ 
   \hline
DR1.hal1 & 0.009 & 0.123 & 0.118 & 93 \\ 
  DR2.hal1 & 0.012 & 0.067 & 0.062 & 92 \\ 
  TMLE1.hal1 & 0.008 & 0.123 & 0.118 & 92 \\ 
  TMLE2.hal1 & 0.011 & 0.067 & 0.063 & 92 \\ 
   \hline
DR1.xgb5 & 0.018 & 0.121 & 0.118 & 94 \\ 
  DR2.xgb5 & 0.021 & 0.065 & 0.062 & 93 \\ 
  TMLE1.xgb5 & 0.017 & 0.121 & 0.119 & 94 \\ 
  TMLE2.xgb5 & 0.020 & 0.065 & 0.063 & 94 \\ 
   \hline
DR1.xgb1 & 0.019 & 0.122 & 0.117 & 92 \\ 
  DR2.xgb1 & 0.022 & 0.065 & 0.060 & 90 \\ 
  TMLE1.xgb1 & 0.017 & 0.122 & 0.118 & 92 \\ 
  TMLE2.xgb1 & 0.019 & 0.065 & 0.061 & 91 \\ 
    \end{tabular}
  \vspace{.8cm}
\caption{Bias, empirical SE (`empSE'), mean estimated SE (`SEhat'), and CI coverage (`cover') for Scenario 7, where $E(n^B) =1000$, $M=50$, $n_{\rm house}=20$ and parametric nuisance models are correctly specified.} 
\label{tab:scenario7}
\end{table}

\begin{table}[ht]
\centering
\begin{tabular}{rrrrr}
  & bias & empSE & SEhat & cover \\ 
  \hline
\hline
HT & 0.010 & 0.208 & 0.220 & 97 \\ 
  Haj & 0.004 & 0.112 & 0.112 & 94 \\ 
  naive & -0.303 & 0.102 &  &  \\ 
   \hline
DR1 & 0.009 & 0.203 & 0.214 & 95 \\ 
  DR2clw & 0.003 & 0.099 & 0.101 & 94 \\ 
  TMLE1 & 0.009 & 0.203 & 0.214 & 95 \\ 
  TMLE2 & 0.003 & 0.099 & 0.101 & 94 \\ 
  KH & 0.009 & 0.202 & 0.213 & 95 \\ 
   \hline
DR1.hal5 & 0.028 & 0.200 & 0.209 & 95 \\ 
  DR2.hal5 & 0.022 & 0.098 & 0.093 & 92 \\ 
  TMLE1.hal5 & 0.027 & 0.201 & 0.209 & 95 \\ 
  TMLE2.hal5 & 0.021 & 0.099 & 0.094 & 92 \\ 
   \hline
DR1.hal1 & 0.025 & 0.201 & 0.209 & 95 \\ 
  DR2.hal1 & 0.019 & 0.098 & 0.093 & 92 \\ 
  TMLE1.hal1 & 0.024 & 0.201 & 0.209 & 95 \\ 
  TMLE2.hal1 & 0.018 & 0.099 & 0.094 & 92 \\ 
   \hline
DR1.xgb5 & 0.037 & 0.203 & 0.210 & 95 \\ 
  DR2.xgb5 & 0.031 & 0.099 & 0.093 & 92 \\ 
  TMLE1.xgb5 & 0.036 & 0.203 & 0.210 & 95 \\ 
  TMLE2.xgb5 & 0.030 & 0.099 & 0.094 & 92 \\ 
   \hline
DR1.xgb1 & 0.036 & 0.203 & 0.210 & 95 \\ 
  DR2.xgb1 & 0.030 & 0.098 & 0.092 & 91 \\ 
  TMLE1.xgb1 & 0.033 & 0.203 & 0.210 & 95 \\ 
  TMLE2.xgb1 & 0.027 & 0.098 & 0.092 & 91 \\ 
    \end{tabular}
  \vspace{.8cm}
\caption{Bias, empirical SE (`empSE'), mean estimated SE (`SEhat'), and CI coverage (`cover') for Scenario 8, where $E(n^B) =1000$, $M=50$, $n_{\rm house}=5$ and parametric nuisance models are correctly specified.} 
\label{tab:scenario8}
\end{table}

\begin{table}[ht]
\centering
\begin{tabular}{rrrrr}
  & bias & empSE & SEhat & cover \\ 
  \hline
\hline
HT & 0.005 & 0.069 & 0.068 & 95 \\ 
  Haj & 0.001 & 0.035 & 0.035 & 94 \\ 
  naive & -0.306 & 0.034 &  &  \\ 
   \hline
DR1 & 0.007 & 0.088 & 0.088 & 95 \\ 
  DR2clw & 0.003 & 0.066 & 0.066 & 94 \\ 
  TMLE1 & 0.007 & 0.088 & 0.088 & 95 \\ 
  TMLE2 & 0.003 & 0.066 & 0.066 & 94 \\ 
  KH & 0.007 & 0.088 & 0.087 & 95 \\ 
   \hline
DR1.hal5 & 0.020 & 0.094 & 0.087 & 94 \\ 
  DR2.hal5 & 0.017 & 0.075 & 0.065 & 92 \\ 
  TMLE1.hal5 & 0.023 & 0.090 & 0.085 & 92 \\ 
  TMLE2.hal5 & 0.020 & 0.069 & 0.062 & 91 \\ 
   \hline
DR1.hal1 & 0.022 & 0.088 & 0.083 & 94 \\ 
  DR2.hal1 & 0.019 & 0.067 & 0.060 & 91 \\ 
  TMLE1.hal1 & 0.020 & 0.088 & 0.083 & 94 \\ 
  TMLE2.hal1 & 0.016 & 0.067 & 0.060 & 91 \\ 
   \hline
DR1.xgb5 & 0.026 & 0.091 & 0.086 & 92 \\ 
  DR2.xgb5 & 0.023 & 0.070 & 0.064 & 90 \\ 
  TMLE1.xgb5 & 0.027 & 0.092 & 0.086 & 91 \\ 
  TMLE2.xgb5 & 0.023 & 0.071 & 0.063 & 90 \\ 
   \hline
DR1.xgb1 & 0.034 & 0.089 & 0.080 & 88 \\ 
  DR2.xgb1 & 0.031 & 0.068 & 0.056 & 86 \\ 
  TMLE1.xgb1 & 0.030 & 0.090 & 0.080 & 90 \\ 
  TMLE2.xgb1 & 0.027 & 0.069 & 0.056 & 88 \\ 
    \end{tabular}
  \vspace{.8cm}
\caption{Bias, empirical SE (`empSE'), mean estimated SE (`SEhat'), and CI coverage (`cover') for Scenario 9, where $E(n^B) =500$, $M=150$, $n_{\rm house}=20$ and parametric nuisance models are correctly specified.} 
\label{tab:scenario9}
\end{table}

\begin{table}[ht]
\centering
\begin{tabular}{rrrrr}
  & bias & empSE & SEhat & cover \\ 
  \hline
\hline
HT & -0.006 & 0.127 & 0.127 & 95 \\ 
  Haj & -0.009 & 0.064 & 0.064 & 96 \\ 
  naive & -0.313 & 0.061 &  &  \\ 
   \hline
DR1 & 0.001 & 0.137 & 0.136 & 94 \\ 
  DR2clw & -0.002 & 0.082 & 0.080 & 93 \\ 
  TMLE1 & 0.002 & 0.137 & 0.136 & 94 \\ 
  TMLE2 & -0.002 & 0.082 & 0.080 & 93 \\ 
  KH & 0.001 & 0.137 & 0.135 & 94 \\ 
   \hline
DR1.hal5 & 0.023 & 0.137 & 0.131 & 92 \\ 
  DR2.hal5 & 0.019 & 0.083 & 0.073 & 90 \\ 
  TMLE1.hal5 & 0.023 & 0.138 & 0.130 & 92 \\ 
  TMLE2.hal5 & 0.019 & 0.085 & 0.073 & 87 \\ 
   \hline
DR1.hal1 & 0.022 & 0.136 & 0.130 & 92 \\ 
  DR2.hal1 & 0.019 & 0.083 & 0.071 & 89 \\ 
  TMLE1.hal1 & 0.019 & 0.137 & 0.130 & 92 \\ 
  TMLE2.hal1 & 0.015 & 0.083 & 0.072 & 89 \\ 
   \hline
DR1.xgb5 & 0.031 & 0.137 & 0.131 & 92 \\ 
  DR2.xgb5 & 0.027 & 0.084 & 0.073 & 88 \\ 
  TMLE1.xgb5 & 0.031 & 0.137 & 0.131 & 91 \\ 
  TMLE2.xgb5 & 0.027 & 0.084 & 0.073 & 88 \\ 
   \hline
DR1.xgb1 & 0.036 & 0.135 & 0.128 & 91 \\ 
  DR2.xgb1 & 0.033 & 0.081 & 0.067 & 84 \\ 
  TMLE1.xgb1 & 0.033 & 0.136 & 0.128 & 92 \\ 
  TMLE2.xgb1 & 0.030 & 0.082 & 0.068 & 85 \\ 
    \end{tabular}
  \vspace{.8cm}
\caption{Bias, empirical SE (`empSE'), mean estimated SE (`SEhat'), and CI coverage (`cover') for Scenario 10, where $E(n^B) =500$, $M=150$, $n_{\rm house}=5$ and parametric nuisance models are correctly specified.} 
\label{tab:scenario10}
\end{table}

\begin{table}[ht]
\centering
\begin{tabular}{rrrrr}
  & bias & empSE & SEhat & cover \\ 
  \hline
\hline
HT & -0.003 & 0.120 & 0.119 & 93 \\ 
  Haj & -0.000 & 0.060 & 0.062 & 94 \\ 
  naive & -0.308 & 0.058 &  &  \\ 
   \hline
DR1 & -0.002 & 0.131 & 0.129 & 94 \\ 
  DR2clw & 0.000 & 0.080 & 0.080 & 96 \\ 
  TMLE1 & -0.002 & 0.130 & 0.129 & 94 \\ 
  TMLE2 & 0.000 & 0.080 & 0.080 & 95 \\ 
  KH & -0.003 & 0.130 & 0.128 & 94 \\ 
   \hline
DR1.hal5 & 0.013 & 0.133 & 0.125 & 93 \\ 
  DR2.hal5 & 0.016 & 0.083 & 0.075 & 92 \\ 
  TMLE1.hal5 & 0.016 & 0.131 & 4.014 & 92 \\ 
  TMLE2.hal5 & 0.019 & 0.081 & 3.964 & 92 \\ 
   \hline
DR1.hal1 & 0.016 & 0.131 & 0.123 & 92 \\ 
  DR2.hal1 & 0.018 & 0.081 & 0.072 & 90 \\ 
  TMLE1.hal1 & 0.013 & 0.131 & 0.123 & 92 \\ 
  TMLE2.hal1 & 0.016 & 0.081 & 0.072 & 90 \\ 
   \hline
DR1.xgb5 & 0.024 & 0.132 & 0.125 & 92 \\ 
  DR2.xgb5 & 0.027 & 0.081 & 0.074 & 91 \\ 
  TMLE1.xgb5 & 0.025 & 0.132 & 0.125 & 92 \\ 
  TMLE2.xgb5 & 0.028 & 0.081 & 0.075 & 91 \\ 
   \hline
DR1.xgb1 & 0.030 & 0.130 & 0.121 & 92 \\ 
  DR2.xgb1 & 0.033 & 0.079 & 0.068 & 87 \\ 
  TMLE1.xgb1 & 0.027 & 0.130 & 0.121 & 92 \\ 
  TMLE2.xgb1 & 0.030 & 0.080 & 0.068 & 88 \\ 
    \end{tabular}
  \vspace{.8cm}
\caption{Bias, empirical SE (`empSE'), mean estimated SE (`SEhat'), and CI coverage (`cover') for Scenario 11, where $E(n^B) =500$, $M=50$, $n_{\rm house}=20$ and parametric nuisance models are correctly specified.} 
\label{tab:scenario11}
\end{table}

\begin{table}[ht]
\centering
\begin{tabular}{rrrrr}
  & bias & empSE & SEhat & cover \\ 
  \hline
\hline
HT & 0.010 & 0.208 & 0.220 & 97 \\ 
  Haj & 0.004 & 0.112 & 0.112 & 94 \\ 
  naive & -0.303 & 0.102 &  &  \\ 
   \hline
DR1 & 0.010 & 0.206 & 0.219 & 95 \\ 
  DR2clw & 0.004 & 0.106 & 0.111 & 95 \\ 
  TMLE1 & 0.010 & 0.205 & 0.219 & 96 \\ 
  TMLE2 & 0.004 & 0.106 & 0.110 & 95 \\ 
  KH & 0.011 & 0.204 & 0.217 & 95 \\ 
   \hline
DR1.hal5 & 0.040 & 0.213 & 0.214 & 96 \\ 
  DR2.hal5 & 0.030 & 0.106 & 0.099 & 91 \\ 
  TMLE1.hal5 & 0.034 & 0.203 & 0.211 & 95 \\ 
  TMLE2.hal5 & 0.028 & 0.106 & 0.100 & 91 \\ 
   \hline
DR1.hal1 & 0.033 & 0.203 & 0.211 & 96 \\ 
  DR2.hal1 & 0.027 & 0.105 & 0.099 & 93 \\ 
  TMLE1.hal1 & 0.030 & 0.204 & 0.211 & 96 \\ 
  TMLE2.hal1 & 0.024 & 0.105 & 0.099 & 93 \\ 
   \hline
DR1.xgb5 & 0.047 & 0.205 & 0.213 & 96 \\ 
  DR2.xgb5 & 0.041 & 0.107 & 0.100 & 92 \\ 
  TMLE1.xgb5 & 0.047 & 0.206 & 0.213 & 96 \\ 
  TMLE2.xgb5 & 0.041 & 0.106 & 0.101 & 92 \\ 
   \hline
DR1.xgb1 & 0.049 & 0.206 & 0.211 & 96 \\ 
  DR2.xgb1 & 0.043 & 0.105 & 0.096 & 91 \\ 
  TMLE1.xgb1 & 0.047 & 0.206 & 0.211 & 96 \\ 
  TMLE2.xgb1 & 0.041 & 0.106 & 0.097 & 91 \\ 
    \end{tabular}
  \vspace{.8cm}
\caption{Bias, empirical SE (`empSE'), mean estimated SE (`SEhat'), and CI coverage (`cover') for Scenario 12, where $E(n^B) =500$, $M=50$, $n_{\rm house}=5$ and parametric nuisance models are correctly specified.} 
\label{tab:scenario12}
\end{table}

\begin{table}[ht]
\centering
\begin{tabular}{rrrrr}
  & bias & empSE & SEhat & cover \\ 
  \hline
\hline
HT & 0.003 & 0.045 & 0.044 & 95 \\ 
  Haj & 0.001 & 0.033 & 0.032 & 95 \\ 
  naive & -0.151 & 0.029 &  &  \\ 
   \hline
DR1 & 0.103 & 0.043 & 0.044 & 33 \\ 
  DR2clw & 0.100 & 0.030 & 0.031 & 10 \\ 
  TMLE1 & 0.099 & 0.043 & 0.044 & 37 \\ 
  TMLE2 & 0.097 & 0.030 & 0.031 & 11 \\ 
  KH & 0.085 & 0.043 & 0.044 & 51 \\ 
   \hline
DR1.hal5 & 0.006 & 0.043 & 0.043 & 94 \\ 
  DR2.hal5 & 0.004 & 0.031 & 0.031 & 96 \\ 
  TMLE1.hal5 & 0.006 & 0.043 & 0.043 & 94 \\ 
  TMLE2.hal5 & 0.005 & 0.031 & 0.031 & 96 \\ 
   \hline
DR1.hal1 & 0.006 & 0.044 & 0.043 & 95 \\ 
  DR2.hal1 & 0.004 & 0.031 & 0.031 & 95 \\ 
  TMLE1.hal1 & 0.006 & 0.044 & 0.043 & 95 \\ 
  TMLE2.hal1 & 0.004 & 0.031 & 0.031 & 95 \\ 
   \hline
DR1.xgb5 & 0.011 & 0.044 & 0.043 & 95 \\ 
  DR2.xgb5 & 0.009 & 0.031 & 0.031 & 92 \\ 
  TMLE1.xgb5 & 0.010 & 0.044 & 0.043 & 94 \\ 
  TMLE2.xgb5 & 0.009 & 0.031 & 0.031 & 94 \\ 
   \hline
DR1.xgb1 & 0.011 & 0.044 & 0.043 & 94 \\ 
  DR2.xgb1 & 0.009 & 0.031 & 0.030 & 93 \\ 
  TMLE1.xgb1 & 0.011 & 0.044 & 0.043 & 94 \\ 
  TMLE2.xgb1 & 0.009 & 0.031 & 0.030 & 93 \\ 
    \end{tabular}
  \vspace{.8cm}
\caption{Bias, empirical SE (`empSE'), mean estimated SE (`SEhat'), and CI coverage (`cover') for Scenario 13, where $E(n^B) =5000$, $M=150$, $n_{\rm house}=20$ and parametric nuisance models are misspecified.} 
\label{tab:scenario13}
\end{table}

\begin{table}[ht]
\centering
\begin{tabular}{rrrrr}
  & bias & empSE & SEhat & cover \\ 
  \hline
\hline
HT & -0.007 & 0.084 & 0.084 & 96 \\ 
  Haj & -0.008 & 0.063 & 0.062 & 94 \\ 
  naive & -0.158 & 0.056 &  &  \\ 
   \hline
DR1 & 0.100 & 0.074 & 0.077 & 75 \\ 
  DR2clw & 0.097 & 0.048 & 0.050 & 49 \\ 
  TMLE1 & 0.095 & 0.074 & 0.077 & 77 \\ 
  TMLE2 & 0.093 & 0.048 & 0.050 & 53 \\ 
  KH & 0.080 & 0.075 & 0.077 & 85 \\ 
   \hline
DR1.hal5 & 0.003 & 0.075 & 0.076 & 96 \\ 
  DR2.hal5 & 0.001 & 0.051 & 0.051 & 96 \\ 
  TMLE1.hal5 & 0.002 & 0.075 & 0.076 & 95 \\ 
  TMLE2.hal5 & 0.001 & 0.051 & 0.051 & 96 \\ 
   \hline
DR1.hal1 & 0.002 & 0.074 & 0.076 & 96 \\ 
  DR2.hal1 & 0.000 & 0.051 & 0.051 & 96 \\ 
  TMLE1.hal1 & 0.002 & 0.074 & 0.076 & 96 \\ 
  TMLE2.hal1 & 0.000 & 0.051 & 0.051 & 96 \\ 
   \hline
DR1.xgb5 & 0.007 & 0.074 & 0.076 & 95 \\ 
  DR2.xgb5 & 0.005 & 0.050 & 0.051 & 96 \\ 
  TMLE1.xgb5 & 0.006 & 0.074 & 0.076 & 96 \\ 
  TMLE2.xgb5 & 0.005 & 0.050 & 0.051 & 96 \\ 
   \hline
DR1.xgb1 & 0.006 & 0.074 & 0.076 & 96 \\ 
  DR2.xgb1 & 0.005 & 0.050 & 0.051 & 96 \\ 
  TMLE1.xgb1 & 0.006 & 0.074 & 0.076 & 96 \\ 
  TMLE2.xgb1 & 0.004 & 0.050 & 0.051 & 96 \\ 
    \end{tabular}
  \vspace{.8cm}
\caption{Bias, empirical SE (`empSE'), mean estimated SE (`SEhat'), and CI coverage (`cover') for Scenario 14, where $E(n^B) =5000$, $M=150$, $n_{\rm house}=5$ and parametric nuisance models are misspecified.} 
\label{tab:scenario14}
\end{table}

\begin{table}[ht]
\centering
\begin{tabular}{rrrrr}
  & bias & empSE & SEhat & cover \\ 
  \hline
\hline
HT & 0.000 & 0.076 & 0.077 & 95 \\ 
  Haj & 0.001 & 0.054 & 0.057 & 96 \\ 
  naive & -0.151 & 0.047 &  &  \\ 
   \hline
DR1 & 0.100 & 0.070 & 0.071 & 73 \\ 
  DR2clw & 0.100 & 0.043 & 0.046 & 41 \\ 
  TMLE1 & 0.095 & 0.071 & 0.071 & 76 \\ 
  TMLE2 & 0.097 & 0.043 & 0.046 & 44 \\ 
  KH & 0.081 & 0.072 & 0.071 & 82 \\ 
   \hline
DR1.hal5 & 0.004 & 0.072 & 0.071 & 94 \\ 
  DR2.hal5 & 0.005 & 0.048 & 0.048 & 95 \\ 
  TMLE1.hal5 & 0.004 & 0.072 & 0.071 & 93 \\ 
  TMLE2.hal5 & 0.006 & 0.047 & 0.048 & 95 \\ 
   \hline
DR1.hal1 & 0.003 & 0.072 & 0.071 & 94 \\ 
  DR2.hal1 & 0.005 & 0.048 & 0.048 & 95 \\ 
  TMLE1.hal1 & 0.003 & 0.072 & 0.071 & 94 \\ 
  TMLE2.hal1 & 0.005 & 0.048 & 0.048 & 95 \\ 
   \hline
DR1.xgb5 & 0.008 & 0.073 & 0.071 & 93 \\ 
  DR2.xgb5 & 0.010 & 0.048 & 0.048 & 95 \\ 
  TMLE1.xgb5 & 0.008 & 0.072 & 0.071 & 93 \\ 
  TMLE2.xgb5 & 0.010 & 0.048 & 0.048 & 95 \\ 
   \hline
DR1.xgb1 & 0.008 & 0.072 & 0.071 & 93 \\ 
  DR2.xgb1 & 0.010 & 0.048 & 0.047 & 95 \\ 
  TMLE1.xgb1 & 0.008 & 0.072 & 0.071 & 93 \\ 
  TMLE2.xgb1 & 0.009 & 0.048 & 0.048 & 95 \\ 
    \end{tabular}
  \vspace{.8cm}
\caption{Bias, empirical SE (`empSE'), mean estimated SE (`SEhat'), and CI coverage (`cover') for Scenario 15, where $E(n^B) =5000$, $M=50$, $n_{\rm house}=20$ and parametric nuisance models are misspecified.} 
\label{tab:scenario15}
\end{table}

\begin{table}[ht]
\centering
\begin{tabular}{rrrrr}
  & bias & empSE & SEhat & cover \\ 
  \hline
\hline
HT & 0.009 & 0.142 & 0.146 & 96 \\ 
  Haj & 0.005 & 0.109 & 0.109 & 94 \\ 
  naive & -0.147 & 0.092 &  &  \\ 
   \hline
DR1 & 0.114 & 0.119 & 0.130 & 90 \\ 
  DR2clw & 0.108 & 0.075 & 0.083 & 78 \\ 
  TMLE1 & 0.105 & 0.122 & 0.130 & 91 \\ 
  TMLE2 & 0.102 & 0.077 & 0.083 & 80 \\ 
  KH & 0.089 & 0.124 & 0.131 & 92 \\ 
   \hline
DR1.hal5 & 0.013 & 0.124 & 0.131 & 96 \\ 
  DR2.hal5 & 0.010 & 0.084 & 0.087 & 94 \\ 
  TMLE1.hal5 & 0.012 & 0.123 & 0.130 & 96 \\ 
  TMLE2.hal5 & 0.009 & 0.083 & 0.086 & 95 \\ 
   \hline
DR1.hal1 & 0.012 & 0.124 & 0.130 & 96 \\ 
  DR2.hal1 & 0.009 & 0.083 & 0.086 & 94 \\ 
  TMLE1.hal1 & 0.012 & 0.124 & 0.130 & 96 \\ 
  TMLE2.hal1 & 0.009 & 0.083 & 0.086 & 94 \\ 
   \hline
DR1.xgb5 & 0.017 & 0.125 & 0.131 & 95 \\ 
  DR2.xgb5 & 0.014 & 0.084 & 0.087 & 95 \\ 
  TMLE1.xgb5 & 0.017 & 0.125 & 0.131 & 95 \\ 
  TMLE2.xgb5 & 0.014 & 0.084 & 0.087 & 95 \\ 
   \hline
DR1.xgb1 & 0.017 & 0.125 & 0.131 & 96 \\ 
  DR2.xgb1 & 0.014 & 0.084 & 0.086 & 94 \\ 
  TMLE1.xgb1 & 0.016 & 0.125 & 0.131 & 96 \\ 
  TMLE2.xgb1 & 0.013 & 0.084 & 0.087 & 95 \\ 
    \end{tabular}
  \vspace{.8cm}
\caption{Bias, empirical SE (`empSE'), mean estimated SE (`SEhat'), and CI coverage (`cover') for Scenario 16, where $E(n^B) =5000$, $M=50$, $n_{\rm house}=5$ and parametric nuisance models are misspecified.} 
\label{tab:scenario16}
\end{table}

\begin{table}[ht]
\centering
\begin{tabular}{rrrrr}
  & bias & empSE & SEhat & cover \\ 
  \hline
\hline
HT & 0.003 & 0.045 & 0.044 & 95 \\ 
  Haj & 0.001 & 0.033 & 0.032 & 95 \\ 
  naive & -0.151 & 0.029 &  &  \\ 
   \hline
DR1 & 0.104 & 0.057 & 0.059 & 57 \\ 
  DR2clw & 0.100 & 0.048 & 0.050 & 48 \\ 
  TMLE1 & 0.099 & 0.056 & 0.059 & 60 \\ 
  TMLE2 & 0.097 & 0.048 & 0.050 & 50 \\ 
  KH & 0.084 & 0.056 & 0.056 & 67 \\ 
   \hline
DR1.hal5 & 0.009 & 0.055 & 0.055 & 96 \\ 
  DR2.hal5 & 0.007 & 0.047 & 0.046 & 95 \\ 
  TMLE1.hal5 & 0.010 & 0.055 & 0.055 & 95 \\ 
  TMLE2.hal5 & 0.008 & 0.047 & 0.046 & 94 \\ 
   \hline
DR1.hal1 & 0.009 & 0.055 & 0.054 & 96 \\ 
  DR2.hal1 & 0.007 & 0.046 & 0.045 & 94 \\ 
  TMLE1.hal1 & 0.008 & 0.055 & 0.054 & 96 \\ 
  TMLE2.hal1 & 0.007 & 0.047 & 0.045 & 94 \\ 
   \hline
DR1.xgb5 & 0.016 & 0.055 & 0.054 & 95 \\ 
  DR2.xgb5 & 0.015 & 0.047 & 0.045 & 94 \\ 
  TMLE1.xgb5 & 0.016 & 0.056 & 0.055 & 95 \\ 
  TMLE2.xgb5 & 0.015 & 0.047 & 0.045 & 95 \\ 
   \hline
DR1.xgb1 & 0.019 & 0.055 & 0.053 & 94 \\ 
  DR2.xgb1 & 0.017 & 0.047 & 0.043 & 93 \\ 
  TMLE1.xgb1 & 0.017 & 0.055 & 0.053 & 94 \\ 
  TMLE2.xgb1 & 0.016 & 0.047 & 0.043 & 93 \\ 
    \end{tabular}
  \vspace{.8cm}
\caption{Bias, empirical SE (`empSE'), mean estimated SE (`SEhat'), and CI coverage (`cover') for Scenario 17, where $E(n^B) =1000$, $M=150$, $n_{\rm house}=20$ and parametric nuisance models are misspecified.} 
\label{tab:scenario17}
\end{table}

\begin{table}[ht]
\centering
\begin{tabular}{rrrrr}
  & bias & empSE & SEhat & cover \\ 
  \hline
\hline
HT & -0.007 & 0.084 & 0.084 & 96 \\ 
  Haj & -0.008 & 0.063 & 0.062 & 94 \\ 
  naive & -0.158 & 0.056 &  &  \\ 
   \hline
DR1 & 0.105 & 0.083 & 0.087 & 79 \\ 
  DR2clw & 0.102 & 0.061 & 0.064 & 66 \\ 
  TMLE1 & 0.099 & 0.084 & 0.087 & 80 \\ 
  TMLE2 & 0.097 & 0.061 & 0.064 & 69 \\ 
  KH & 0.083 & 0.084 & 0.085 & 87 \\ 
   \hline
DR1.hal5 & 0.011 & 0.083 & 0.083 & 94 \\ 
  DR2.hal5 & 0.009 & 0.063 & 0.061 & 93 \\ 
  TMLE1.hal5 & 0.011 & 0.083 & 0.083 & 94 \\ 
  TMLE2.hal5 & 0.009 & 0.062 & 0.061 & 93 \\ 
   \hline
DR1.hal1 & 0.009 & 0.083 & 0.083 & 94 \\ 
  DR2.hal1 & 0.007 & 0.062 & 0.060 & 93 \\ 
  TMLE1.hal1 & 0.009 & 0.083 & 0.083 & 94 \\ 
  TMLE2.hal1 & 0.007 & 0.062 & 0.060 & 93 \\ 
   \hline
DR1.xgb5 & 0.018 & 0.083 & 0.083 & 94 \\ 
  DR2.xgb5 & 0.016 & 0.062 & 0.061 & 94 \\ 
  TMLE1.xgb5 & 0.017 & 0.083 & 0.083 & 94 \\ 
  TMLE2.xgb5 & 0.015 & 0.062 & 0.061 & 94 \\ 
   \hline
DR1.xgb1 & 0.018 & 0.083 & 0.082 & 94 \\ 
  DR2.xgb1 & 0.016 & 0.063 & 0.059 & 92 \\ 
  TMLE1.xgb1 & 0.016 & 0.083 & 0.082 & 94 \\ 
  TMLE2.xgb1 & 0.015 & 0.063 & 0.059 & 93 \\ 
    \end{tabular}
  \vspace{.8cm}
\caption{Bias, empirical SE (`empSE'), mean estimated SE (`SEhat'), and CI coverage (`cover') for Scenario 18, where $E(n^B) =1000$, $M=150$, $n_{\rm house}=5$ and parametric nuisance models are misspecified.} 
\label{tab:scenario18}
\end{table}

\begin{table}[ht]
\centering
\begin{tabular}{rrrrr}
  & bias & empSE & SEhat & cover \\ 
  \hline
\hline
HT & 0.000 & 0.076 & 0.077 & 95 \\ 
  Haj & 0.001 & 0.054 & 0.057 & 96 \\ 
  naive & -0.151 & 0.047 &  &  \\ 
   \hline
DR1 & 0.103 & 0.079 & 0.082 & 76 \\ 
  DR2clw & 0.103 & 0.056 & 0.061 & 62 \\ 
  TMLE1 & 0.098 & 0.079 & 0.081 & 78 \\ 
  TMLE2 & 0.099 & 0.056 & 0.061 & 65 \\ 
  KH & 0.082 & 0.079 & 0.080 & 83 \\ 
   \hline
DR1.hal5 & 0.009 & 0.078 & 0.079 & 94 \\ 
  DR2.hal5 & 0.011 & 0.058 & 0.059 & 95 \\ 
  TMLE1.hal5 & 0.011 & 0.077 & 0.079 & 94 \\ 
  TMLE2.hal5 & 0.012 & 0.056 & 0.059 & 96 \\ 
   \hline
DR1.hal1 & 0.009 & 0.077 & 0.078 & 94 \\ 
  DR2.hal1 & 0.010 & 0.056 & 0.057 & 95 \\ 
  TMLE1.hal1 & 0.009 & 0.077 & 0.078 & 94 \\ 
  TMLE2.hal1 & 0.010 & 0.056 & 0.058 & 95 \\ 
   \hline
DR1.xgb5 & 0.017 & 0.079 & 0.078 & 94 \\ 
  DR2.xgb5 & 0.019 & 0.058 & 0.058 & 93 \\ 
  TMLE1.xgb5 & 0.017 & 0.079 & 0.079 & 94 \\ 
  TMLE2.xgb5 & 0.018 & 0.058 & 0.059 & 94 \\ 
   \hline
DR1.xgb1 & 0.019 & 0.079 & 0.077 & 92 \\ 
  DR2.xgb1 & 0.020 & 0.057 & 0.056 & 92 \\ 
  TMLE1.xgb1 & 0.017 & 0.079 & 0.077 & 94 \\ 
  TMLE2.xgb1 & 0.018 & 0.057 & 0.056 & 92 \\ 
    \end{tabular}
  \vspace{.8cm}
\caption{Bias, empirical SE (`empSE'), mean estimated SE (`SEhat'), and CI coverage (`cover') for Scenario 19, where $E(n^B) =1000$, $M=50$, $n_{\rm house}=20$ and parametric nuisance models are misspecified.} 
\label{tab:scenario19}
\end{table}

\begin{table}[ht]
\centering
\begin{tabular}{rrrrr}
  & bias & empSE & SEhat & cover \\ 
  \hline
\hline
HT & 0.009 & 0.142 & 0.146 & 96 \\ 
  Haj & 0.005 & 0.109 & 0.109 & 94 \\ 
  naive & -0.147 & 0.092 &  &  \\ 
   \hline
DR1 & 0.119 & 0.129 & 0.138 & 90 \\ 
  DR2clw & 0.112 & 0.087 & 0.093 & 80 \\ 
  TMLE1 & 0.109 & 0.131 & 0.137 & 91 \\ 
  TMLE2 & 0.106 & 0.088 & 0.093 & 82 \\ 
  KH & 0.092 & 0.133 & 0.136 & 92 \\ 
   \hline
DR1.hal5 & 0.022 & 0.135 & 0.136 & 95 \\ 
  DR2.hal5 & 0.018 & 0.098 & 0.094 & 93 \\ 
  TMLE1.hal5 & 0.022 & 0.131 & 0.135 & 96 \\ 
  TMLE2.hal5 & 0.019 & 0.093 & 0.093 & 94 \\ 
   \hline
DR1.hal1 & 0.020 & 0.130 & 0.134 & 95 \\ 
  DR2.hal1 & 0.017 & 0.092 & 0.091 & 94 \\ 
  TMLE1.hal1 & 0.020 & 0.130 & 0.134 & 95 \\ 
  TMLE2.hal1 & 0.017 & 0.092 & 0.091 & 94 \\ 
   \hline
DR1.xgb5 & 0.031 & 0.132 & 0.135 & 95 \\ 
  DR2.xgb5 & 0.027 & 0.094 & 0.092 & 92 \\ 
  TMLE1.xgb5 & 0.030 & 0.132 & 0.135 & 95 \\ 
  TMLE2.xgb5 & 0.027 & 0.094 & 0.092 & 94 \\ 
   \hline
DR1.xgb1 & 0.029 & 0.132 & 0.134 & 95 \\ 
  DR2.xgb1 & 0.026 & 0.093 & 0.091 & 93 \\ 
  TMLE1.xgb1 & 0.028 & 0.132 & 0.134 & 95 \\ 
  TMLE2.xgb1 & 0.024 & 0.093 & 0.091 & 93 \\ 
    \end{tabular}
  \vspace{.8cm}
\caption{Bias, empirical SE (`empSE'), mean estimated SE (`SEhat'), and CI coverage (`cover') for Scenario 20, where $E(n^B) =1000$, $M=50$, $n_{\rm house}=5$ and parametric nuisance models are misspecified.} 
\label{tab:scenario20}
\end{table}

\begin{table}[ht]
\centering
\begin{tabular}{rrrrr}
  & bias & empSE & SEhat & cover \\ 
  \hline
\hline
HT & 0.003 & 0.045 & 0.044 & 95 \\ 
  Haj & 0.001 & 0.033 & 0.032 & 95 \\ 
  naive & -0.151 & 0.029 &  &  \\ 
   \hline
DR1 & 0.105 & 0.070 & 0.074 & 73 \\ 
  DR2clw & 0.102 & 0.063 & 0.067 & 69 \\ 
  TMLE1 & 0.100 & 0.069 & 0.073 & 74 \\ 
  TMLE2 & 0.098 & 0.063 & 0.066 & 70 \\ 
  KH & 0.084 & 0.068 & 0.069 & 79 \\ 
   \hline
DR1.hal5 & 0.016 & 0.068 & 0.068 & 95 \\ 
  DR2.hal5 & 0.014 & 0.061 & 0.060 & 94 \\ 
  TMLE1.hal5 & 0.015 & 0.067 & 0.067 & 94 \\ 
  TMLE2.hal5 & 0.014 & 0.061 & 0.060 & 94 \\ 
   \hline
DR1.hal1 & 0.012 & 0.065 & 0.065 & 95 \\ 
  DR2.hal1 & 0.010 & 0.059 & 0.057 & 94 \\ 
  TMLE1.hal1 & 0.011 & 0.067 & 0.065 & 95 \\ 
  TMLE2.hal1 & 0.010 & 0.061 & 0.058 & 94 \\ 
   \hline
DR1.xgb5 & 0.020 & 0.068 & 0.067 & 94 \\ 
  DR2.xgb5 & 0.019 & 0.062 & 0.059 & 93 \\ 
  TMLE1.xgb5 & 0.020 & 0.069 & 0.067 & 93 \\ 
  TMLE2.xgb5 & 0.019 & 0.063 & 0.059 & 93 \\ 
   \hline
DR1.xgb1 & 0.024 & 0.066 & 0.062 & 92 \\ 
  DR2.xgb1 & 0.022 & 0.060 & 0.055 & 91 \\ 
  TMLE1.xgb1 & 0.021 & 0.066 & 0.062 & 92 \\ 
  TMLE2.xgb1 & 0.020 & 0.060 & 0.055 & 92 \\ 
    \end{tabular}
  \vspace{.8cm}
\caption{Bias, empirical SE (`empSE'), mean estimated SE (`SEhat'), and CI coverage (`cover') for Scenario 21, where $E(n^B) =500$, $M=150$, $n_{\rm house}=20$ and parametric nuisance models are misspecified.} 
\label{tab:scenario21}
\end{table}

\begin{table}[ht]
\centering
\begin{tabular}{rrrrr}
  & bias & empSE & SEhat & cover \\ 
  \hline
\hline
HT & -0.007 & 0.084 & 0.084 & 96 \\ 
  Haj & -0.008 & 0.063 & 0.062 & 94 \\ 
  naive & -0.158 & 0.056 &  &  \\ 
   \hline
DR1 & 0.109 & 0.095 & 0.098 & 84 \\ 
  DR2clw & 0.105 & 0.076 & 0.079 & 74 \\ 
  TMLE1 & 0.103 & 0.095 & 0.097 & 84 \\ 
  TMLE2 & 0.101 & 0.075 & 0.078 & 76 \\ 
  KH & 0.086 & 0.095 & 0.094 & 88 \\ 
   \hline
DR1.hal5 & 0.019 & 0.095 & 0.091 & 94 \\ 
  DR2.hal5 & 0.018 & 0.076 & 0.072 & 92 \\ 
  TMLE1.hal5 & 0.019 & 0.095 & 0.091 & 93 \\ 
  TMLE2.hal5 & 0.017 & 0.076 & 0.071 & 92 \\ 
   \hline
DR1.hal1 & 0.016 & 0.094 & 0.090 & 94 \\ 
  DR2.hal1 & 0.014 & 0.075 & 0.070 & 92 \\ 
  TMLE1.hal1 & 0.015 & 0.095 & 0.091 & 94 \\ 
  TMLE2.hal1 & 0.013 & 0.078 & 0.071 & 92 \\ 
   \hline
DR1.xgb5 & 0.026 & 0.096 & 0.091 & 94 \\ 
  DR2.xgb5 & 0.024 & 0.077 & 0.071 & 91 \\ 
  TMLE1.xgb5 & 0.026 & 0.097 & 0.091 & 94 \\ 
  TMLE2.xgb5 & 0.025 & 0.078 & 0.071 & 91 \\ 
   \hline
DR1.xgb1 & 0.029 & 0.094 & 0.088 & 93 \\ 
  DR2.xgb1 & 0.027 & 0.076 & 0.067 & 90 \\ 
  TMLE1.xgb1 & 0.027 & 0.094 & 0.088 & 93 \\ 
  TMLE2.xgb1 & 0.025 & 0.076 & 0.067 & 90 \\ 
    \end{tabular}
  \vspace{.8cm}
\caption{Bias, empirical SE (`empSE'), mean estimated SE (`SEhat'), and CI coverage (`cover') for Scenario 22, where $E(n^B) =500$, $M=150$, $n_{\rm house}=5$ and parametric nuisance models are misspecified.} 
\label{tab:scenario22}
\end{table}

\begin{table}[ht]
\centering
\begin{tabular}{rrrrr}
  & bias & empSE & SEhat & cover \\ 
  \hline
\hline
HT & 0.000 & 0.076 & 0.077 & 95 \\ 
  Haj & 0.001 & 0.054 & 0.057 & 96 \\ 
  naive & -0.151 & 0.047 &  &  \\ 
   \hline
DR1 & 0.102 & 0.091 & 0.093 & 82 \\ 
  DR2clw & 0.101 & 0.073 & 0.076 & 74 \\ 
  TMLE1 & 0.096 & 0.090 & 0.093 & 84 \\ 
  TMLE2 & 0.097 & 0.072 & 0.075 & 76 \\ 
  KH & 0.080 & 0.089 & 0.090 & 86 \\ 
   \hline
DR1.hal5 & 0.011 & 0.087 & 0.087 & 94 \\ 
  DR2.hal5 & 0.013 & 0.070 & 0.069 & 94 \\ 
  TMLE1.hal5 & 0.014 & 0.087 & 0.088 & 94 \\ 
  TMLE2.hal5 & 0.016 & 0.070 & 0.070 & 95 \\ 
   \hline
DR1.hal1 & 0.010 & 0.087 & 0.086 & 95 \\ 
  DR2.hal1 & 0.012 & 0.069 & 0.068 & 94 \\ 
  TMLE1.hal1 & 0.010 & 0.087 & 0.086 & 95 \\ 
  TMLE2.hal1 & 0.011 & 0.069 & 0.068 & 94 \\ 
   \hline
DR1.xgb5 & 0.019 & 0.088 & 0.087 & 94 \\ 
  DR2.xgb5 & 0.021 & 0.070 & 0.069 & 94 \\ 
  TMLE1.xgb5 & 0.020 & 0.087 & 0.088 & 94 \\ 
  TMLE2.xgb5 & 0.022 & 0.070 & 0.070 & 94 \\ 
   \hline
DR1.xgb1 & 0.024 & 0.087 & 0.084 & 94 \\ 
  DR2.xgb1 & 0.025 & 0.069 & 0.065 & 92 \\ 
  TMLE1.xgb1 & 0.022 & 0.087 & 0.084 & 94 \\ 
  TMLE2.xgb1 & 0.023 & 0.069 & 0.065 & 92 \\ 
    \end{tabular}
  \vspace{.8cm}
\caption{Bias, empirical SE (`empSE'), mean estimated SE (`SEhat'), and CI coverage (`cover') for Scenario 23, where $E(n^B) =500$, $M=50$, $n_{\rm house}=20$ and parametric nuisance models are misspecified.} 
\label{tab:scenario23}
\end{table}

\begin{table}[ht]
\centering
\begin{tabular}{rrrrr}
  & bias & empSE & SEhat & cover \\ 
  \hline
\hline
HT & 0.009 & 0.142 & 0.146 & 96 \\ 
  Haj & 0.005 & 0.109 & 0.109 & 94 \\ 
  naive & -0.147 & 0.092 &  &  \\ 
   \hline
DR1 & 0.118 & 0.134 & 0.146 & 91 \\ 
  DR2clw & 0.111 & 0.095 & 0.105 & 82 \\ 
  TMLE1 & 0.108 & 0.135 & 0.145 & 92 \\ 
  TMLE2 & 0.105 & 0.095 & 0.104 & 84 \\ 
  KH & 0.091 & 0.137 & 0.142 & 92 \\ 
   \hline
DR1.hal5 & 0.027 & 0.138 & 0.139 & 94 \\ 
  DR2.hal5 & 0.024 & 0.103 & 0.098 & 93 \\ 
  TMLE1.hal5 & 0.026 & 0.138 & 0.139 & 94 \\ 
  TMLE2.hal5 & 0.023 & 0.102 & 0.099 & 92 \\ 
   \hline
DR1.hal1 & 0.024 & 0.137 & 0.138 & 93 \\ 
  DR2.hal1 & 0.021 & 0.101 & 0.097 & 93 \\ 
  TMLE1.hal1 & 0.023 & 0.137 & 0.138 & 93 \\ 
  TMLE2.hal1 & 0.020 & 0.101 & 0.098 & 93 \\ 
   \hline
DR1.xgb5 & 0.039 & 0.140 & 0.139 & 94 \\ 
  DR2.xgb5 & 0.036 & 0.103 & 0.099 & 92 \\ 
  TMLE1.xgb5 & 0.038 & 0.139 & 0.140 & 94 \\ 
  TMLE2.xgb5 & 0.035 & 0.103 & 0.100 & 92 \\ 
   \hline
DR1.xgb1 & 0.038 & 0.139 & 0.137 & 92 \\ 
  DR2.xgb1 & 0.034 & 0.102 & 0.096 & 91 \\ 
  TMLE1.xgb1 & 0.036 & 0.139 & 0.137 & 93 \\ 
  TMLE2.xgb1 & 0.033 & 0.102 & 0.096 & 91 \\ 
    \end{tabular}
  \vspace{.8cm}
\caption{Bias, empirical SE (`empSE'), mean estimated SE (`SEhat'), and CI coverage (`cover') for Scenario 24, where $E(n^B) =500$, $M=50$, $n_{\rm house}=5$ and parametric nuisance models are misspecified.} 
\label{tab:scenario24}
\end{table}

\begin{table}[ht]
\centering
\begin{tabular}{rrrrr}
  & bias & empSE & SEhat & cover \\ 
  \hline
\hline
HT & 0.03 & 0.63 & 0.66 & 96 \\ 
  Haj & 0.03 & 0.61 & 0.65 & 96 \\ 
  naive & -1.94 & 0.48 &  &  \\ 
   \hline
DR1 & 0.02 & 0.49 & 0.51 & 96 \\ 
  DR2clw & 0.02 & 0.46 & 0.48 & 95 \\ 
  TMLE1 & 0.02 & 0.49 & 0.51 & 96 \\ 
  TMLE2 & 0.02 & 0.46 & 0.48 & 95 \\ 
  KH & 0.02 & 0.49 & 0.51 & 96 \\ 
   \hline
DR1.hal5 & 0.08 & 0.49 & 0.50 & 96 \\ 
  DR2.hal5 & 0.08 & 0.47 & 0.48 & 96 \\ 
  TMLE1.hal5 & 0.08 & 0.49 & 0.50 & 96 \\ 
  TMLE2.hal5 & 0.08 & 0.47 & 0.48 & 96 \\ 
   \hline
DR1.hal1 & 0.06 & 0.49 & 0.50 & 96 \\ 
  DR2.hal1 & 0.06 & 0.47 & 0.48 & 95 \\ 
  TMLE1.hal1 & 0.06 & 0.49 & 0.50 & 96 \\ 
  TMLE2.hal1 & 0.06 & 0.47 & 0.48 & 95 \\ 
   \hline
DR1.xgb5 & 0.17 & 0.49 & 0.52 & 96 \\ 
  DR2.xgb5 & 0.17 & 0.47 & 0.50 & 96 \\ 
  TMLE1.xgb5 & 0.21 & 0.49 & 0.52 & 95 \\ 
  TMLE2.xgb5 & 0.21 & 0.46 & 0.50 & 94 \\ 
   \hline
DR1.xgb1 & 0.14 & 0.49 & 0.51 & 96 \\ 
  DR2.xgb1 & 0.14 & 0.47 & 0.49 & 95 \\ 
  TMLE1.xgb1 & 0.15 & 0.49 & 0.51 & 95 \\ 
  TMLE2.xgb1 & 0.16 & 0.47 & 0.49 & 95 \\ 
    \end{tabular}
  \vspace{.8cm}
\caption{Bias $\times 100$, empirical SE $\times 100$ (`empSE'), mean estimated SE $\times 100$ (`SEhat') and CI coverage (`cover') for Scenario 1b, where $E(n^B) =5000$, $M=150$, $n_{\rm house}=20$ and parametric nuisance models are correctly specified.} 
\label{tab:scenario1b}
\end{table}

\begin{table}[ht]
\centering
\begin{tabular}{rrrrr}
  & bias & empSE & SEhat & cover \\ 
  \hline
\hline
HT & -0.07 & 1.28 & 1.30 & 94 \\ 
  Haj & -0.07 & 1.25 & 1.27 & 94 \\ 
  naive & -2.03 & 0.98 &  &  \\ 
   \hline
DR1 & -0.03 & 0.73 & 0.72 & 93 \\ 
  DR2clw & -0.03 & 0.67 & 0.67 & 94 \\ 
  TMLE1 & -0.03 & 0.73 & 0.72 & 93 \\ 
  TMLE2 & -0.03 & 0.67 & 0.67 & 94 \\ 
  KH & -0.03 & 0.73 & 0.72 & 93 \\ 
   \hline
DR1.hal5 & 0.06 & 0.73 & 0.71 & 94 \\ 
  DR2.hal5 & 0.06 & 0.67 & 0.66 & 94 \\ 
  TMLE1.hal5 & 0.06 & 0.73 & 0.71 & 94 \\ 
  TMLE2.hal5 & 0.06 & 0.67 & 0.66 & 94 \\ 
   \hline
DR1.hal1 & 0.04 & 0.72 & 0.71 & 94 \\ 
  DR2.hal1 & 0.04 & 0.66 & 0.66 & 94 \\ 
  TMLE1.hal1 & 0.04 & 0.73 & 0.71 & 93 \\ 
  TMLE2.hal1 & 0.04 & 0.66 & 0.66 & 94 \\ 
   \hline
DR1.xgb5 & 0.20 & 0.74 & 0.73 & 94 \\ 
  DR2.xgb5 & 0.19 & 0.68 & 0.67 & 94 \\ 
  TMLE1.xgb5 & 0.26 & 0.73 & 0.72 & 94 \\ 
  TMLE2.xgb5 & 0.26 & 0.67 & 0.67 & 94 \\ 
   \hline
DR1.xgb1 & 0.16 & 0.73 & 0.72 & 93 \\ 
  DR2.xgb1 & 0.16 & 0.67 & 0.67 & 94 \\ 
  TMLE1.xgb1 & 0.19 & 0.73 & 0.72 & 93 \\ 
  TMLE2.xgb1 & 0.19 & 0.67 & 0.66 & 95 \\ 
    \end{tabular}
  \vspace{.8cm}
\caption{Bias $\times 100$, empirical SE $\times 100$ (`empSE'), mean estimated SE $\times 100$ (`SEhat') and CI coverage (`cover') for Scenario 2b, where $E(n^B) =5000$, $M=150$, $n_{\rm house}=5$ and parametric nuisance models are correctly specified.} 
\label{tab:scenario2b}
\end{table}

\begin{table}[ht]
\centering
\begin{tabular}{rrrrr}
  & bias & empSE & SEhat & cover \\ 
  \hline
\hline
HT & 0.02 & 1.17 & 1.14 & 95 \\ 
  Haj & 0.03 & 1.14 & 1.12 & 96 \\ 
  naive & -1.93 & 0.89 &  &  \\ 
   \hline
DR1 & 0.00 & 0.73 & 0.70 & 94 \\ 
  DR2clw & 0.01 & 0.68 & 0.65 & 94 \\ 
  TMLE1 & 0.00 & 0.73 & 0.70 & 94 \\ 
  TMLE2 & 0.01 & 0.68 & 0.65 & 94 \\ 
  KH & 0.00 & 0.73 & 0.70 & 94 \\ 
   \hline
DR1.hal5 & 0.09 & 0.73 & 0.68 & 93 \\ 
  DR2.hal5 & 0.10 & 0.68 & 0.64 & 94 \\ 
  TMLE1.hal5 & 0.10 & 0.73 & 0.69 & 94 \\ 
  TMLE2.hal5 & 0.10 & 0.68 & 0.64 & 94 \\ 
   \hline
DR1.hal1 & 0.07 & 0.73 & 0.69 & 93 \\ 
  DR2.hal1 & 0.07 & 0.68 & 0.64 & 93 \\ 
  TMLE1.hal1 & 0.07 & 0.73 & 0.68 & 93 \\ 
  TMLE2.hal1 & 0.07 & 0.68 & 0.64 & 93 \\ 
   \hline
DR1.xgb5 & 0.22 & 0.73 & 0.70 & 94 \\ 
  DR2.xgb5 & 0.22 & 0.68 & 0.65 & 93 \\ 
  TMLE1.xgb5 & 0.28 & 0.73 & 0.70 & 94 \\ 
  TMLE2.xgb5 & 0.28 & 0.68 & 0.66 & 93 \\ 
   \hline
DR1.xgb1 & 0.18 & 0.74 & 0.70 & 95 \\ 
  DR2.xgb1 & 0.19 & 0.69 & 0.65 & 93 \\ 
  TMLE1.xgb1 & 0.20 & 0.74 & 0.69 & 94 \\ 
  TMLE2.xgb1 & 0.21 & 0.69 & 0.64 & 93 \\ 
    \end{tabular}
  \vspace{.8cm}
\caption{Bias $\times 100$, empirical SE $\times 100$ (`empSE'), mean estimated SE $\times 100$ (`SEhat') and CI coverage (`cover') for Scenario 3b, where $E(n^B) =5000$, $M=50$, $n_{\rm house}=20$ and parametric nuisance models are correctly specified.} 
\label{tab:scenario3b}
\end{table}

\begin{table}[ht]
\centering
\begin{tabular}{rrrrr}
  & bias & empSE & SEhat & cover \\ 
  \hline
\hline
HT & 0.07 & 2.21 & 2.25 & 95 \\ 
  Haj & 0.08 & 2.16 & 2.20 & 95 \\ 
  naive & -1.90 & 1.74 &  &  \\ 
   \hline
DR1 & 0.03 & 1.11 & 1.13 & 94 \\ 
  DR2clw & 0.04 & 1.02 & 1.03 & 94 \\ 
  TMLE1 & 0.03 & 1.11 & 1.13 & 94 \\ 
  TMLE2 & 0.04 & 1.02 & 1.03 & 94 \\ 
  KH & 0.03 & 1.11 & 1.12 & 94 \\ 
   \hline
DR1.hal5 & 0.15 & 1.12 & 1.11 & 94 \\ 
  DR2.hal5 & 0.16 & 1.03 & 1.00 & 94 \\ 
  TMLE1.hal5 & 0.16 & 1.11 & 1.10 & 95 \\ 
  TMLE2.hal5 & 0.17 & 1.02 & 1.00 & 94 \\ 
   \hline
DR1.hal1 & 0.13 & 1.10 & 1.10 & 94 \\ 
  DR2.hal1 & 0.14 & 1.02 & 1.00 & 94 \\ 
  TMLE1.hal1 & 0.13 & 1.11 & 1.10 & 94 \\ 
  TMLE2.hal1 & 0.14 & 1.02 & 1.00 & 94 \\ 
   \hline
DR1.xgb5 & 0.32 & 1.12 & 1.11 & 95 \\ 
  DR2.xgb5 & 0.33 & 1.02 & 1.00 & 94 \\ 
  TMLE1.xgb5 & 0.40 & 1.10 & 1.10 & 95 \\ 
  TMLE2.xgb5 & 0.42 & 1.00 & 1.00 & 94 \\ 
   \hline
DR1.xgb1 & 0.27 & 1.10 & 1.11 & 96 \\ 
  DR2.xgb1 & 0.28 & 1.01 & 1.00 & 94 \\ 
  TMLE1.xgb1 & 0.29 & 1.10 & 1.09 & 95 \\ 
  TMLE2.xgb1 & 0.31 & 1.00 & 0.99 & 94 \\ 
    \end{tabular}
  \vspace{.8cm}
\caption{Bias $\times 100$, empirical SE $\times 100$ (`empSE'), mean estimated SE $\times 100$ (`SEhat') and CI coverage (`cover') for Scenario 4b, where $E(n^B) =5000$, $M=50$, $n_{\rm house}=5$ and parametric nuisance models are correctly specified.} 
\label{tab:scenario4b}
\end{table}

\begin{table}[ht]
\centering
\begin{tabular}{rrrrr}
  & bias & empSE & SEhat & cover \\ 
  \hline
\hline
HT & 0.03 & 0.63 & 0.66 & 96 \\ 
  Haj & 0.03 & 0.61 & 0.65 & 96 \\ 
  naive & -1.94 & 0.48 &  &  \\ 
   \hline
DR1 & 0.01 & 0.93 & 0.91 & 96 \\ 
  DR2clw & 0.01 & 0.92 & 0.90 & 96 \\ 
  TMLE1 & 0.01 & 0.93 & 0.91 & 96 \\ 
  TMLE2 & 0.01 & 0.92 & 0.90 & 96 \\ 
  KH & 0.01 & 0.93 & 0.91 & 96 \\ 
   \hline
DR1.hal5 & 0.11 & 0.97 & 0.93 & 95 \\ 
  DR2.hal5 & 0.11 & 0.96 & 0.92 & 95 \\ 
  TMLE1.hal5 & 0.13 & 0.93 & 0.92 & 95 \\ 
  TMLE2.hal5 & 0.13 & 0.92 & 0.91 & 95 \\ 
   \hline
DR1.hal1 & 0.09 & 0.93 & 0.90 & 95 \\ 
  DR2.hal1 & 0.09 & 0.93 & 0.89 & 95 \\ 
  TMLE1.hal1 & 0.08 & 0.93 & 0.90 & 95 \\ 
  TMLE2.hal1 & 0.08 & 0.92 & 0.89 & 95 \\ 
   \hline
DR1.xgb5 & 0.25 & 0.99 & 0.99 & 95 \\ 
  DR2.xgb5 & 0.26 & 0.98 & 0.98 & 96 \\ 
  TMLE1.xgb5 & 0.36 & 0.96 & 0.98 & 95 \\ 
  TMLE2.xgb5 & 0.36 & 0.95 & 0.97 & 95 \\ 
   \hline
DR1.xgb1 & 0.24 & 0.96 & 0.94 & 94 \\ 
  DR2.xgb1 & 0.24 & 0.96 & 0.93 & 94 \\ 
  TMLE1.xgb1 & 0.20 & 0.95 & 0.93 & 95 \\ 
  TMLE2.xgb1 & 0.20 & 0.95 & 0.92 & 94 \\ 
    \end{tabular}
  \vspace{.8cm}
\caption{Bias $\times 100$, empirical SE $\times 100$ (`empSE'), mean estimated SE $\times 100$ (`SEhat') and CI coverage (`cover') for Scenario 5b, where $E(n^B) =1000$, $M=150$, $n_{\rm house}=20$ and parametric nuisance models are correctly specified.} 
\label{tab:scenario5b}
\end{table}

\begin{table}[ht]
\centering
\begin{tabular}{rrrrr}
  & bias & empSE & SEhat & cover \\ 
  \hline
\hline
HT & -0.07 & 1.28 & 1.30 & 94 \\ 
  Haj & -0.07 & 1.25 & 1.27 & 94 \\ 
  naive & -2.03 & 0.98 &  &  \\ 
   \hline
DR1 & 0.01 & 1.07 & 1.06 & 94 \\ 
  DR2clw & 0.01 & 1.03 & 1.02 & 94 \\ 
  TMLE1 & 0.01 & 1.07 & 1.06 & 94 \\ 
  TMLE2 & 0.01 & 1.03 & 1.02 & 94 \\ 
  KH & 0.01 & 1.07 & 1.06 & 94 \\ 
   \hline
DR1.hal5 & 0.17 & 1.08 & 1.05 & 94 \\ 
  DR2.hal5 & 0.17 & 1.04 & 1.01 & 94 \\ 
  TMLE1.hal5 & 0.18 & 1.10 & 1.05 & 94 \\ 
  TMLE2.hal5 & 0.18 & 1.06 & 1.01 & 94 \\ 
   \hline
DR1.hal1 & 0.13 & 1.07 & 1.03 & 94 \\ 
  DR2.hal1 & 0.13 & 1.03 & 1.00 & 94 \\ 
  TMLE1.hal1 & 0.12 & 1.07 & 1.03 & 93 \\ 
  TMLE2.hal1 & 0.12 & 1.03 & 1.00 & 94 \\ 
   \hline
DR1.xgb5 & 0.43 & 1.14 & 1.14 & 94 \\ 
  DR2.xgb5 & 0.43 & 1.10 & 1.10 & 94 \\ 
  TMLE1.xgb5 & 0.58 & 1.13 & 1.13 & 92 \\ 
  TMLE2.xgb5 & 0.58 & 1.09 & 1.09 & 92 \\ 
   \hline
DR1.xgb1 & 0.37 & 1.11 & 1.09 & 94 \\ 
  DR2.xgb1 & 0.37 & 1.07 & 1.05 & 93 \\ 
  TMLE1.xgb1 & 0.38 & 1.10 & 1.08 & 94 \\ 
  TMLE2.xgb1 & 0.38 & 1.06 & 1.05 & 93 \\ 
    \end{tabular}
  \vspace{.8cm}
\caption{Bias $\times 100$, empirical SE $\times 100$ (`empSE'), mean estimated SE $\times 100$ (`SEhat') and CI coverage (`cover') for Scenario 6b, where $E(n^B) =1000$, $M=150$, $n_{\rm house}=5$ and parametric nuisance models are correctly specified.} 
\label{tab:scenario6b}
\end{table}

\begin{table}[ht]
\centering
\begin{tabular}{rrrrr}
  & bias & empSE & SEhat & cover \\ 
  \hline
\hline
HT & 0.02 & 1.17 & 1.14 & 95 \\ 
  Haj & 0.03 & 1.14 & 1.12 & 96 \\ 
  naive & -1.93 & 0.89 &  &  \\ 
   \hline
DR1 & -0.01 & 1.09 & 1.03 & 94 \\ 
  DR2clw & -0.00 & 1.05 & 1.00 & 94 \\ 
  TMLE1 & -0.01 & 1.09 & 1.03 & 94 \\ 
  TMLE2 & -0.00 & 1.05 & 1.00 & 94 \\ 
  KH & -0.01 & 1.09 & 1.03 & 94 \\ 
   \hline
DR1.hal5 & 0.17 & 1.13 & 1.03 & 93 \\ 
  DR2.hal5 & 0.17 & 1.10 & 1.00 & 94 \\ 
  TMLE1.hal5 & 0.18 & 1.11 & 1.04 & 93 \\ 
  TMLE2.hal5 & 0.19 & 1.07 & 1.01 & 93 \\ 
   \hline
DR1.hal1 & 0.12 & 1.11 & 1.01 & 92 \\ 
  DR2.hal1 & 0.13 & 1.07 & 0.98 & 93 \\ 
  TMLE1.hal1 & 0.11 & 1.11 & 1.01 & 92 \\ 
  TMLE2.hal1 & 0.11 & 1.07 & 0.98 & 93 \\ 
   \hline
DR1.xgb5 & 0.39 & 1.14 & 1.12 & 93 \\ 
  DR2.xgb5 & 0.39 & 1.10 & 1.08 & 93 \\ 
  TMLE1.xgb5 & 0.54 & 1.12 & 1.12 & 93 \\ 
  TMLE2.xgb5 & 0.54 & 1.08 & 1.09 & 94 \\ 
   \hline
DR1.xgb1 & 0.34 & 1.14 & 1.06 & 93 \\ 
  DR2.xgb1 & 0.34 & 1.10 & 1.03 & 92 \\ 
  TMLE1.xgb1 & 0.34 & 1.13 & 1.05 & 92 \\ 
  TMLE2.xgb1 & 0.34 & 1.09 & 1.02 & 93 \\ 
    \end{tabular}
  \vspace{.8cm}
\caption{Bias $\times 100$, empirical SE $\times 100$ (`empSE'), mean estimated SE $\times 100$ (`SEhat') and CI coverage (`cover') for Scenario 7b, where $E(n^B) =1000$, $M=50$, $n_{\rm house}=20$ and parametric nuisance models are correctly specified.} 
\label{tab:scenario7b}
\end{table}

\begin{table}[ht]
\centering
\begin{tabular}{rrrrr}
  & bias & empSE & SEhat & cover \\ 
  \hline
\hline
HT & 0.07 & 2.21 & 2.25 & 95 \\ 
  Haj & 0.08 & 2.16 & 2.20 & 95 \\ 
  naive & -1.90 & 1.74 &  &  \\ 
   \hline
DR1 & 0.05 & 1.38 & 1.39 & 95 \\ 
  DR2clw & 0.06 & 1.31 & 1.31 & 94 \\ 
  TMLE1 & 0.05 & 1.38 & 1.39 & 95 \\ 
  TMLE2 & 0.06 & 1.31 & 1.31 & 94 \\ 
  KH & 0.05 & 1.38 & 1.38 & 94 \\ 
   \hline
DR1.hal5 & 0.30 & 1.42 & 1.34 & 94 \\ 
  DR2.hal5 & 0.31 & 1.35 & 1.26 & 94 \\ 
  TMLE1.hal5 & 0.33 & 1.39 & 1.35 & 95 \\ 
  TMLE2.hal5 & 0.34 & 1.32 & 1.27 & 94 \\ 
   \hline
DR1.hal1 & 0.24 & 1.38 & 1.33 & 94 \\ 
  DR2.hal1 & 0.26 & 1.32 & 1.24 & 93 \\ 
  TMLE1.hal1 & 0.23 & 1.38 & 1.32 & 93 \\ 
  TMLE2.hal1 & 0.24 & 1.32 & 1.24 & 94 \\ 
   \hline
DR1.xgb5 & 0.62 & 1.46 & 1.45 & 94 \\ 
  DR2.xgb5 & 0.64 & 1.39 & 1.36 & 94 \\ 
  TMLE1.xgb5 & 0.81 & 1.40 & 1.43 & 93 \\ 
  TMLE2.xgb5 & 0.83 & 1.34 & 1.35 & 94 \\ 
   \hline
DR1.xgb1 & 0.52 & 1.40 & 1.39 & 95 \\ 
  DR2.xgb1 & 0.53 & 1.33 & 1.30 & 94 \\ 
  TMLE1.xgb1 & 0.54 & 1.39 & 1.36 & 94 \\ 
  TMLE2.xgb1 & 0.55 & 1.32 & 1.28 & 94 \\ 
    \end{tabular}
  \vspace{.8cm}
\caption{Bias $\times 100$, empirical SE $\times 100$ (`empSE'), mean estimated SE $\times 100$ (`SEhat') and CI coverage (`cover') for Scenario 8b, where $E(n^B) =1000$, $M=50$, $n_{\rm house}=5$ and parametric nuisance models are correctly specified.} 
\label{tab:scenario8b}
\end{table}

\begin{table}[ht]
\centering
\begin{tabular}{rrrrr}
  & bias & empSE & SEhat & cover \\ 
  \hline
\hline
HT & 0.03 & 0.63 & 0.66 & 96 \\ 
  Haj & 0.03 & 0.61 & 0.65 & 96 \\ 
  naive & -1.94 & 0.48 &  &  \\ 
   \hline
DR1 & 0.02 & 1.25 & 1.25 & 94 \\ 
  DR2clw & 0.02 & 1.25 & 1.24 & 94 \\ 
  TMLE1 & 0.02 & 1.25 & 1.25 & 94 \\ 
  TMLE2 & 0.02 & 1.25 & 1.24 & 94 \\ 
  KH & 0.02 & 1.25 & 1.24 & 94 \\ 
   \hline
DR1.hal5 & 0.14 & 1.30 & 1.29 & 94 \\ 
  DR2.hal5 & 0.15 & 1.29 & 1.28 & 94 \\ 
  TMLE1.hal5 & 0.21 & 1.27 & 1.29 & 94 \\ 
  TMLE2.hal5 & 0.21 & 1.26 & 1.28 & 94 \\ 
   \hline
DR1.hal1 & 0.16 & 1.27 & 1.23 & 94 \\ 
  DR2.hal1 & 0.16 & 1.27 & 1.22 & 94 \\ 
  TMLE1.hal1 & 0.13 & 1.27 & 1.23 & 94 \\ 
  TMLE2.hal1 & 0.13 & 1.26 & 1.22 & 94 \\ 
   \hline
DR1.xgb5 & 0.31 & 1.36 & 1.40 & 96 \\ 
  DR2.xgb5 & 0.32 & 1.35 & 1.39 & 96 \\ 
  TMLE1.xgb5 & 0.50 & 1.33 & 1.37 & 96 \\ 
  TMLE2.xgb5 & 0.50 & 1.32 & 1.36 & 96 \\ 
   \hline
DR1.xgb1 & 0.34 & 1.32 & 1.28 & 94 \\ 
  DR2.xgb1 & 0.35 & 1.31 & 1.27 & 94 \\ 
  TMLE1.xgb1 & 0.21 & 1.28 & 1.27 & 94 \\ 
  TMLE2.xgb1 & 0.21 & 1.27 & 1.26 & 95 \\ 
    \end{tabular}
  \vspace{.8cm}
\caption{Bias $\times 100$, empirical SE $\times 100$ (`empSE'), mean estimated SE $\times 100$ (`SEhat') and CI coverage (`cover') for Scenario 9b, where $E(n^B) =500$, $M=150$, $n_{\rm house}=20$ and parametric nuisance models are correctly specified.} 
\label{tab:scenario9b}
\end{table}

\begin{table}[ht]
\centering
\begin{tabular}{rrrrr}
  & bias & empSE & SEhat & cover \\ 
  \hline
\hline
HT & -0.07 & 1.28 & 1.30 & 94 \\ 
  Haj & -0.07 & 1.25 & 1.27 & 94 \\ 
  naive & -2.03 & 0.98 &  &  \\ 
   \hline
DR1 & 0.08 & 1.40 & 1.37 & 93 \\ 
  DR2clw & 0.08 & 1.36 & 1.34 & 94 \\ 
  TMLE1 & 0.08 & 1.40 & 1.37 & 93 \\ 
  TMLE2 & 0.08 & 1.36 & 1.34 & 94 \\ 
  KH & 0.08 & 1.40 & 1.36 & 93 \\ 
   \hline
DR1.hal5 & 0.30 & 1.41 & 1.38 & 94 \\ 
  DR2.hal5 & 0.29 & 1.37 & 1.35 & 94 \\ 
  TMLE1.hal5 & 0.35 & 1.41 & 1.37 & 92 \\ 
  TMLE2.hal5 & 0.35 & 1.38 & 1.35 & 93 \\ 
   \hline
DR1.hal1 & 0.24 & 1.40 & 1.33 & 93 \\ 
  DR2.hal1 & 0.24 & 1.36 & 1.31 & 93 \\ 
  TMLE1.hal1 & 0.21 & 1.40 & 1.33 & 93 \\ 
  TMLE2.hal1 & 0.21 & 1.36 & 1.30 & 93 \\ 
   \hline
DR1.xgb5 & 0.62 & 1.53 & 1.53 & 95 \\ 
  DR2.xgb5 & 0.61 & 1.49 & 1.50 & 95 \\ 
  TMLE1.xgb5 & 0.86 & 1.48 & 1.50 & 94 \\ 
  TMLE2.xgb5 & 0.86 & 1.44 & 1.48 & 95 \\ 
   \hline
DR1.xgb1 & 0.55 & 1.45 & 1.42 & 93 \\ 
  DR2.xgb1 & 0.55 & 1.41 & 1.39 & 93 \\ 
  TMLE1.xgb1 & 0.49 & 1.42 & 1.40 & 94 \\ 
  TMLE2.xgb1 & 0.49 & 1.38 & 1.38 & 93 \\ 
    \end{tabular}%
  \vspace{.8cm}
\caption{Bias $\times 100$, empirical SE $\times 100$ (`empSE'), mean estimated SE $\times 100$ (`SEhat') and CI coverage (`cover') for Scenario 13b, where $E(n^B) =5000$, $M=150$, $n_{\rm house}=20$ and parametric nuisance models are misspecified.} 
\label{tab:scenario13b}
\end{table}

\begin{table}[ht]
\centering
\begin{tabular}{rrrrr}
  & bias & empSE & SEhat & cover \\ 
  \hline
\hline
HT & -0.05 & 1.23 & 1.28 & 95 \\ 
  Haj & -0.05 & 1.21 & 1.25 & 95 \\ 
  naive & -1.56 & 0.99 &  &  \\ 
   \hline
DR1 & 0.63 & 0.79 & 0.76 & 88 \\ 
  DR2clw & 0.62 & 0.73 & 0.72 & 89 \\ 
  TMLE1 & 0.64 & 0.79 & 0.76 & 88 \\ 
  TMLE2 & 0.64 & 0.73 & 0.71 & 88 \\ 
  KH & 0.73 & 0.79 & 0.77 & 86 \\ 
   \hline
DR1.hal5 & 0.12 & 0.83 & 0.78 & 92 \\ 
  DR2.hal5 & 0.12 & 0.78 & 0.74 & 92 \\ 
  TMLE1.hal5 & 0.13 & 0.83 & 0.78 & 93 \\ 
  TMLE2.hal5 & 0.12 & 0.77 & 0.74 & 93 \\ 
   \hline
DR1.hal1 & 0.10 & 0.83 & 0.79 & 94 \\ 
  DR2.hal1 & 0.10 & 0.78 & 0.75 & 92 \\ 
  TMLE1.hal1 & 0.10 & 0.82 & 0.79 & 94 \\ 
  TMLE2.hal1 & 0.10 & 0.77 & 0.75 & 93 \\ 
   \hline
DR1.xgb5 & 0.19 & 0.83 & 0.79 & 94 \\ 
  DR2.xgb5 & 0.19 & 0.77 & 0.75 & 94 \\ 
  TMLE1.xgb5 & 0.23 & 0.82 & 0.78 & 92 \\ 
  TMLE2.xgb5 & 0.23 & 0.77 & 0.74 & 94 \\ 
   \hline
DR1.xgb1 & 0.17 & 0.82 & 0.78 & 92 \\ 
  DR2.xgb1 & 0.17 & 0.77 & 0.74 & 92 \\ 
  TMLE1.xgb1 & 0.18 & 0.82 & 0.78 & 92 \\ 
  TMLE2.xgb1 & 0.18 & 0.77 & 0.74 & 92 \\ 
    \end{tabular}
  \vspace{.8cm}
\caption{Bias $\times 100$, empirical SE $\times 100$ (`empSE'), mean estimated SE $\times 100$ (`SEhat') and CI coverage (`cover') for Scenario 14b, where $E(n^B) =5000$, $M=150$, $n_{\rm house}=5$ and parametric nuisance models are misspecified.} 
\label{tab:scenario14b}
\end{table}

\begin{table}[ht]
\centering
%
  \vspace{.8cm}
\caption{Bias $\times 100$, empirical SE $\times 100$ (`empSE'), mean estimated SE $\times 100$ (`SEhat') and CI coverage (`cover') for Scenario 15b, where $E(n^B) =5000$, $M=50$, $n_{\rm house}=20$ and parametric nuisance models are misspecified.} 
\label{tab:scenario15b}
\end{table}

\begin{table}[ht]
\centering
%
  \vspace{.8cm}
\caption{Bias $\times 100$, empirical SE $\times 100$ (`empSE'), mean estimated SE $\times 100$ (`SEhat') and CI coverage (`cover') for Scenario 16b, where $E(n^B) =5000$, $M=50$, $n_{\rm house}=5$ and parametric nuisance models are misspecified.} 
\label{tab:scenario16b}
\end{table}

\clearpage

\begin{table}[ht]
\centering
%
  \vspace{.8cm}
\caption{Bias $\times 100$, empirical SE $\times 100$ (`empSE'), mean estimated SE $\times 100$ (`SEhat') and CI coverage (`cover') for Scenario 24b, where $E(n^B) =500$, $M=50$, $n_{\rm house}=5$ and parametric nuisance models are misspecified.} 
\label{tab:scenario24b}
\end{table}

\end{document}